\documentclass{aa}  

\usepackage{color}

\usepackage{soul}
\usepackage{graphicx}
\usepackage{txfonts}
\usepackage{hyperref}
\usepackage{booktabs}
\usepackage{adjustbox}
\usepackage{siunitx}
\usepackage{threeparttable}
\usepackage{gensymb}
\begin{document}

\title{Galaxies OBserved as Low-luminosity Identified Nebulae (GOBLIN): a catalog of 43,000 high-probability dwarf galaxy candidates in the UNIONS survey}
   \subtitle{}

   \author{Nick Heesters
        \inst{1}
        \and
        David Chemaly
        \inst{2}
        \and
        Oliver Müller
        \inst{1, 2, 3}
        \and
        Elisabeth Sola
        \inst{2}
        \and
        S\'ebastien Fabbro
        \inst{4}
        \and
        Ashley Ferreira
        \inst{5}
        \and
        Alan W. McConnachie
        \inst{4}
        \and
        Eugene Magnier
        \inst{6}
        \and
        Michael J. Hudson
        \inst{5, 7, 8}
        \and
        Kenneth Chambers
        \inst{6}
        \and
        François Hammer
        \inst{9}
        \and
        Ruben Sanchez-Janssen
        \inst{10}
        }

   \institute{Institute of Physics, Laboratory of Astrophysics, Ecole Polytechnique F\'ed\'erale de Lausanne (EPFL), 1290 Sauverny, Switzerland\\
   \email{nick.heesters@epfl.ch}
   \and
    Institute of Astronomy, Madingley Road, Cambridge, CB3 0HA, UK
    \and
    Visiting Fellow, Clare Hall, University of Cambridge, Cambridge, UK
    \and
    National Research Council of Canada, Herzberg Astronomy \& Astrophysics Research Centre, 5071 West Saanich Road, Victoria, BC V9E 2E7, Canada
    \and
    Department of Physics and Astronomy, University of Waterloo, 200 University Avenue West, Waterloo, ON N2L 3G1, Canada
    \and
    Institute for Astronomy, University of Hawaii, 2680 Woodlawn Drive, Honolulu HI 96822
    \and
    Waterloo Centre for Astrophysics, University of Waterloo, Waterloo, Ontario N2L 3G1, Canada
    \and
    Perimeter Institute for Theoretical Physics, 31 Caroline St. North, Waterloo, ON N2L 2Y5, Canada
    \and
    LIRA, Observatoire de Paris, Universite PSL, CNRS, Place Jules Janssen, 92195 Meudon, France
    \and
    UK Astronomy Technology Centre, Royal Observatory, Blackford Hill, Edinburgh EH9 3HJ, UK
    }

   \date{Received 12 March 2025 / Accepted 23 May 2025}

 
  \abstract
   {The detection of low surface brightness galaxies beyond the Local Group poses significant observational challenges, yet these faint systems are fundamental to our understanding of dark matter, hierarchical galaxy formation, and cosmic structure. Their abundance and distribution provide crucial tests for cosmological models, particularly regarding the small-scale predictions of $\Lambda$CDM. We present a systematic detection and classification framework for unresolved dwarf galaxy candidates in the large-scale Ultraviolet Near Infrared Optical Northern Survey (UNIONS) imaging data. The main survey region covers 4,861\,deg$^{2}$. Our pipeline preprocesses UNIONS data in three (\emph{gri}) of the five bands (\emph{ugriz}), including binning, artifact removal, and stellar masking before employing the software \textsc{MTObjects} (\textsc{MTO}) to detect low surface brightness objects. After parameter cuts using known dwarf galaxies from the literature and cross-matching between the three bands, we are left with an average of $\sim$360\,candidates per deg$^{2}$. With $\sim$4,000\,deg$^{2}$ in \emph{g}, \emph{r} and \emph{i}, this amounts to $\sim$1.5\,million candidates that form our GOBLIN (Galaxies OBserved as Low-luminosity Identified Nebulae) catalog. For the final classification of these candidates, we fine-tuned the deep learning model \textsc{Zoobot}, which was pre-trained based on labels from the Galaxy Zoo project. We created our training dataset by visually inspecting dwarf galaxy candidates from existing literature catalogs within our survey area and assigning probability labels based on averaged expert assessments. This approach captures both consensus and uncertainty among experts. Applied to all detected \textsc{MTO} objects, our method identifies 42,965 dwarf galaxy candidates with probability scores > 0.8, of which 23,072 have probabilities exceeding 0.9. The spatial distribution of high-probability candidates reveals a correlation with the locations of massive galaxies (log\,$(M_{*}/M_\odot) \geq$ 10) within 120\,Mpc. While some of these objects may have been previously identified in other surveys, we present this extensive catalog of candidates, including their positions, structural parameter estimates, and classification probabilities, as a resource for the community to enable studies of galaxy formation, evolution, and the distribution of dwarf galaxies in different environments.
   }

    \titlerunning {Galaxies OBserved as Low-luminosity Identified Nebulae (GOBLIN)}
   \maketitle
%

\section{Introduction}

Dwarf galaxies occupy the low-mass, low-luminosity end of the galaxy distribution while representing the oldest and most numerous galaxy type in the Universe \citep{binggeli1990abundance,ferguson1994dwarf}. Dwarfs are small systems that contain gas, dust, dark matter, and up to a billion stars with total stellar masses $\leq$ 10$^{9}M_\odot$ \citep{bullock2017small}. They are the most dark matter dominated galaxy type \citep[e.g.,][]{mateo1991kinematic,walker2009universal,2020MNRAS.491.3496C}, making them excellent laboratories for studying this puzzling matter component. In the $\Lambda$CDM standard model of cosmology, they are thought of as the fundamental building blocks of all galaxies, contributing to their formation via hierarchical merger \citep{frenk2012dark}. Through the study of dwarf galaxies, we can thus gain important insights into how galaxies form and evolve over time \citep{revaz2018pushing}. 

Even though dwarfs are the most abundant galaxies in the Universe, their small size, low surface brightness, and faint nature pose a unique set of challenges for the detection of these elusive objects. Before the year 2004, nearly all then-known dwarf galaxies had been discovered and characterized via visual inspection of deep, wide-field photographic plates. Significant advancements in this area came from dedicated surveys utilizing the unique wide-field and high-resolution capabilities of instruments like the 2.5\,m du Pont telescope at Las Campanas. These systematic investigations of clusters such as Virgo \citep[e.g.,][]{1984AJ.....89..919S,1985AJ.....90.1681B}, Fornax \citep{1988AJ.....96.1520F,1989AJ.....98..367F}, and Centaurus \citep{1997A&AS..124....1J}, along with several nearby galaxy groups \citep{1990AJ....100....1F}, were instrumental in establishing morphological classifications, substantially expanding the census of low-surface-brightness dwarfs, and providing foundational datasets for analyzing their populations and distributions (see also \citet{1994A&ARv...6...67F} for a review). As notable exceptions to this primary reliance on visual plate inspection, the dwarf galaxies Sextans and Sagittarius were found as stellar over-densities in automated surveys in 1990 and 1994, respectively \citep{irwin1990new,ibata1994dwarf}. A stark discrepancy in the number count of Local Group (LG) dwarf galaxies between simulations and observations -- the so-called "missing satellites" problem \citep{klypin1999missing,moore1999dark} -- enforced the previously noted need to gain a complete census of the nearby dwarfs \citep{willman2010pursuit,macgillivray1987automated}. The advent of the Sloan Digital Sky Survey \citep[SDSS;][]{york2000sloan}, a large-scale imaging and spectroscopic survey, enabled this pursuit and revolutionized the field of dwarf galaxy research by leading to the discovery of 14 new Milky Way (MW) dwarf galaxies and thus doubling the number of known dwarfs in the MW \citep{bullock2017small}. Around the same time, similar success was achieved in regards to our massive neighbor galaxy M31, where 11 new dwarfs were discovered, mostly through dedicated surveys using the Isaac Newton Telescope (INT) and the Canada–France–Hawaii Telescope (CFHT) \citep[e.g.,][]{ferguson2002evidence,martin2006discovery,2007ApJ...671.1591I,irwin2008andromeda}. These numbers were further expanded as a result of the Pan-Andromeda Archaeological Survey \citep[PAndAS;][]{2009Natur.461...66M} which led to the detection of a similar number of new M31 dwarfs \citep{mcconnachie2008trio,2009ApJ...705..758M,richardson2011pandas,2013ApJ...776...80M}. Following further discoveries from surveys such as the Pan-STARRS1 surveys \citep{chambers2016pan}, the Dark Energy Survey \citep[DES][]{2018ApJS..239...18A}, the Hyper Suprime-Cam Subaru Strategic Program \citep[HSC-SSP;][]{2018PASJ...70S...4A,2018PASJ...70S...8A}, and the Dark Energy Local Volume Explorer \citep[DELVE;][]{2021ApJS..256....2D,2022ApJS..261...38D} and most recently, the UNIONS survey \citep{2025arXiv250313783G}, to date, we know of $\sim$60 and $\sim$40 dwarf satellites of the MW and M31, respectively \citep{Pace2024arXiv241107424P,2025arXiv250206948D}. Most LG dwarf galaxies in the SDSS era and after were discovered as overdensities of resolved or partially resolved stars. This was made possible on the one hand through wide and deep multi-band photometry, which enabled the detection of fainter stars, and on the other hand through improvements in handling contamination by MW stars and star-galaxy separation \citep[see][for a review on LG dwarf galaxy detection]{willman2010pursuit}. 

It is important to study dwarf galaxies in a range of different systems and environments to gain a complete picture of their nature, as the dwarfs in the LG might not represent the object class overall. Various isolated and grouped host environments and their satellite populations were studied in great detail \citep[e.g.,][]{chiboucas2013confirmation,2016A&A...588A..89J,danieli2017dragonfly,2017A&A...597A...7M,2017A&A...602A.119M,crnojevic2019faint,carlsten2019using,bennet2020satellite,2020MNRAS.491.1901H,2021MNRAS.506.5494P,2022ApJ...933...47C,2023MNRAS.521.4009C,2024MNRAS.527.9118C,2024arXiv240503769M}. As we explore satellite systems out to farther distances beyond the LG, the dominant dwarf detection method reverts to visual inspection of photometric data, since the objects no longer appear as a collection of individual, resolved stars, but as faint and diffuse light. Automatic processing pipelines usually play an important role in the detection efforts of faint, distant dwarfs. They are used to filter detections as much as possible and to produce a list of possible dwarf candidates. The final dwarf catalog of highly confident detections, however, is typically presented only after visual cleaning by several experts \citep[e.g.,][]{2017ApJ...850..109B,2018ApJ...857..104G,2019ApJ...875..155D,2020A&A...644A..91M,2020MNRAS.491.1901H,2022ApJ...933...47C}. 

In recent years, machine learning (ML) techniques have been employed for various tasks in Astronomy \citep[e.g.,][]{kim2016star,dominguez2018improving,lanusse2018cmu,zhu2019galaxy,jacobs2019finding,davies2019using,bom2019deep,ribli2019weak,caldeira2019deepcmb,paranjpye2020eliminating,cheng2020optimizing,ciprijanovic2020deepmerge}. In particular, the citizen science project Galaxy Zoo \citep{lintott2008galaxy,lintott2011galaxy,willett2013galaxy} has been utilized to generate a large database of labeled galaxy images. This database can be used as a training set for ML models that can be applied to replicate the visual classifications of Galaxy Zoo participants \citep{dieleman2015rotation}, increasing the number of high-quality labels for astronomical research. Beyond these general applications, ML approaches have been specifically adapted for the challenging task of detecting low surface brightness (LSB) galaxies. \citet{2021aApJS..252...18T} conducted a pioneering study using data from the first three years of the Dark Energy Survey \citep[DES;][]{2018ApJS..239...18A}. Starting with objects detected by \textsc{Source Extractor} \citep[][]{1996A&AS..117..393B} across $\sim$5,000\,deg$^{2}$, they significantly reduced the number of candidates through both photometric parameter cuts and ML techniques such as random forests and support vector machines \citep[SVM;][]{cortes1995support}. Finally, a visual inspection of $\sim$45k objects led to $\sim$23k high-confidence LSB objects (i.e., $\sim$50\% false positives after ML classification). \citet{2021bA&C....3500469T} used convolutional neural networks \citep[CNNs;][]{lecun1998gradient} to study LSB galaxies for the first time. They used the labels they generated through their work in \citet{2021aApJS..252...18T} to train a CNN classifier to differentiate between true LSB galaxies and artifacts such as cirrus, diffuse light from stars or galaxies, arms of spiral galaxies, etc., which often have similar photometric parameters as LSB galaxies. By using the actual images instead of photometric parameters generated by detection software, they could significantly boost the classification accuracy from $\sim$80\% (SVM) to 92\% (CNN). \citet{2021OJAp....4E...3M} built on this work and developed a CNN classifier to differentiate between spiral, elliptical, and irregular dwarf galaxies detected in DES with an accuracy of 85\%, 94\%, and 52\%, respectively. These pioneering works show that ML can aid the detection and classification of these elusive objects. With ongoing and future large-scale ground and space-based surveys such as Euclid \citep{2025A&A...697A...1E}, LSST \citep{2019ApJ...873..111I}, and Roman \citep{2020JATIS...6d6001M,2022SPIE12180E..1OD}, that produce terabytes of data each night, developing reliable ML algorithms becomes imperative. Visual inspection, albeit the most robust detection method for LSB galaxies, becomes infeasible at modern scales. 

The Ultraviolet Near Infrared Optical Northern Survey \citep[UNIONS\footnote{https://www.skysurvey.cc/};][]{2025arXiv250313783G} falls into this large-scale category and delivers almost half a petabyte of deep, multi-band, wide-field imaging data. One of the main aims of UNIONS is to complement the Euclid space mission by providing ground-based photometry for photometric redshift determination. However, UNIONS is also a standalone project that will be the reference optical survey in the northern sky for the next decade. It has a wide variety of scientific activities ranging from the MW assembly \citep[e.g.,][]{ibata2017chemical,2018MNRAS.481.5223T,thomas2019type,thomas2019dwarfs,fantin2019canada,fantin2021mass,smith2023discovery,smith2024discovery} over galaxy evolution \citep{thomas2020hidden,bickley2021convolutional,bickley2022star,sola2022characterization,roberts2022ram,2023A&A...673A.100C,2023MNRAS.525.1443L} and mergers \citep{wilkinson2022merger,ellison2022galaxy,2024MNRAS.533.2547F}, clusters of galaxies \citep{2024MNRAS.532.2521M,2025arXiv250109147M}, high redshift galaxies \citep{payerne2024high}, and AGN \citep{ellison2019definitive,bickley2023agns} to strong \citep{savary2022strong,chan2022discovery} and weak lensing \citep{2022A&A...666A.162G,2023A&A...671A..17A,2023MNRAS.523.1614R,2024ApJ...969L..25L,2024A&A...691A..75Z,2024arXiv241214666G}. In this paper, we present GOBLIN (Galaxies OBserved as Low-luminosity Identified Nebulae), an extensive catalog of objects that we detected in the UNIONS footprint with their probability of being a dwarf galaxy. In a similar fashion as \citet{2021aApJS..252...18T} and \citet{2021bA&C....3500469T} we started with the catalog of a detection software, made cuts in parameter space, and trained a CNN classifier by using known dwarf galaxies from various publications in the literature \citep{2014ApJ...787L..37M,duc2015atlas3d,2017ApJ...847....4G,2019ApJS..240....1Z,bilek2020census,2020MNRAS.491.1901H,2021ApJ...907...85M,2022ApJ...933...47C,2023AJ....166..185G,2023ApJS..265...57P,2024ApJ...976..117M}. Finally, we applied the trained model to our filtered catalog of candidates and thus assigned a probability of being a dwarf galaxy to every object. The code used for the detection and classification pipeline described in this paper is publicly available\footnote{https://gitlab.com/nick-main-group/dwarforge}.

This paper is structured as follows. In Section \ref{sec:data} we elaborate on the UNIONS survey. Then Section \ref{sec:methods} describes the different methods we used to produce the final catalog from image preprocessing to the ML-classified candidates. In Section \ref{sec:results} we present and discuss our catalog, followed by a summary and conclusions in Section \ref{sec:sum_concl}.

\section{Data}
\label{sec:data}

The Ultraviolet Near Infrared Optical Northern Survey (UNIONS) is a collaboration between four projects that gather \emph{u}, \emph{g}, \emph{r}, \emph{i}, and \emph{z}-band observations using three different telescopes in Hawaii. The Canada-France Imaging Survey \citep[CFIS\footnote{https://www.cfht.hawaii.edu/Science/CFIS/};][]{ibata2017canada} is using the MegaCam camera on the Canada-France-Hawaii Telescope (CFHT) to provide \emph{u}-band and \emph{r}-band data. The Panoramic Survey Telescope And Rapid Response System \citep[Pan-STARRS;][]{kaiser2002pan,chambers2016pan} team covers the \emph{i}-band using the Pan-STARRS1 and Pan-STARRS2 telescopes. The Wide Imaging with Subaru HSC of the Euclid Sky (WISHES) survey gathers \emph{z}-band data, and the Waterloo-Hawaii-IfA G-band Survey (WHIGS) contributes the \emph{g}-band using the Subaru Hyper Suprime-Cam (HSC). The main UNIONS footprint >30\,$^{\circ}$ in declination and a galactic latitude |b|\,>\,25\,$^{\circ}$ encloses an area of 4,861\,deg$^{2}$ and is covered by all three telescopes. These boundaries were initially set to match the northernmost coverage of the Euclid survey and the anticipated declination limit of 30\,$^{\circ}$ for LSST in the southern sky. However, with LSST now expected to cover only up to 15\,$^{\circ}$ in declination, UNIONS will extend its coverage to 15\,$^{\circ}$ through a series of approved extension programs beginning in 2025 \citep{2025arXiv250313783G}. In addition to the main footprint, CFIS will exploit the exceptional blue sensitivity of CFHT to map nearly the entire northern hemisphere in the \emph{u}-band, excluding the Galactic plane.

In part motivated by the Euclid requirements for deep observations, UNIONS will reach 10-$\sigma$ point source depths in a 2\,arcsec diameter aperture of 23.7, 24.5, 24.2, 23.8, 23.3\,mag in \emph{u}, \emph{g}, \emph{r}, \emph{i} and \emph{z}, respectively \citep{2025arXiv250313783G}. The \emph{r}-band is of particular interest for dwarf galaxy studies as the data reduction process is specifically optimized for LSB science. Both CFIS bands are reduced using the \textsc{MegaPipe} pipeline \citep{gwyn2008megapipe,gwyn2019megapipe}. In the standard reduction, the local sky background is estimated and subtracted from the images via \textsc{SWarp} \citep{bertin2010swarp}. In addition, the \emph{r}-band data are also processed with the Elixir-LSB methodology \citep{ferrarese2012next}, which leverages exposures taken immediately before and after the target observation to construct a robust sky background model. This approach is specifically designed to preserve the faint, low-surface-brightness outskirts of astronomical objects by accurately modeling and subtracting the sky without removing intrinsic extended features. The \emph{r}-band also has an excellent median seeing of 0.69\,arcsec and reaches a surface brightness limit of 28.4\,mag/arcsec$^{2}$ for extended sources. This limit signifies the survey's capability to directly detect a 10\,arcsec$\times$10\,arcsec extended feature at a 1-$\sigma$ significance level above the sky background. The HSC data (\emph{g}, \emph{z}) is processed using the LSST Science Pipeline \citep{2018PASJ...70S...5B,2019ASPC..523..521B}, developed for the Vera C. Rubin Observatory \citep{2019ApJ...873..111I}. The Pan-STARRS data is processed with the Pan-STARRS Image Processing Pipeline (IPP) \citep{2020ApJS..251....3M}. Astrometric calibrations for all bands are performed using Gaia DR2 \citep{collaboration2018gaia} sources as a reference frame. All but the \emph{u}-band are photometrically calibrated using the Pan-STARRS 3$\pi$ catalog \citep{chambers2016pan,2020ApJS..251....6M}. The \emph{u}-band is calibrated using the Gaia DR3  spectra \citep{2023A&A...674A...1G,de2023gaia} to create synthetic \emph{u}-band magnitudes \citep{2025arXiv250313783G}. After photometric calibration, the data from the individual exposures are stacked onto evenly spaced, square 0.51\,deg image tiles and resampled such that the data from the three different telescopes share a common pixel scale. The stacked tiles each have a size of 10k$\times$10k pixels with a pixel scale of 0.1857 arcsec/pixel (approximately the native pixel scale of MegaCam; 0.187).

In this work, we used the $\sim$16.5k tiles that are currently (as of January 2025) observed in \emph{g}, \emph{r}, and \emph{i}. We chose these bands because they have the largest overlapping coverage, their depth is sufficient for LSB detection, and the model we fine-tuned for dwarf classification requires three bands. 

\section{Methods}
\label{sec:methods}
The following describes our methods to preprocess the data, automatically detecting potential dwarf candidates, and using an ML classifier to assign a probability of being a dwarf galaxy to these candidates. We carried out all preprocessing, detection, and postprocessing steps for each of the three utilized bands \emph{g}, \emph{r}, and \emph{i} independently. 

\subsection{Preprocessing}

\subsubsection{Binning}

Before being passed to the actual detection algorithm, the image tiles underwent heavy preprocessing to enhance low surface-brightness features — such as ultra-diffuse galaxies \citep[][UDGs]{1984AJ.....89..919S,2015ApJ...798L..45V} — and suppress unlikely dwarf candidates, elements of contamination, and data corruption. The first step was to bin the image tiles by averaging over a group of 4$\times$4 pixels\footnote{openCV.resize}, a choice motivated by the need to reliably detect a set of UDGs near the early-type galaxy NGC\,5485 \citep{2014ApJ...787L..37M,2016ApJ...833..168M}, which are among the faintest dwarfs we gathered from the literature. Trial and error showed that we could detect these faint, diffuse galaxies only after applying this level of binning. We illustrate the gain in detectability of these four UDGs using this binning scheme in Figure \ref{fig:appendix_udgs} in Appendix \ref{sec:appendix_udgs}. Moreover, this 4$\times$4 binning strikes a balance, enhancing the signal-to-noise ratio for LSB detection while preserving sensitivity to dwarf candidates at the lower end of the size distribution. This process also reduced the image size by a factor of 16, significantly accelerating processing times across all subsequent steps. Although binning degrades image resolution, this is not a concern for our purposes, as the detection step prioritizes diffuse LSB structures over sharp features.

\subsubsection{Masking data corruptions}

In the second step, we automatically detected data corruptions such as readout errors (vertical stripes), saturated pixels, or patches containing unusual flux gradients. Typically, such defects are removed before binning, as their influence can spread during averaging and introduce artifacts in the binned image tiles. However, we chose to identify and mask these corruptions after binning because performing this step on full-resolution images would significantly increase processing times. We first replaced all NaN values, saturated pixels, and highly negative values ($\leq$ -5\,ADU for g and r-bands, $\leq$ -2\,ADU for i-band) with zeros. Next, we applied a single-level wavelet decomposition to the binned image tiles using a discrete wavelet transform with Haar wavelets. This decomposition separates the image tile into three detail coefficient maps that highlight different directional features: (1) horizontal detail coefficients that capture intensity changes between rows and are sensitive to horizontal edges and artifacts, (2) vertical detail coefficients that detect intensity changes between columns and are particularly effective at identifying vertical stripes from readout errors, and (3) diagonal detail coefficients that highlight changes across diagonal directions. We identified corrupted patches in each coefficient map where the variation between neighboring values was nearly zero (indicating uniform artifacts rather than natural image features), thresholded these regions, and merged them into a single binary mask by taking their union (logical OR). This mask was then dilated with one binary iteration to include adjacent bad pixels on the patch outskirts, and all masked pixels were set to zero. This helped avoid false detections and aided the detection algorithm by stabilizing the background, which is crucial for the detection of LSB objects since they can be barely above the estimated background level.

\subsubsection{Masking hot pixels}

Next, we estimated the background in the image tile using the \textsc{photutils} \citep{larry_bradley_2024_12585239} function \textsc{Background2D} with the `MedianBackground' estimator. We then used it to identify and mask hot pixels, i.e., small groups of pixels with unusually high values that are not obviously related to bright sources such as stars or galaxies. Dwarf candidates were selected via object parameters produced by the detection algorithm, such as the effective radius and the surface brightness. Hot pixels can thus distort these parameters if they overlap with a dwarf galaxy and elude its detection. We preselected groups of pixels ($\leq$ 3) with values $\geq$ 70 (determined via trial and error). To qualify as hot pixels at least 20\% of the nearest neighbor pixels should be $\leq 3\sigma$ of the estimated background level, defined as the background median plus three times the background RMS. Pixel groups associated with bright objects would be surrounded by other pixels well above $3\sigma$ of the background level. The median pixel value of the image tile replaced the identified hot pixels.

\subsubsection{Correcting background over-subtraction for the \emph{g}-band}

In the next step, we used the detection software SEP\footnote{https://github.com/kbarbary/sep} \citep[][]{Barbary2016}, a Python wrapper of the original \textsc{Source-Extractor} \citep{1996A&AS..117..393B}, to detect all objects that are $1\sigma$ above the background level. These detections and the corresponding segmentation map produced by SEP were used in the following preprocessing steps. In the g-band data, the aggressive local background subtraction during data reduction left behind a halo of negative pixel values around large objects in the image tiles. We found that these values have a negative impact on the detection of LSB objects. We thus attempted to correct this over-subtraction by picking out the largest objects found in the segmentation map and iterating binary dilations of their segments until the halo of negative values was covered. Binary dilation in image processing expands foreground regions in a binary image (i.e., the objects of interest) by adding pixels to their boundaries, effectively thickening the objects. We then created a background map that contained summary statistics (mean, median, standard deviation) in regular grid cells over the field, not including pixels that belong to objects detected in the SEP run. We replaced the difference between the binary dilated segments and the original segmentation map, i.e., the outskirts of large objects, with pixels sampled from a normal distribution with the median and standard deviation of the nearest background cell.

\subsubsection{Masking small objects and bright stars}

Next, we used the SEP detections to mask small objects in the field and replaced them with local background noise as before. The smallest known dwarf in the footprint (40 pixels) informed the upper size limit for this masking procedure. We also masked slightly larger objects ($\leq$ 100\,pixels), if their effective radius was smaller than that of the smallest dwarf (1.5\,arcsec).

We used the Gaia DR3 catalog \citep{2016A&A...595A...1G, 2023A&A...674A...1G, 2023A&A...674A..32B} to identify very bright stars in the image tile and matched them with the SEP detections. To find a relationship between the apparent size of a star in our images and its Gaia G-band magnitude $Gmag$, we ran SEP on 250 image tiles, matched the stars in the field taken from the Gaia catalog with the SEP detections, and fit an empirical function between the magnitude and the number of pixels SEP assigned to the detection. We also found a connection between the magnitude and the length and thickness of possible diffraction spikes. These relationships were used to mask all but the brightest stars in the image tiles ($Gmag \leq$ 10.5). We left brighter stars unmasked because they cover a significant portion of the image tile and feature large diffraction spikes and stellar halos. By masking them, we would risk removing nearby LSB objects of interest. Furthermore, clean masking became increasingly difficult for these brighter stars due to their irregular shapes. Consequently, LSB artifacts from diffuse starlight may appear on the outskirts of the mask, which would lead to false detections.

The star masking process works as follows: first, we analyzed the SEP segmentation map to differentiate between five situations stars can be in. They can be isolated from other SEP detections, adjacent to one or more detections, not deblended, embedded in another detection, or not detected. We masked bright stars that were not embedded in another detection by probing the background statistics in four annular segments around the star (excludes diffraction spikes). Other nearby detections were excluded and did not influence this calculation. Background noise replaces the full star with its diffraction spikes and diffuse stellar halo. We made sure not to mask any other sources in the vicinity of bright stars if they were detected by SEP. For stars that are isolated or adjacent, we performed a binary dilation of their segment with a single iteration and replaced it with the local background as sampled from the background grid. In case a star was embedded in a bright source such as another star or a galaxy, we did not replace it with background values because this would introduce a sharp discontinuity in the object and could lead to fragmented detections and therefore inaccurate photometry in the following steps. Instead, we replaced it with values sampled from a positive truncated normal distribution using the pixel distribution in a small annulus around the star. In the special instance where a star may be overlapping with an LSB object and the detection algorithm is not capable of deblending the two, we could not mask the star via the SEP segment. To test for this, we compared the expected size of the star from the Gaia magnitude relation with the size from SEP. If there was a discrepancy such that the SEP size was larger than expected or if the star was on the bright end of the distribution ($Gmag \leq 13.5$\,mag) and the flux peak of the segment was significantly offset from its center, we considered a star non-deblended. We then shrank the expected star mask and replaced it analogous to the embedded case. We skipped undetected stars, which can occur when parts of the image tile are corrupt or data is missing.

\begin{figure*}[!htb]
\centering
\includegraphics[width=\linewidth]{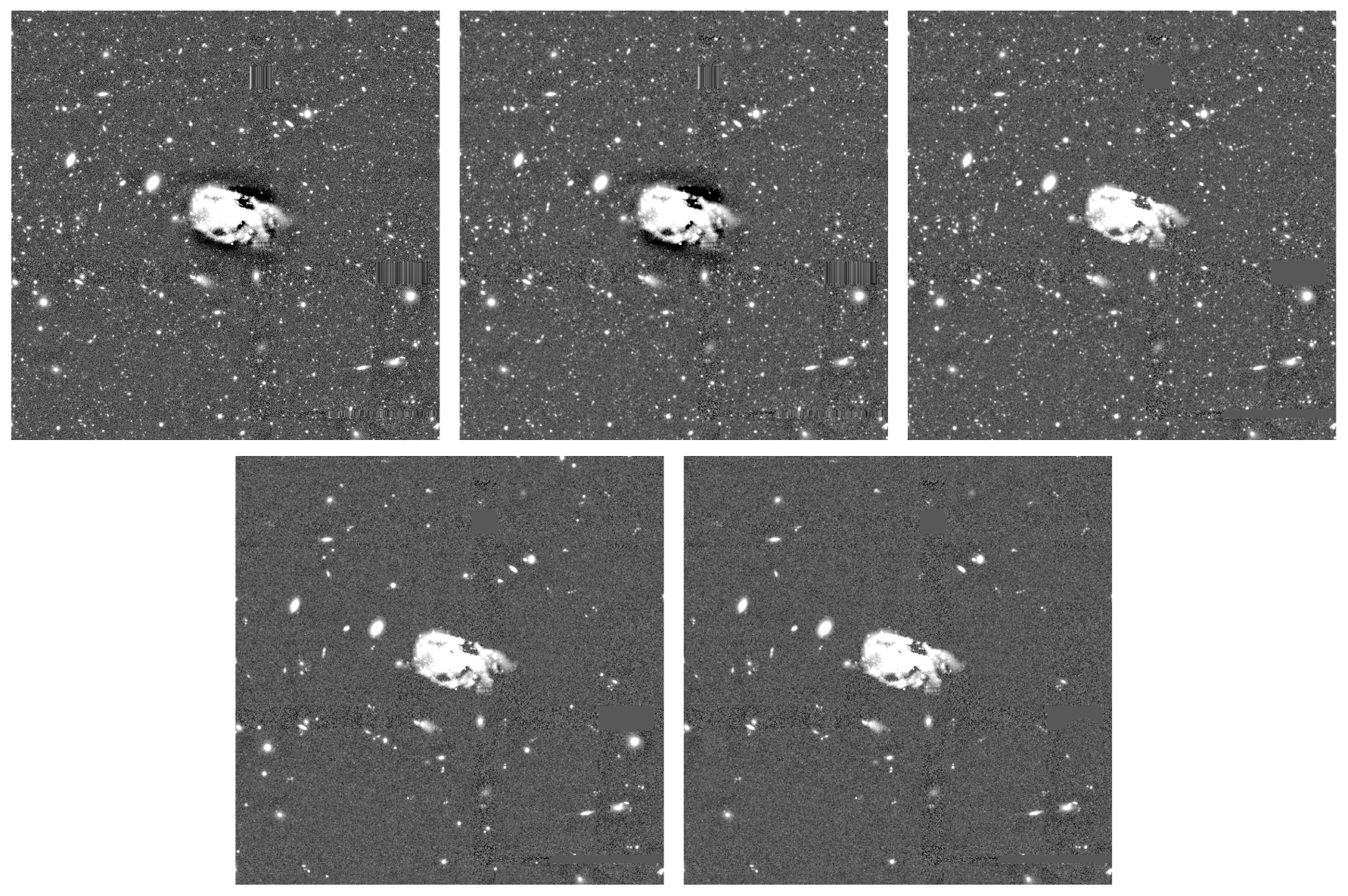}
\caption{Illustration of the main preprocessing steps on a \emph{g}-band image cutout. Top left: original full-resolution image. Top center: image after 4$\times$4 pixel binning. Top right: corrected for background over-subtraction near bright objects (spiral galaxy at the center) and set anomalies, such as vertical stripes, to zero. Bottom left: replaced small objects with nearby background noise. Bottom right: replaced MW stars with the local background.}
\label{fig:prep}
\end{figure*}

In the final preprocessing step -- if there was a bright star in the image tile ($Gmag \leq 9$\,mag) -- we subtracted the background using the function \textsc{Background2D} with the 'MedianBackground' estimator. This was a necessary step for bright stars because the detection algorithm struggled to accurately estimate the background on its own in these instances and missed faint LSB objects. Under good image conditions, i.e., no very bright objects in the field, the algorithm performs best without or only coarse background subtraction. Therefore, we adapted the mesh size used to estimate the background according to the magnitude of the brightest star and used smaller mesh sizes for brighter stars. We illustrate the main preprocessing steps on an example \emph{g}-band image cutout in Figure \ref{fig:prep}.

\subsection{Detection with MTObjects}

\textsc{MTObjects} (Max-Tree Objects; henceforth \textsc{MTO}) \citep{teeninga2015improved, teeninga2016statistical} is a detection software similar to \textsc{SEP} or \textsc{Source-Extractor}, designed for the study of LSB objects. In contrast to \textsc{SEP}, which estimates and subtracts the local background, \textsc{MTO} calculates a global background value from areas devoid of objects and removes it. In \textsc{SEP}, a fixed threshold is chosen (an integer multiple of the standard deviation of the local background level) to differentiate between objects and background. In \textsc{MTO}, on the other hand, a so-called max-tree is constructed, which is a hierarchical representation of the image. The leaf nodes represent the image maxima, and parent nodes are progressively fainter regions of a given object. In a series of statistical tests using the estimated global background level and local node characteristics, the algorithm decides between real objects and noise. Compared to \textsc{SEP}, \textsc{MTO} excels at detecting LSB objects and correctly captures the faint outskirts of extended galaxies. It also performs better at identifying nested objects, and its background estimation is not biased by large objects in the image. 

We used a Python implementation of \textsc{MTO}\footnote{https://github.com/CarolineHaigh/mtobjects} in our pipeline. We left default values for all but one parameter. The so-called `move factor' controls how many faint pixels are included on the outskirts of objects. With a lower value, we extend object boundaries to faint outer regions but also risk including noise. Since we are particularly interested in capturing LSB features, we moved this parameter from the default value of 0.5 to 0.39. We found this final value by trial and error, focusing on striking a balance between capturing most of the LSB outskirts of the objects and avoiding object blending. Since we chose a lower value for the move factor, careful preprocessing such as removing small sources beforehand was required to avoid including neighboring sources that are not physically connected to the object of interest. It was important to obtain real object boundaries as well as possible, such that \textsc{MTO} can accurately calculate structural parameters for the objects, which were used to filter the detections and select likely LSB candidates. We used \textsc{SEP} rather than \textsc{MTO} to detect point-like objects that should be masked in the preprocessing step because it is significantly faster and better at deblending point sources. 

\subsection{Postprocessing}
\label{sec:postprocessing}

After running MTO on the preprocessed image tiles we used the output parameter file to derive structural parameters for all objects in arcseconds. By default the \textsc{MTO} implementation calculates the effective radius $R_{e}$, the radius at full-width-half-max of the light distribution $R_{fwhm}$, and the radii from which 10 ($R_{10}$) and 90 ($R_{90}$) percent of the total light is emitted. It is important to note that these radius calculations are based on a simplified approach where pixels belonging to an object are sorted by their brightness values rather than by their spatial distribution. The calculation does not consider the distance of pixels from the object's center. Instead, it determines how many of the brightest pixels contribute to specific percentages of the total flux, and then converts these pixel counts to equivalent circular radii using the formula $r = \sqrt{area/\pi}$, effectively assuming a circular profile. The parameter $R_{fwhm}$ is calculated as the radius of a circle that would contain all pixels that are at least half as bright as the brightest pixel in the distribution. $R_{e}$, on the other hand, is the radius of a circle containing the brightest pixels that together account for 50\% of the total flux in the segment. After initial experimentation using a binary random forest classifier to distinguish between dwarf galaxy and everything else, we analyzed the feature importance scores of the model. This analysis revealed that the radius parameters -- in particular $R_{10}$ -- had the highest importance values, contributing most significantly to the classifier's ability to discriminate between dwarf and non-dwarf objects. Therefore, we decided to add calculations for $R_{25}$, $R_{75}$, and $R_{100}$ to the \textsc{MTO} source code. $R_{100}$ represents the radius of a circle with an area equivalent to the total area of all pixels assigned to an object by \textsc{MTO}. We initially applied a coarse filter to the \textsc{MTO} detections based on $R_{e}$ > 1.6\,arcsec and mean effective surface brightness $<\mu>_{e}$ > 19\,mag/arcsec$^{2}$ and removed possible satellite streaks by requiring an object axis ratio > 0.17 for large detections, which is based on a set of streaks in the image tiles. This initial filter leaves about 400-600 candidates for every image tile.

In the next step, we added labels to the detected candidates by cross-matching their positions with a catalog of known dwarfs we compiled from the literature. The catalog contains dwarfs from different surveys namely, ELVES \citep{2022ApJ...933...47C}, MATLAS \citep{duc2015atlas3d,bilek2020census,2020MNRAS.491.1901H,2021MNRAS.506.5494P}, SAGA \citep{2017ApJ...847....4G,2021ApJ...907...85M,2024ApJ...976..117M}, SMUDGES \citep{2019ApJS..240....1Z,2023AJ....166..185G,2023ApJS..267...27Z}, UDGs detected in the field around M101 \citep{2014ApJ...787L..37M} and early-type dwarfs that were visually identified and presented in \citet{2023ApJS..265...57P} using the DESI Legacy Imaging Surveys \citep{2019AJ....157..168D}. Table \ref{tab:literature_dwarfs} details, for each source catalog, the number of objects that fall within the UNIONS footprint and their corresponding measured or estimated distance ranges. For every match, we added the literature ID of the known dwarf galaxy and a dwarf label to the filtered \textsc{MTO} detection catalog. This catalog of known dwarf galaxies serves as the foundation for the training dataset we later use to fine-tune our machine learning classifier, as described in Section \ref{sec:label_and_classification}.

\begin{table}[!htb]
\centering
\caption{Number of dwarf galaxy candidates within the UNIONS footprint and their distance ranges, compiled from the listed literature catalogs.}
\vspace{-10pt}
\begin{threeparttable}
\label{tab:literature_dwarfs}
\begin{tabular}{ccc}
\toprule
Catalog & Number of Objects & Distance \\
 & & (Mpc) \\
\midrule
ELVES & 49 & < 12 \\
MATLAS & 315 & $\sim$10 - 45 \\
SAGA & 27 & 25 - 40.75 \\
SMUDGES & 1212 & $\lesssim$ 143 \\
NGC5485 UDGs & 4 & $\sim$27 \\
dEs Local Universe & 402 & $\lesssim$ 40 \\
\bottomrule
\end{tabular}
\begin{tablenotes}[flushleft]
\footnotesize
\item References for these are ELVES \citep{2022ApJ...933...47C}, MATLAS \citep{duc2015atlas3d,bilek2020census,2020MNRAS.491.1901H,2021MNRAS.506.5494P}, SAGA \citep{2017ApJ...847....4G,2021ApJ...907...85M,2024ApJ...976..117M}, SMUDGES \citep{2019ApJS..240....1Z,2023AJ....166..185G,2023ApJS..267...27Z}, NGC5485 UDGs \citep{2014ApJ...787L..37M} and dEs Local Universe \citet{2023ApJS..265...57P}. Note: for the majority of objects (except ELVES and SAGA), distances are predominantly estimated based on their assumed association with a host galaxy for which spectroscopic redshift or direct distance measurements are available.
\end{tablenotes}
\end{threeparttable}
\end{table}

To further refine our selection and minimize false positives, we sought a clear separation between dwarf candidates and likely non-dwarfs. To achieve this, we gathered detections from all image tiles containing known dwarfs and, for each known dwarf, selected the 20 nearest candidates as non-dwarf examples. We acknowledge that our assumption — that candidates closest to known dwarfs from the literature are likely non-dwarfs that would have been classified as such in previous surveys — has limitations. Our data may be deeper and of higher quality than that of previous surveys, potentially revealing dwarf galaxies that were previously undetectable. Nevertheless, this approach provided a reasonable initial approximation to establish a working filter. For every band individually, we created a series of plots between every combination of two derived object parameters, as well as some other derived products, such as ratios between two parameters. For the \emph{g} and \emph{r} bands, we found the clearest separation between dwarf and non-dwarf in the mean effective surface brightness $<\mu>_{e}$ vs. apparent magnitude $m$ plane. In the \emph{i} band, the separation is less clear in this plane but is more apparent by plotting ratios between radius parameters vs. the magnitude. In particular, $R_{90}/R_{75}$ vs. $m_i$ offers the clearest separation in \emph{i}. With these filters, we estimate to remove $\sim$ 32\%, 42\%, and 28\% of the non-dwarfs from the catalog of possible dwarf candidates for the \emph{g}, \emph{r}, and \emph{i} bands, respectively. We would have also respectively removed $\sim$ 1.3\%, 1.2\%, and 1.5\% of the real dwarfs but left them in the catalog since they are validated in the literature and the calculated \textsc{MTO} structural parameters can be faulty due to any overlapping sources or possible errors during object segmentation. We provide the exact filtering criteria and parameter values used for this process in Appendix \ref{sec:appendix_obj_filters}. 

After filtering the list of candidates for every band, we cross-matched the remaining detections using a coordinate matching approach. We first compiled all objects detected across the three bands, then identified potential matches within 10\,arcsec of each other and built connected components using a Union-Find algorithm, which efficiently groups all related detections (if object A matches B and B matches C, all three are grouped). Generally, we required objects to be present in at least two bands, which eliminated many false positives such as residual light from stellar halos, corrupt data patches that are particular to one band, or \textsc{MTO} segmentation errors. However, we implemented an exception for known dwarf galaxies from the literature catalogs: these were retained even if detected in only one band, provided that at least two bands had valid data overall. In total, 48 of the dwarfs from the literature were only detected in a single band. We added this exception because some confirmed low surface brightness dwarfs might only be detectable in the \emph{r} band, which is optimized for LSB science and probes to deeper surface brightnesses. Further, even though an object may not be detectable by \textsc{MTO} in some of the bands, the machine learning classifier we used later on may still be able to extract useful information from the image. If an object passed our matching criteria, we created a 256$\times$256\,pixel cutout around the object coordinates from the first band the candidate is detected in and stacked the cutouts of the three bands. This way, the objects are aligned precisely, which is crucial for the following classification step. This cross-matching filter left on average $\sim$90 candidates for each image tile. 16519 tiles have been observed in \emph{g}, \emph{r}, and \emph{i}, which results in approximately 1.5 million candidates out of which 2009 are dwarf galaxies following the above-mentioned sources from the literature.

\subsection{Dwarf labeling and classification with Galaxy Zoo}
\label{sec:label_and_classification}

Initial feature-based binary classification tests using random forests and gradient-boosted decision trees (eXtreme Gradient Boosting; XGBoost) on the MTO object parameters revealed that we would expect about 5\% false positive and false negative predictions. This would result in about 160k positive predictions that need to be cleaned visually. This seemed like an unfeasible task, and therefore we attempted a different approach. As demonstrated in \citet{2021aApJS..252...18T}, image-based deep learning models can lead to a significant performance boost over feature-based approaches for this particular task. We thus decided to fine-tune a deep neural network trained on a large database of annotated galaxy images. 

We initially used a set of known dwarf galaxies and their nearest unclassified candidates as positive and negative examples, respectively, to fine-tune the deep learning model \textsc{Zoobot}\footnote{\url{https://github.com/mwalmsley/zoobot}} \citep{walmsley2023zoobot}. Rather than training the model from scratch, we adopted a fine-tuning approach given our limited number of positive labels (i.e., dwarf galaxies). \textsc{Zoobot} was originally trained on galaxy morphology classification tasks using data from the Dark Energy Survey \citep[DES][]{2018ApJS..239...18A}, which benefited from millions of images labeled by Galaxy Zoo volunteers. Consequently, the pretrained weights already capture a robust representation of astronomical data, making \textsc{Zoobot} well suited for adaptation to our dwarf classification problem.

Initial tests revealed that the image-based model did not significantly outperform our previous feature-based one, with both methods showing a false-positive rate of $\sim$5\%. Visual inspection of the training data showed significant label noise. Some "dwarf galaxies" from the literature were non-dwarfs, such as artifacts, cirrus, or diffuse light around bright stars, objects that could mislead automatic detection pipelines. Conversely, we found that some of the candidates we had initially assumed to be non-dwarfs (based on their proximity to known dwarfs from the literature) were actually good dwarf candidates that had been missed in previous surveys. This further confirmed the limitations of our initial assumption and highlighted the need for careful visual classification. To address these issues, we undertook a comprehensive re-labeling of our training data, visually classifying both literature dwarfs and their nearest neighbors to establish reliable positive and negative examples. 

For visual classification, we used the original full-resolution image tiles rather than the binned versions used in \textsc{MTO} detection, as the lower resolution can obscure important morphological features, causing background spiral galaxies to appear as featureless, dwarf-like objects. We created 256$\times$256\,pixel cutouts centered on each object, and processed them through several steps to create RGB color images. First, we computed the intensity $I$ as the average of the three bands. When data in one band was missing or corrupted, we synthesized the missing band by averaging the remaining two bands, ensuring we maintained the three-band image required by \textsc{Zoobot}. In the final RGB composition, we mapped the longest wavelength to the red channel, the intermediate to green, and the shortest to blue. We then applied an inverse hyperbolic sine stretch to each channel using:

\begin{equation}
    channel_{scaled} = channel \times \frac{arcsinh(\frac{Q \cdot I}{stretch})}{Q \cdot I}
\end{equation}

\noindent based on \citet{2004PASP..116..133L} where $Q$ is a softening parameter that controls the relative scale between linear and logarithmic behavior of the function. The $stretch$ factor determines the intensity at which the logarithmic compression becomes significant. This scaling allows simultaneous visualization of faint LSB features and bright objects such as stars. Next, we applied gamma correction to each scaled band to enhance the perceived brightness of the final image. Given that the pixel distribution after scaling the channels includes negative values, we applied the gamma correction in a sign-preserving way:

\begin{equation}
    channel = sign(channel) \times |channel|^{\gamma}.
\end{equation}

\begin{figure*}[!htb]
\centering
\includegraphics[width=1\linewidth]{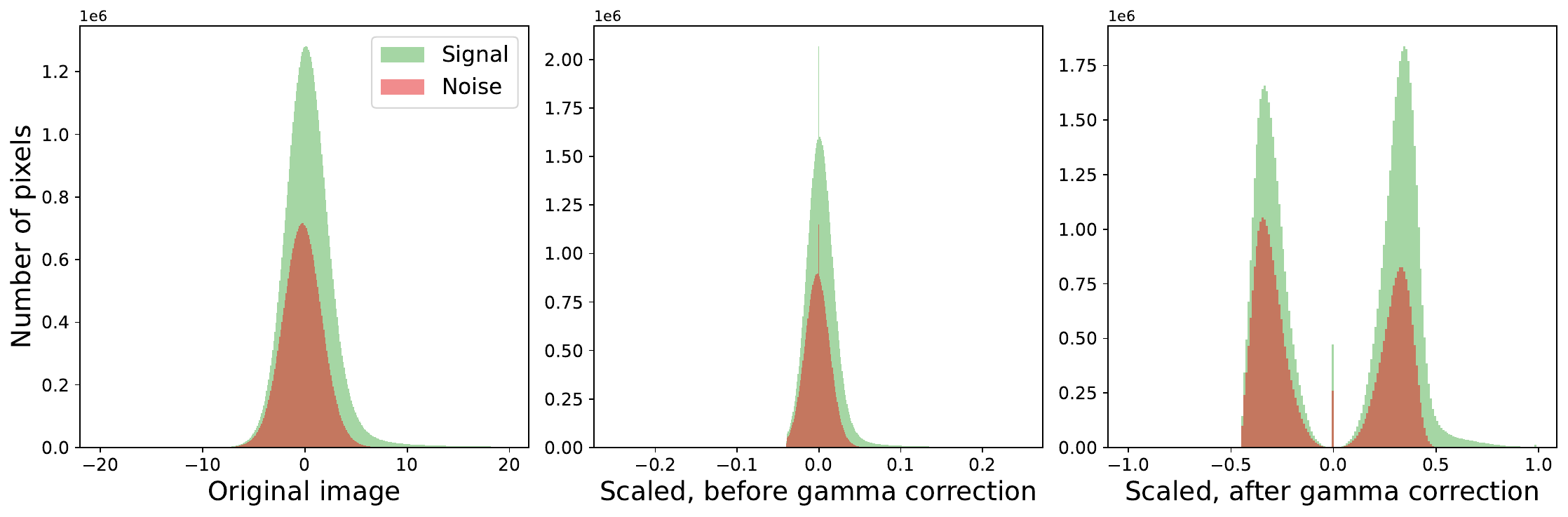}
\caption{Pixel value distribution in the \emph{r}-band for different stages of processing. Left: flux distribution in the original image tile. Middle: data after scaling via the inverse hyperbolic sine function. Right: bimodal distribution of the scaled data after gamma correction. Pixel values attributed to the signal (astronomical objects) are shown in green, and the ones coming from areas devoid of objects (background noise) are shown in red.}
\label{fig:bimodal}
\end{figure*}

\noindent This step pushes positive and negative values apart, turning an approximately Gaussian distribution before gamma correction, to a bimodal one after, as illustrated in Figure \ref{fig:bimodal}. After extensive testing, we found optimal visual results with $Q$ = 7, $stretch$ = 125, and $\gamma$ = 0.25. Finally, we stacked the scaled and gamma-corrected channels to create RGB images.

Four of our team members independently classified the 2,009 dwarf candidates from the literature. Given the challenging nature of dwarf galaxy classification due to their diverse sizes, morphological types, and distances, we implemented a three-tier classification system: `yes' (1), `no' (0), and `unsure' (0.5). This system accommodates ambiguous cases, such as dwarf irregulars that can be difficult to distinguish from more massive late-type galaxies at greater distances. For example, candidates showing irregular features with hints of spiral structure were typically assigned to the `unsure' category. We performed visual classification by inspecting six different cutout versions for a given object in a grid: the full resolution 256$\times$256\,pixel RGB image cutout, a 2$\times$2 binned version, a version that was first binned 2$\times$2 and then smoothed with a Gaussian kernel\footnote{openCV.GaussianBlur with $\sigma$ = 0.9}, the 2$\times$2 binned \emph{r}-band cutout and two RGB cutout versions from the Legacy Survey \citep{2019AJ....157..168D}. The first of these two is a 512$\times$512 pixel image cutout downloaded from the data release 10 of the Legacy Surveys as it appears on the Legacy Survey Sky Viewer\footnote{https://www.legacysurvey.org/viewer}. The second is a 2$\times$2 binned and enhanced version of the first cutout, using contrast-limited adaptive histogram equalization (CLAHE) to better visualize LSB features. After initial testing, we found that classifying an object based on a single RGB image cutout is not ideal when inspecting a large number of objects in quick succession that have vastly different properties, from bright and star-forming to extremely faint and LSB, due to the limited speed at which our eyes can adjust. We picked different degrees of binning and smoothing to aid in visualizing extremely faint and diffuse objects. The \emph{r}-band cutout further promotes the visibility of particularly diffuse objects due to its LSB optimization. Such objects might not be discernible in the stacked RGB image due to the local background removal in the other two bands. The Legacy Survey images were chosen significantly larger than the UNIONS cutouts to provide critical context in the case of large or interacting objects. We used a classification tool\footnote{https://github.com/heesters-nick/DwarfClass} specifically developed for this task to label our training data. A snapshot of the tool's user interface is shown in Figure \ref{fig:appendix_class_ui} in Appendix \ref{sec:appendix_class_ui}.

\begin{figure}[!htb]
\centering
\includegraphics[width=1\linewidth]{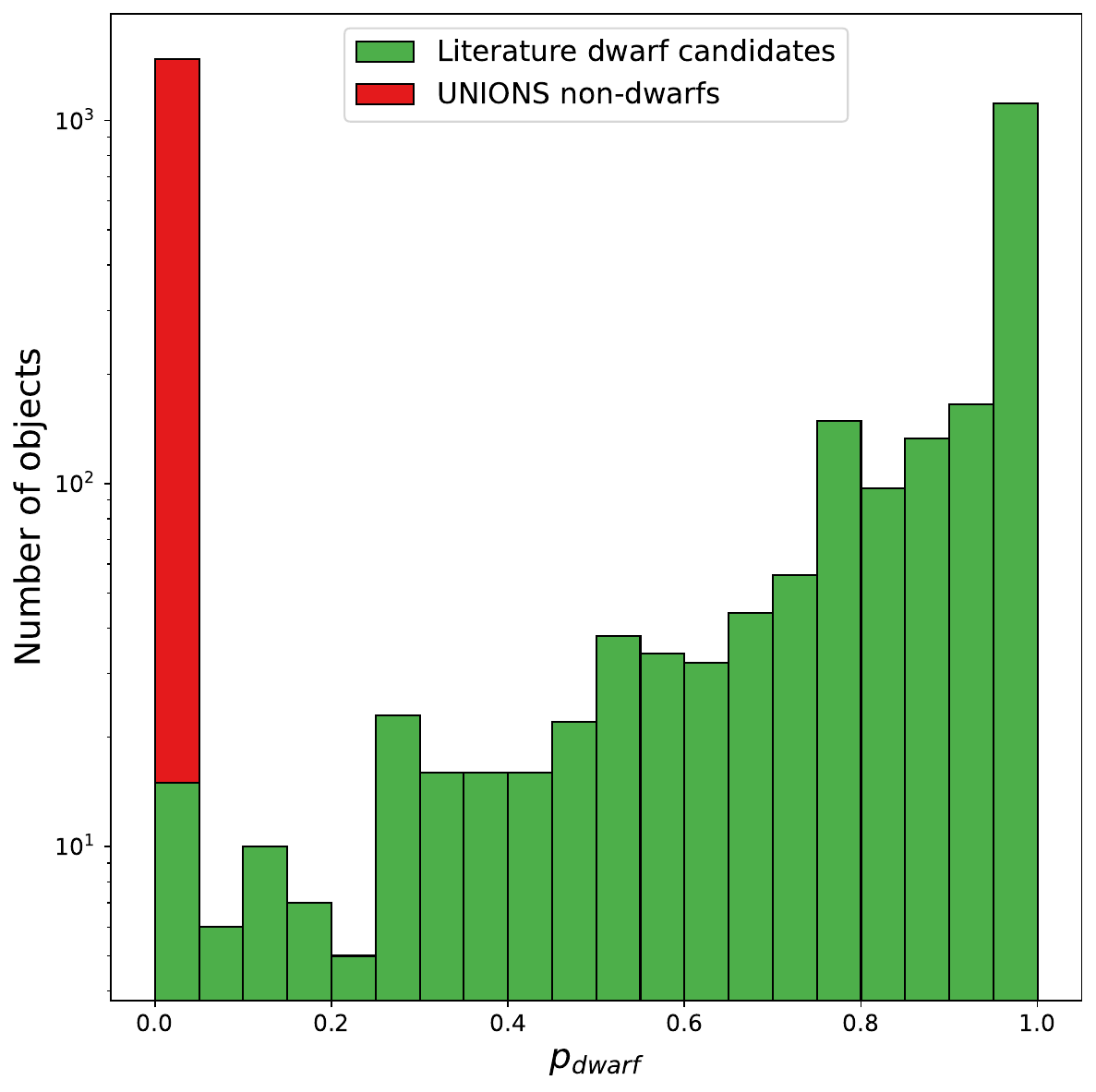}
\caption{Stacked histogram showing the label distribution in the dataset used for training. Dwarf candidates from the literature are shown in green, and non-dwarf objects are in red. The y-axis is shown on a logarithmic scale.}
\label{fig:label_dist}
\end{figure}

To account for both inter-rater variation (different experts classifying the same image differently) and intra-rater variation (the same expert assigning different classifications to the same image across multiple viewings), each of the four raters classified every candidate three times. The final classification for each candidate was derived by averaging all 12 resulting labels. This average represents the probability that a given object is a dwarf galaxy. We can utilize this information to train our model and teach it to reproduce expert uncertainty when classifying objects. After visual classification, 1998 out of 2,009 objects had a non-zero averaged label (indicating potential dwarfs), leaving 11 with a zero label (indicating unanimous non-dwarfs). 4 out of these 11 were already flagged after visual inspection in the SMUDGES survey \citep{2019ApJS..240....1Z,2023AJ....166..185G,2023ApJS..267...27Z}. For reference, we show the remaining 7 objects we visually classified as non-dwarfs in Appendix \ref{sec:appendix_non_dwarfs}. To balance the training dataset, we had to add more negative examples. Thus, two team members independently classified the same set of additional candidates, which were drawn from the pool of objects nearest to known dwarf galaxies. We took samples from this pool because we had initially hypothesized that these were the best candidates for non-dwarfs as they had likely already been inspected in other surveys and not classified as dwarfs (see Sec. \ref{sec:postprocessing}). We continued classifying these candidates until we had enough agreed-upon non-dwarf objects (both raters assigned a zero label) to achieve a dataset with an overall mean label of 0.5. The final training dataset includes 1998 non-zero labels and 1465 zero labels. We show the distribution of these labels in Figure \ref{fig:label_dist} and highlight labeled literature dwarfs vs classified non-dwarfs in the UNIONS data. We applied the same RGB transformation to the training dataset as we did for visual classification. 

To ensure robust model development and evaluation, the full dataset of labeled examples (see Figure \ref{fig:label_dist}) was first partitioned. We reserved 10\% of the data as a stratified, held-out test set, which was not used during any model training or hyperparameter optimization. The remaining 90\% constituted the training-validation pool. Hyperparameters for the \textsc{Zoobot} model were optimized using this training-validation pool.

Following hyperparameter selection, an ensemble of 10 \textsc{Zoobot} models was constructed to provide the final dwarf galaxy classifications. To build this ensemble and maximize the use of our labeled data for training the individual models, the training-validation pool was first divided into 10 equally sized, stratified segments, so-called folds. We then trained 10 distinct \textsc{Zoobot} models. For the training of each of these models, one unique segment (representing 10\% of the training-validation pool) was set aside as a validation set, used to monitor its training. The remaining nine segments (constituting 90\% of the training-validation pool) were combined and used as the training data for that particular model. This procedure was repeated 10 times, with each of the 10 segments systematically serving as the validation set exactly once. This approach ensures that every data point within the training-validation pool contributed to the training process across nine of the models and was used for validation in one, thereby effectively utilizing the entire pool for developing the ensemble members, rather than reserving a fixed portion solely for validation. The final probability assigned to a candidate is the average of the predictions from the 10 trained models, and the overall performance of this ensemble is evaluated on the initially reserved test set.

Our chosen architecture is the ConvNeXT-Nano variant of \textsc{Zoobot}, based on the ConvNeXt architecture \citep{liu2022convnet2020s}. For our application, we used the `FinetuneableZoobotClassifier' class with two classes but instead of rounding our mean classifications to be binary (dwarf or no dwarf), we introduced them to the model as so-called soft labels that take discrete values between 0 and 1, representing the probability $p_{dwarf}$ that a given object is a dwarf galaxy. We framed the problem such that the model predicts the probability distribution [1-$p_{dwarf}$, $p_{dwarf}$]. The model outputs two logits that, after applying the Softmax function, can be interpreted as the probability distributions [1-$p_{pred}$, $p_{pred}$]. We used the Kullback-Leibler (KL) divergence as our loss function:

\begin{equation}
L_{KL_{Div}} = y_{\text{true}} \cdot \log \left(\frac{y_{\text{true}}}{y_{\text{pred}}}\right) + (1-y_{\text{true}}) \cdot \log \left(\frac{1-y_{\text{true}}}{1-y_{\text{pred}}}\right)
\end{equation}

\noindent where $y_{\text{true}}$ is our soft label and $y_{\text{pred}}$ the model prediction. This function measures the difference between the labeled and predicted probability distributions. 

\textsc{Zoobot} is divided into five layer blocks. Earlier blocks generally learn simple features such as the overall shape of the galaxy, while later blocks, towards the output layer, learn increasingly complex and task-specific features. We therefore used a base learning rate for the last block and classification head, and decreased this learning rate by 25\% for each earlier block. With this, we made sure to utilize the pretrained model's ability to extract general features and gave it more flexibility to adjust to more complex ones, specific to our task and dataset. 

Even though we fine-tuned earlier blocks with low learning rates, our small set of labeled data compared to the number of model parameters quickly led to over-fitting. To mitigate this, we applied data augmentations, including random rotations (an integer multiple of 90 degrees), random vertical and horizontal flips, random small variations in contrast and brightness, and the addition of noise. As mentioned before, the data follows a bimodal distribution due to gamma correction. We estimated the noise level in our images by taking 10 random image tiles, creating a final segmentation map for each image tile by fusing the individual segmentation maps of \textsc{Source Extractor} and \textsc{MTO}. We consider any pixel assigned to an object by either detection software to be signal, and all other pixels to be noise. As illustrated in Figure \ref{fig:bimodal}, the noise follows a similar distribution to the signal. We estimated the location and standard deviations of the two noise peaks using a Gaussian Mixture Model. During training, we augmented the data with noise sampled from this bimodal distribution with a tunable noise scale parameter to avoid dominating the signal. We selected a noise scale of 0.3 as a hyperparameter, which means we added noise with a standard deviation of 0.3 times the estimated one.

As a way to additionally improve generalization and to prevent the model from being over-confident in its predictions, we introduced so-called label smoothing \citep{szegedy2016rethinking}. While label smoothing is conventionally applied to hard (e.g., one-hot encoded) labels, we deemed its application appropriate as our label generation process, despite producing soft probabilities, frequently results in `hard' 0 or 1 values at the extreme ends of the [0,1] scale. These extreme target values can promote model overconfidence. With a smoothing parameter of $\alpha$ = 0.01, this technique adjusts our labels by moving them slightly towards 0.5, according to the equation:

\begin{equation}
    label_{smooth} = label \cdot (1 - \alpha) + 0.5 \cdot \alpha
    \label{eq:label_smoothing}
\end{equation}

\noindent This modification of the target labels helps to mitigate the strong pull of the original `hard' 0 and 1 values, and encourages the model to generate less extreme probability outputs.

We trained with a batch size of 128, using it as an additional regularization strategy, and the \textsc{ADAMW} optimizer \citep{Kingma2014AdamAM,loshchilov2018decoupled} with an initial learning rate of $5\times10^{-5}$, a weight decay factor of 0.05 and the so-called ReduceLROnPlateau scheduler. This scheduler dynamically reduces the learning rate during training if no improvement is achieved for a certain number of epochs. We reduced the learning rate by 25\% if the training loss did not improve for five consecutive epochs. We trained until convergence, which took around 15 minutes for each of our 10 models on a single A100 GPU. The resulting model has approximately $15 \times 10^6$ parameters.

\begin{figure}[!htb]
\centering
\includegraphics[width=1\linewidth]{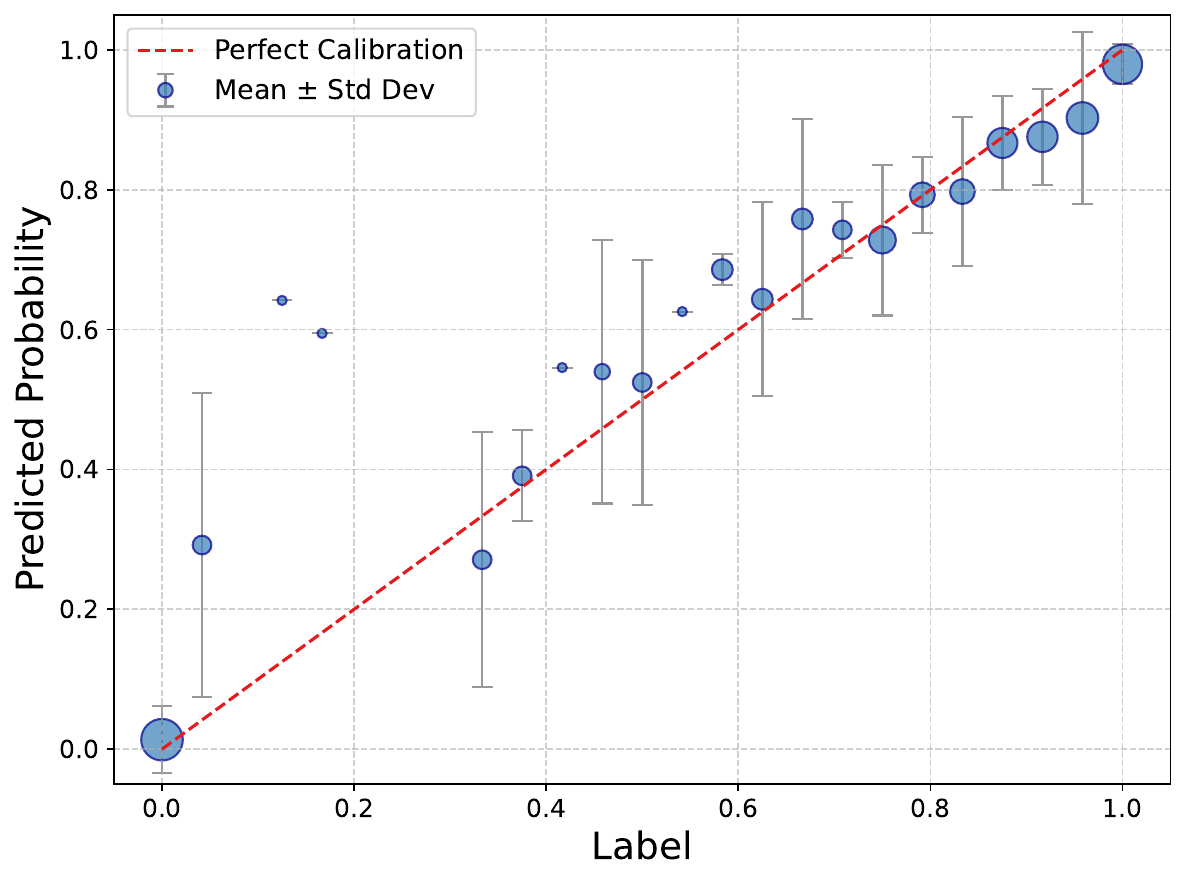}
\caption{Prediction vs label performance of the network on the test set consisting of 347 objects. On the x-axis, we show the 21 soft labels from our visual classification. On the y-axis, we show the mean predictions of the model in the 21 soft-label bins. The data point size reflects the number of examples in each bin, with the smallest points (e.g., two points between 0.1 and 0.2) representing bins containing only a single data point. Error bars represent the standard deviation of predictions within each bin. We assumed a Gaussian distribution to compute the mean and standard deviation. The red dashed line shows the one-to-one relation, indicating ideal calibration.}
\label{fig:pred_vs_true}
\end{figure}

\begin{figure}[!htb]
\centering
\includegraphics[width=1\linewidth]{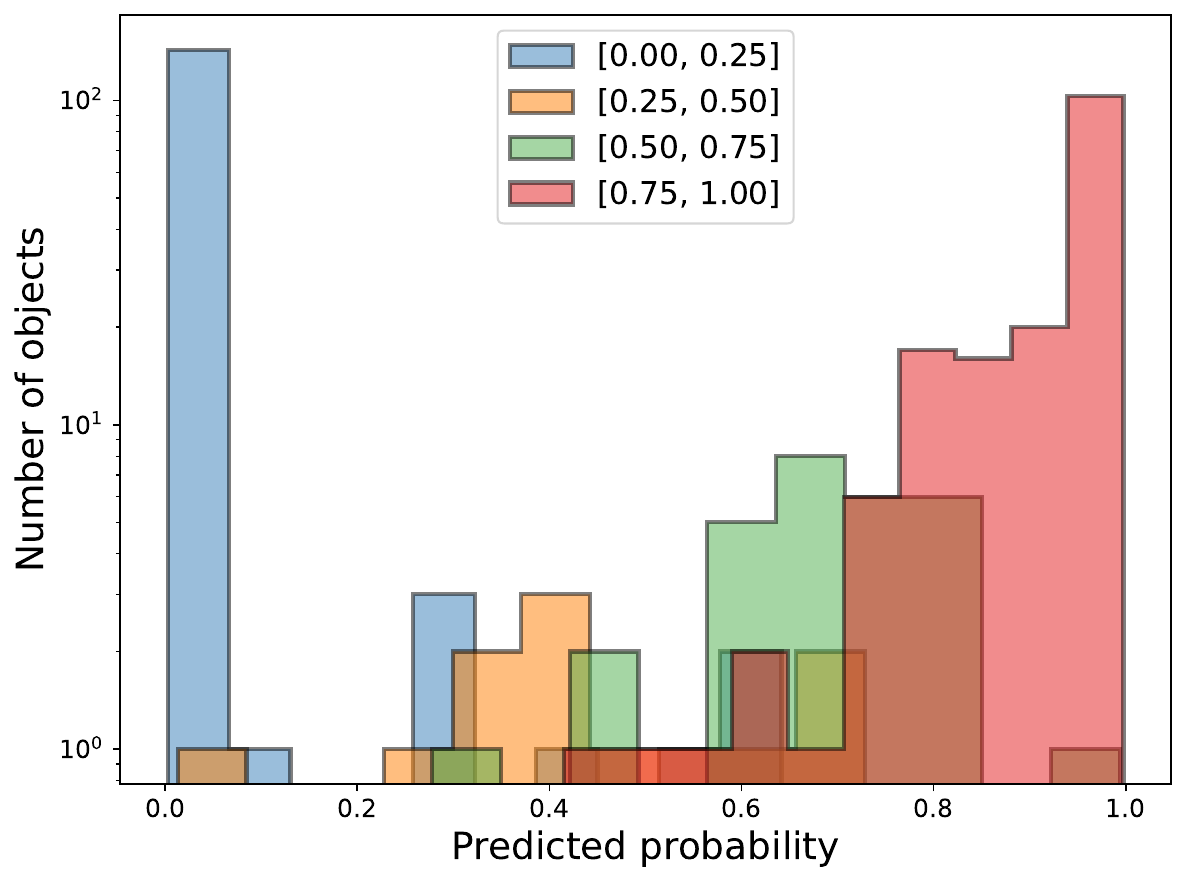}
\caption{Probability distribution of model predictions on the test dataset, color-coded by labels in four bins. The thickness of each bin distribution illustrates the level of accuracy in classifying objects of a given category. Note that the peaks of the distribution are distinct and show a consistent progression.}
\label{fig:pred_distribution}
\end{figure}

\section{Results}
\label{sec:results}

\subsection{Model performance}

\begin{figure*}[!htb]
\centering
\includegraphics[width=1\linewidth]{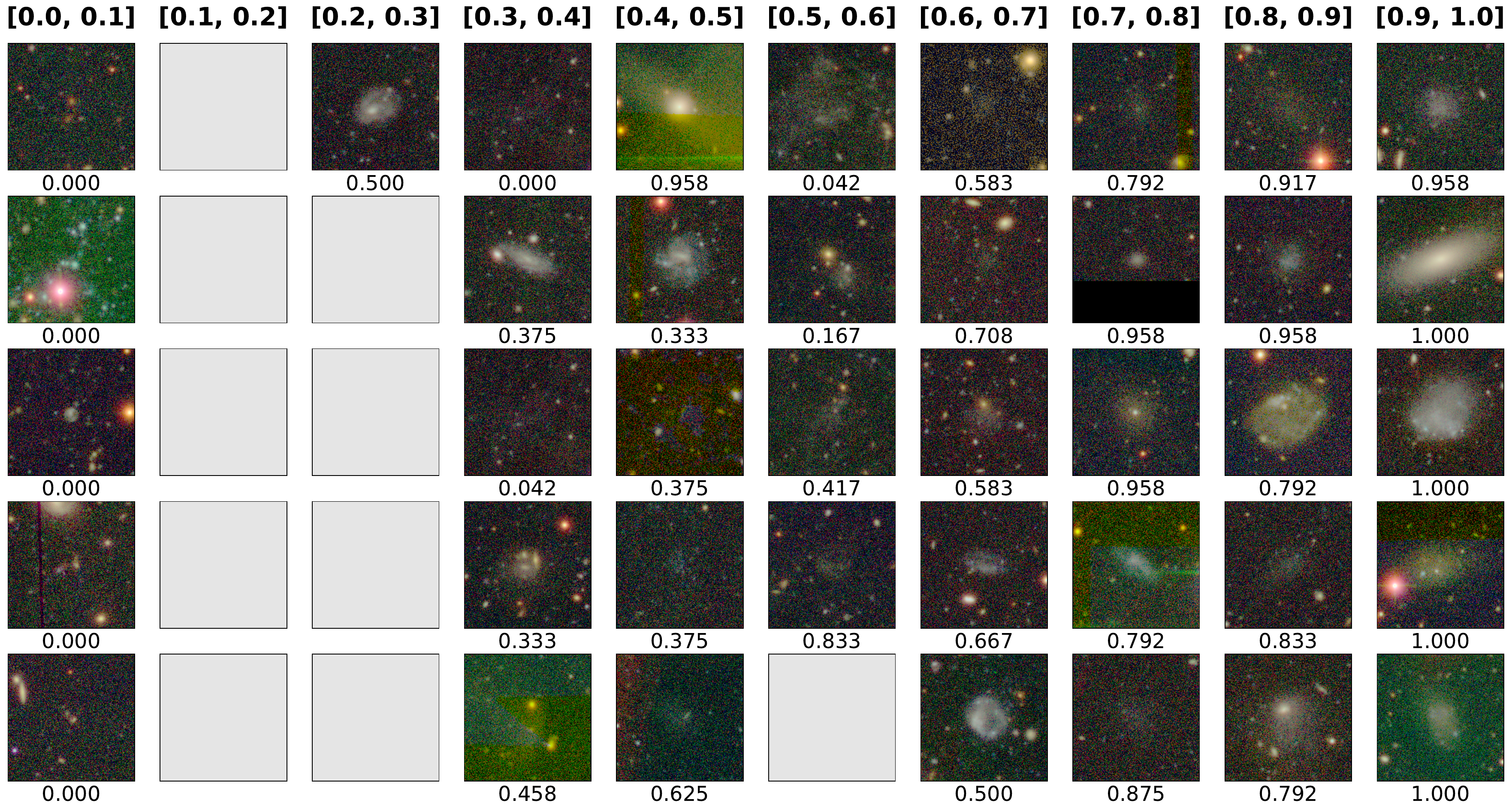}
\caption{Gallery of multiple different cutout examples from the test dataset. The columns are divided into prediction probability bins from the network. Under every cutout, we show the soft label from our visual classification. Gray squares indicate missing examples in a given probability bin.}
\label{fig:cutout_examples}
\end{figure*}

Our models reached their lowest validation loss on average after 30 epochs, and we used this ensemble of models for all further analyses. Figure \ref{fig:pred_vs_true} illustrates the model ensemble's performance on the test set. The final predictions are calculated as the average prediction of our 10 models. For simplicity, we are henceforth referring to our ensemble of models as one model. The model demonstrates generally robust performance across the soft label range, achieving a Brier score \citep{brier1950verification} of $BS$ = 0.0076. The Brier score ranges from 0 to 1, with 0 indicating perfect prediction and higher values indicating worse performance. For soft labels like ours, which range between 0 and 1, this low score indicates that our model's probabilistic predictions align well with our assigned probabilities. The Brier score, mathematically equivalent to the mean squared error between predicted and labeled probabilities, not only reflects overall prediction accuracy but also serves as a measure of the calibration quality between labels and predictions. To further quantify this calibration, the model achieves an Expected Calibration Error ($ECE$) of 0.018. The $ECE$, ranging from 0 for perfect calibration to 1, represents the weighted average of the absolute difference between the average model-predicted probability and the average value of the corresponding soft labels within each probability bin. An $ECE$ of 0.018 signifies that this weighted average deviation is 1.8 percentage points. This low $ECE$, much like the Brier score, indicates that the model's confidence levels are well-aligned with the assigned soft label probabilities. The model shows a slight systematic under-confidence for objects with labels $\geq$ 0.8 and is least confident for labels in the range of roughly 0.05 to 0.5. This is unsurprising since few objects fall into this range in the training data (see Figure \ref{fig:label_dist}). In Figure \ref{fig:pred_distribution} we illustrate the distribution of predicted probabilities color-coded by their labeled probability in bins covering intervals of 0.25 between 0 and 1. We see a clear progression of the four different probability distributions where their spread shows the level of accuracy in predicting objects of a given category. In Figure \ref{fig:cutout_examples} we show a gallery of five random example predictions on the test dataset in ten equally sized bins from 0 to 1. The column the object appears in indicates the probability bin assigned by the model, and under every image cutout, we note the object's soft label from our visual classification. We note a clear probability trend with predictions > 0.8 showing good examples of dwarf galaxies and examples in the lowest bin showing clear non-dwarfs, such as distant massive galaxies. 

\begin{figure}[!htb]
\centering
\includegraphics[width=1\linewidth]{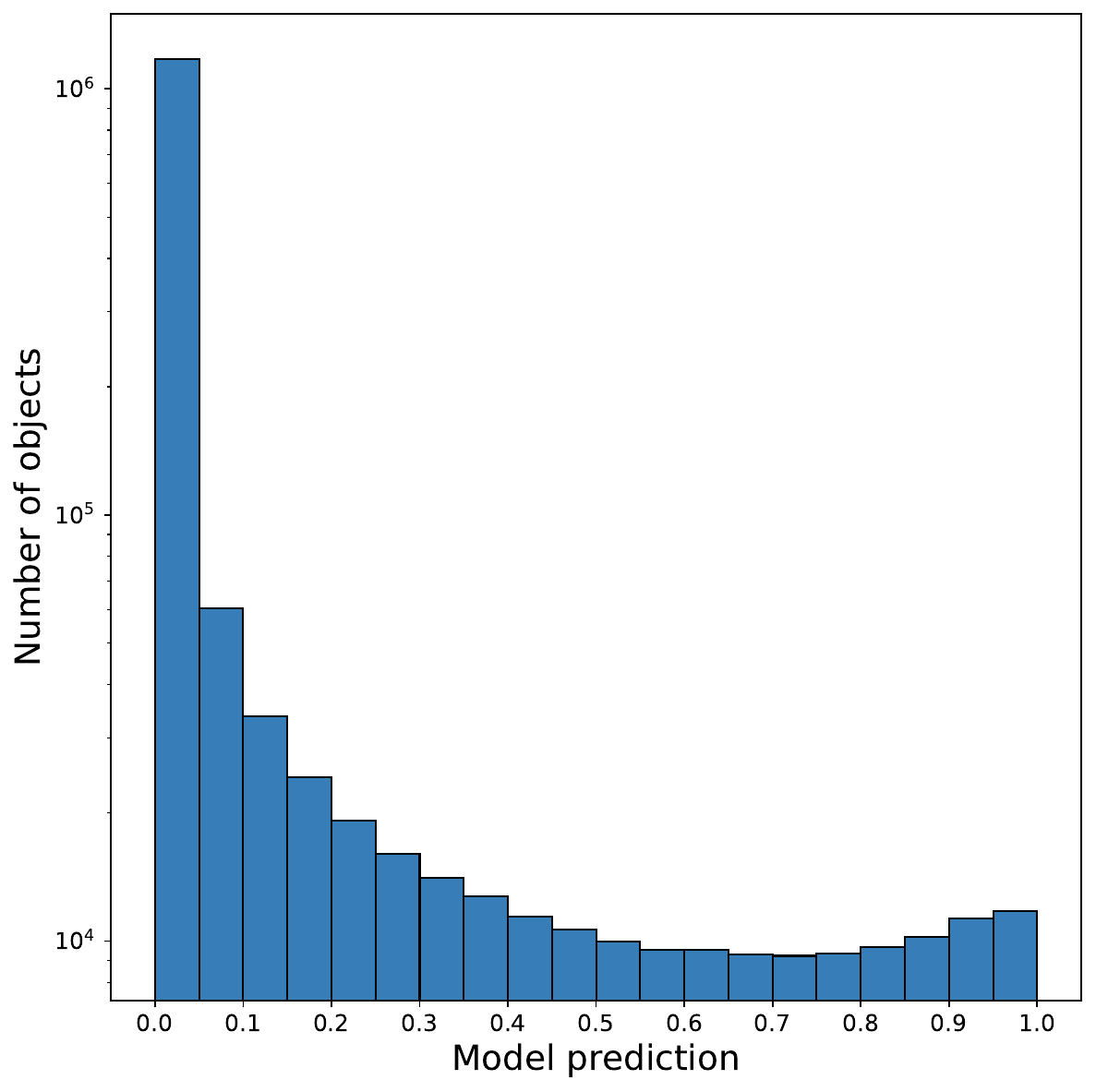}
\caption{Probability distribution of model predictions of the full de-duplicated catalog of \textsc{MTO} candidates. The y-axis is shown on a logarithmic scale.}
\label{fig:dist_all_pred}
\end{figure}

\begin{table}[!htb]
\centering
\caption{Distribution of objects across the different probability bins.}
\vspace{-10pt}
\begin{threeparttable}
\label{tab:num_preds_all}
\begin{tabular}{cc}
\toprule
Probability bin & Number of Objects \\
\midrule
0.00 -- 0.05 & 1176597 \\
0.05 -- 0.10 & 60271 \\
0.10 -- 0.15 & 33702 \\
0.15 -- 0.20 & 24251 \\
0.20 -- 0.25 & 19172 \\
0.25 -- 0.30 & 16005 \\
0.30 -- 0.35 & 14066 \\
0.35 -- 0.40 & 12737 \\
0.40 -- 0.45 & 11395 \\
0.45 -- 0.50 & 10655 \\
0.50 -- 0.55 & 9980 \\
0.55 -- 0.60 & 9533 \\
0.60 -- 0.65 & 9524 \\
0.65 -- 0.70 & 9295 \\
0.70 -- 0.75 & 9233 \\
0.75 -- 0.80 & 9352 \\
0.80 -- 0.85 & 9669 \\
0.85 -- 0.90 & 10224 \\
0.90 -- 0.95 & 11325 \\
0.95 -- 1.00 & 11747 \\
\bottomrule
\end{tabular}
\begin{tablenotes}[flushleft]
\footnotesize
\item Note: these numbers result from the complete de-duplicated catalog with detections from $\sim$16.5k image tiles.
\end{tablenotes}
\end{threeparttable}
\end{table}

\begin{figure*}[htb]
\centering
\includegraphics[width=1\linewidth]{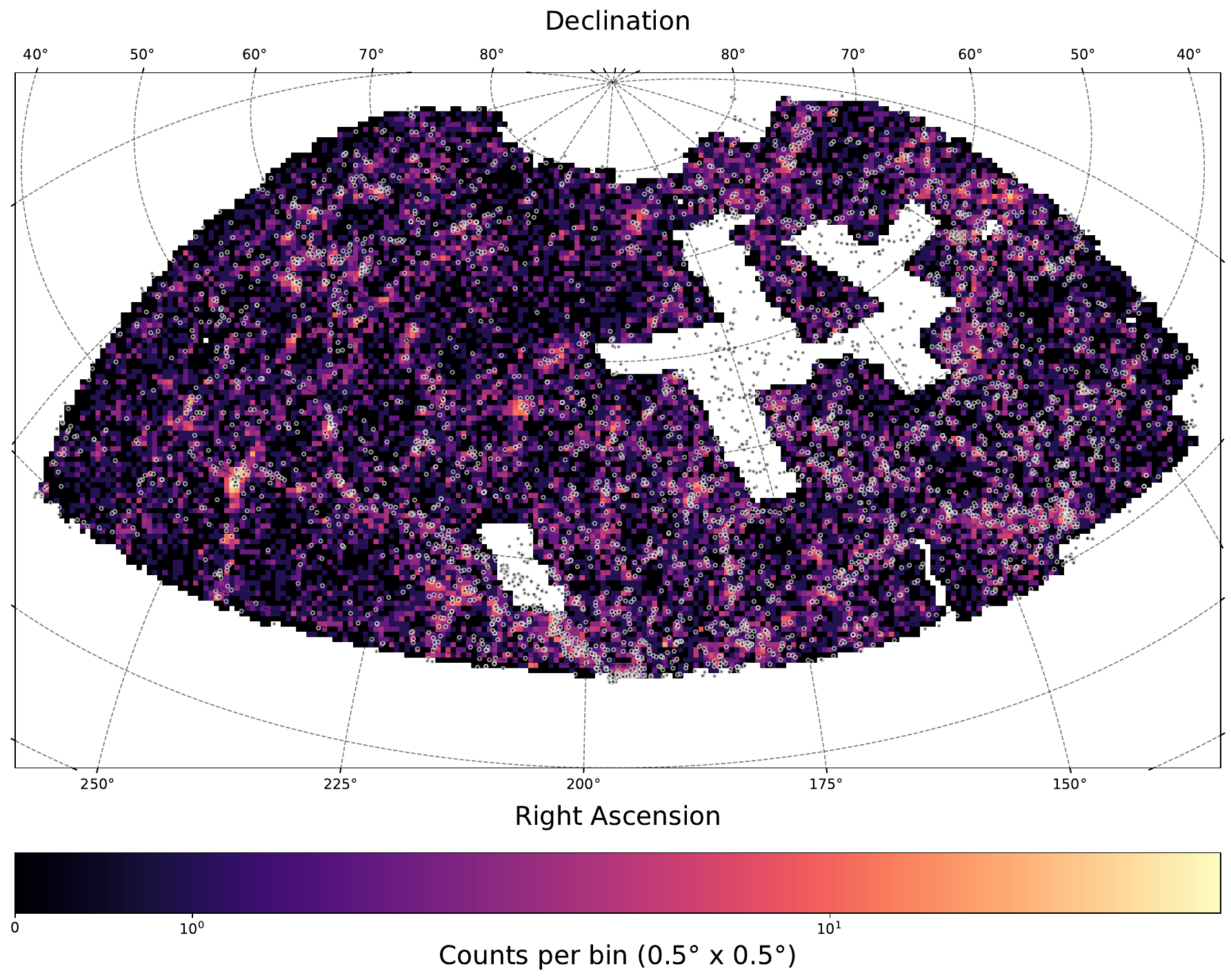}
\caption{Distribution of high-confidence dwarf galaxy candidates (prediction score > 0.9) from the GOBLIN catalog on a Lambert azimuthal equal-area projection. Each pixel (or bin) represents an equal size of 0.5$\times$0.5 degrees, regardless of the location on the map. Black dots with white outlines show massive galaxies up to 120\,Mpc with stellar masses $log\,(M_{*}/M_\odot) \geq$ 10. The color bar shows counts per bin on a logarithmic scale. Colored areas show survey coverage in \emph{g}, \emph{r}, and \emph{i}.}
\label{fig:density}
\end{figure*}

\begin{table*}[htb]
\caption{Sample of the full GOBLIN catalog with a subset of the available columns.}
\vspace{-10pt}
\begin{threeparttable}
\setlength{\tabcolsep}{5pt}
\label{tab:master_cat}
\begin{tabular}{ccccccccccccccc}
\toprule
ID & R.A. & DEC & Flux & R$_{e}$ & R$_{10}$ & R$_{90}$ & A & B & PA & m & $<\mu>_{e}$ & $p_{model}$ & $p_{label}$ & Tile \\
   & (deg) & (deg) & (ADU) & (") & (") & (") & (") & (") & (deg) & (mag) & (mag/arcsec$^{2}$) &  &  &  \\
(1) & (2) & (3) & (4) & (5) & (6) & (7) & (8) & (9) & (10) & (11) & (12) & (13) & (14) & (15) \\
\midrule
-- & 256.9975 & 30.3519 & 9021 & 2.22 & 0.73 & 4.44 & 2.6 & 1.8 & 170.8 & 20.1 & 23.4 & 0.97 & -- & (443, 241) \\
-- & 156.1404 & 47.8375 & 12007 & 2.48 & 0.84 & 4.71 & 3.4 & 1.6 & 60.1 & 19.8 & 23.0 & 0.99 & -- & (209, 276) \\
-- & 155.9319 & 48.0479 & 42099 & 3.63 & 1.11 & 7.21 & 4.6 & 2.9 & 164.0 & 18.4 & 22.7 & 0.97 & -- & (209, 276) \\
-- & 153.4467 & 76.2892 & 20454 & 3.70 & 1.26 & 7.82 & 6.2 & 4.2 & 165.4 & 19.2 & 23.7 & 0.99 & -- & (072, 333) \\
-- & 138.9832 & 79.0316 & 153146 & 4.13 & 1.62 & 8.73 & 4.9 & 3.4 & 123.4 & 17.0 & 21.7 & 1.00 & -- & (053, 338) \\
\bottomrule
\end{tabular}
\begin{tablenotes}[flushleft]
\item Columns: (1) ID if the object was in one of the dwarf catalogs from the literature, (2, 3) coordinates in degrees in the J2000 frame. (4) total object flux within the \textsc{MTO} segment, (5) effective radius in arcseconds, (6, 7) radii enclosing 10\% and 90\% of the total light, respectively, (8) major axis in arcseconds, (9) minor axis in arcseconds, (10) major axis position angle in degrees $\in$ [0, 180]\,deg, measured from north to east, (11) apparent magnitude, (12) mean effective surface brightness, (13) model prediction, (14) soft label, if the object was used to train the model and (15) UNIONS tile number the object was detected in. Note: this sample table only contains a subset of the columns in the full catalog. The \textsc{MTO} parameters (columns 4-12) are only shown for the \emph{r}-band but are also available in \emph{g} and \emph{i} in the full catalog.
\end{tablenotes}
\end{threeparttable}
\end{table*}

\subsection{The GOBLIN catalog}

Using the trained model, we ran inference on all ($\sim$1.5\,million) \textsc{MTO} candidates that passed the filtering and band cross-matching step described in Section \ref{sec:postprocessing}. Due to the tiling architecture of the UNIONS footprint, there is a slight overlap ($\sim$3\% total) between neighboring image tiles. Therefore, we removed any duplicate detections stemming from this overlap and from extended sources featuring regions of strongly varying brightness that may have been detected as two or more distinct objects. To achieve this efficiently, we use a Friends-of-Friends (FoF) clustering approach with a union-find algorithm to group objects that are close in angular separation. We used the object's mean $R_{e}$ in the available bands as estimated by \textsc{MTO} to determine whether they are in the same group. Two objects are considered part of the same group if the angular separation between them is smaller than 1.5 times the effective radius of the larger object. We chose this factor to account for errors in the estimation of $R_{e}$. We kept only the object with the highest model prediction from each group. This procedure removed $\sim$56k objects from the catalog, where $\sim$55k of the duplicates are due to the overlapping nature of the footprint and $\sim$1k are same-tile duplicates, mostly due to multiple detections of the same extended object.

We show the distribution of the model predictions on the de-duplicated catalog in Figure \ref{fig:dist_all_pred}. Table \ref{tab:num_preds_all} shows the corresponding detailed number of predictions in probability bins of 0.05. We note the stark difference between the populations in the first (0\,-\,0.05) and second (0.05\,-\,0.1) bins, illustrating that the vast majority ($\sim$80\%) of \textsc{MTO} candidates are non-dwarfs with a high confidence. Our catalog includes 42,965 objects with a model prediction score > 0.8. We report this number because our visual inspection of the test set (see Figure \ref{fig:cutout_examples}) suggests that objects in these highest probability bins are consistently good dwarf candidates. However, this 0.8 threshold serves merely as a reference point — the full catalog contains objects across the entire probability range, and many genuine dwarf galaxies likely exist at lower probability scores. Users of the GOBLIN catalog may select different thresholds depending on their specific requirements for completeness versus contamination. After removing objects that were in the training data, 41,462 candidates remain with probabilities > 0.8.

In Figure \ref{fig:density} we show a 2D density map of all objects in the catalog with a model prediction > 0.9 (23,072). We overlay the distribution of massive galaxies up to 120\,Mpc and a stellar mass $log\,(M_{*}/M_\odot) \geq$ 10 that fall within the footprint. We gather the information about nearby massive galaxies from the Heraklion Extragalactic CATaloguE \citep[HECATE;][]{2021MNRAS.506.1896K}, a value-added catalog of galaxies up to a distance of 200\,Mpc from the Milky Way. As expected, we see that peaks in the density map coincide with the presence of massive host galaxies. Blank regions within the footprint have not yet been covered by the three bands we used in this work.

It is important to note that while this spatial correlation provides circumstantial evidence supporting many of our candidates, our catalog has a fundamental limitation: we lack distance measurements for the vast majority of these objects. Even galaxies with high probability scores should still be considered candidates, as background spiral galaxies at larger distances can appear morphologically similar to nearby dwarf galaxies when their detailed features (such as spiral arms) become unresolvable due to distance. The angular resolution limits of our observations mean that distant galaxies lose their distinctive morphological characteristics and may be classified as dwarf-like based on their apparent size, surface brightness, and diffuse appearance. Therefore, although the correlation with massive galaxies is encouraging, individual confirmation of these candidates requires dedicated follow-up observations.

We present a small sample of our extensive GOBLIN catalog in Table \ref{tab:master_cat}. The full version will be available in machine-readable form online. The catalog contains the object's coordinates right ascension and declination in degrees, the literature ID if it is in one of the dwarf catalogs we used for training, a column indicating whether or not the object was used to train our model, its soft label from our visual classification, the UNIONS tile it appears in, the model prediction and several photometric parameters provided by \textsc{MTO} and derived properties such as magnitude and mean effective surface brightness. We provide these parameters in \emph{g}, \emph{r}, and \emph{i}. If a given object was not detected in one of the bands, the corresponding columns are filled with NaN values. We note that these \textsc{MTO} parameters are derived from the binned images. The magnitudes and therefore mean effective surface brightnesses are calculated by multiplying the total flux measured on the binned image tiles by a factor of 16 to account for the 4$\times$4 binning scheme. They should not be regarded as accurate measurements but rather as rough estimates. To gain accurate structural parameters, 2D model fitting via software such as \textsc{GALFIT} \citep{peng2010detailed} should be used.

To illustrate and quantify the quality of the parameters obtained through \textsc{MTO}, we gathered photometric measurements for the dwarf galaxies from the literature that we used to train our model. Different parameters are available for the source catalogs we used. We gathered the effective radius, the magnitude, as well as the mean effective surface brightness in the \emph{g}-band and \emph{r}-band, where available. We show parameter comparisons between \textsc{MTO} and literature measurements in Figures \ref{fig:appendix_param_comp_reff} and \ref{fig:appendix_param_comp_mag_mu} in Appendix \ref{sec:appendix_param_comp}. We note that \textsc{MTO} largely underestimates the effective radius, in particular towards larger values and for objects from the SMUDGES survey \citep{2019ApJS..240....1Z,2023AJ....166..185G,2023ApJS..267...27Z}. This discrepancy is caused by the method by which \textsc{MTO} calculates the effective radius (see Sec. \ref{sec:postprocessing}). The magnitudes are approximately consistent, albeit with a positive bias for the \emph{g}-band and a scaling bias for the \emph{r}-band. We can correct for these biases by fitting linear relationships to the data. The standard deviations of the residuals for these linear fits are $\sigma_{res} (g)$ = 0.59 and $\sigma_{res} (r)$ = 0.42. In the final catalog, we report the measured magnitudes as well as the corrected relationships for the \emph{g}-band and \emph{r}-band. There are no corresponding \emph{i}-band measurements available in the literature. As a consequence of the discrepancy in the effective radius measurement, the mean effective surface brightness is largely underestimated by \textsc{MTO}. We do not attempt to correct the effective radius and the mean effective surface brightness since the discrepancy cannot be accounted for by a simple linear shift.

\section{Summary and conclusions}
\label{sec:sum_concl}

In this work, we have conducted an extensive search for dwarf galaxies in the UNIONS footprint, a deep and wide optical survey in the northern hemisphere. We have developed an automatic method to robustly detect these objects in an elaborate multi-step process that heavily relies on image preprocessing and uses machine learning in the final classification step. We ran our pipeline on all UNIONS tiles that have been observed and reduced in the \emph{g}, \emph{r}, and \emph{i} bands at the time of writing. These bands are most relevant for detecting LSB objects, cover most of the final UNIONS footprint, and show a good overlap. For each band, there are $\sim$20k 0.51$\times$0.51 deg image tiles available, and $\sim$16.5k are covered by all three. After a preprocessing routine that involves binning the image, detecting and masking corrupted data patches, small sources, and stars, we employed the software \textsc{MTO} to detect LSB objects. We filtered the resulting detections by making parameter cuts using known dwarfs from the literature that reside in the UNIONS footprint. Cross-matching the detections between the three bands and requiring a detection to be present in at least two out of three bands led to the final catalog of candidates ($\sim$90 detections per tile). 

We fine-tuned the deep-learning model \textsc{Zoobot}, which was pre-trained on galaxy morphology classification tasks using labeled data from the Galaxy Zoo project. After initial attempts using dwarf galaxies from various catalogs from the literature yielded poor results, we re-labeled $\sim$2,000 literature dwarfs plus $\sim$1,500 non-dwarf examples from the UNIONS data. The average of these labels represents the probability that a given object is a dwarf galaxy. We used this information to train our dwarf galaxy classifier, and rather than outputting a binary classification (dwarf or no dwarf), the model learned to predict the probability that an object is a dwarf. 

We applied the trained model on the $\sim$1.5$\times$10$^{6}$ candidates and thus assigned a dwarf probability to each of them. After removing duplicate detections due to the small overlap between neighboring tiles and large objects being split into multiple detections, 1,478,733 objects remain in the catalog. Following our model predictions, the vast majority ($\sim$80\%) of these objects are non-dwarfs ($p_{model}$\,<\,0.05). 42,965 objects reach a probability > 0.8 and 41,462 of these were not used to train the model. 

A density map of the highest probability candidates reveals that the detected objects follow the projected distribution of massive galaxies in the sky. In future work, we will conduct a detailed study of the distribution of these identified objects in relation to their potential host galaxies. 

While our detection approach has identified a vast number of dwarf galaxy candidates, we acknowledge that some of these objects may have been previously detected in other surveys. Given the distributed nature of astronomical databases and the continuous updates to the literature, we present this as an independent catalog rather than claiming first detections. We provide the complete de-duplicated catalog of $\sim$1.5 million objects along with their classification probabilities, allowing researchers to select candidates based on probability thresholds that best suit their specific science goals. Whether studying the most secure candidates ($p_{model}$ > 0.9) or investigating a larger sample with more relaxed probability cuts, the GOBLIN catalog's flexible nature supports a wide range of scientific investigations. We anticipate that this comprehensive dataset will serve as a valuable resource for the community, enabling various follow-up studies and complementing existing catalogs of LSB galaxies.

\begin{acknowledgements}
      We thank the referee for the constructive report, which helped to clarify and improve the manuscript.
      O.M. and N.H. are grateful to the Swiss National Science Foundation for financial support under the grant number PZ00P2\_202104. N.H. thanks Stephen Gwyn for the help provided with questions regarding the UNIONS data. N.H. also thanks Jean-Charles Cuillandre for clarifying the nature of the surface brightness limit for the UNIONS \emph{r}-band data. E.S. is grateful to the Leverhulme Trust for funding under the grant number RPG-2021-205. D.C. is grateful for the financial support provided by the Harding Distinguished Postgraduate Scholars Programme. MJH acknowledges support from NSERC through a Discovery Grant. We are honored and grateful for the opportunity of observing the Universe from Maunakea and Haleakala, which both have cultural, historical and natural significance in Hawaii. This work is based on data obtained as part of the Canada-France Imaging Survey, a CFHT large program of the National Research Council of Canada and the French Centre National de la Recherche Scientifique. Based on observations obtained with MegaPrime/MegaCam, a joint project of CFHT and CEA Saclay, at the Canada-France-Hawaii Telescope (CFHT) which is operated by the National Research Council (NRC) of Canada, the Institut National des Science de l’Univers (INSU) of the Centre National de la Recherche Scientifique (CNRS) of France, and the University of Hawaii. This research used the facilities of the Canadian Astronomy Data Centre operated by the National Research Council of Canada with the support of the Canadian Space Agency. This research is based in part on data collected at Subaru Telescope, which is operated by the National Astronomical Observatory of Japan. Pan-STARRS is a project of the Institute for Astronomy of the University of Hawaii, and is supported by the NASA SSO Near Earth Observation Program under grants 80NSSC18K0971, NNX14AM74G, NNX12AR65G, NNX13AQ47G, NNX08AR22G, 80NSSC21K1572 and by the State of Hawaii. This work has made use of data from the European Space Agency (ESA) mission {\it Gaia} (\url{https://www.cosmos.esa.int/gaia}), processed by the {\it Gaia} Data Processing and Analysis Consortium (DPAC, \url{https://www.cosmos.esa.int/web/gaia/dpac/consortium}). Funding for the DPAC has been provided by national institutions, in particular the institutions participating in the {\it Gaia} Multilateral  Agreement. This research made use of Photutils, an Astropy package for detection and photometry of astronomical sources \citep{larry_bradley_2024_12585239}. 
\end{acknowledgements}

%
\bibliographystyle{aa} 
\bibliography{aa.bib} 

\begin{thebibliography}{156}
\expandafter\ifx\csname natexlab\endcsname\relax\def\natexlab#1{#1}\fi

\bibitem[{{Abbott} {et~al.}(2018){Abbott}, {Abdalla}, {Allam}, {Amara}, {Annis}, {Asorey}, {Avila}, {Ballester}, {Banerji}, {Barkhouse}, {Baruah}, {Baumer}, {Bechtol}, {Becker}, {Benoit-L{\'e}vy}, {Bernstein}, {Bertin}, {Blazek}, {Bocquet}, {Brooks}, {Brout}, {Buckley-Geer}, {Burke}, {Busti}, {Campisano}, {Cardiel-Sas}, {Carnero Rosell}, {Carrasco Kind}, {Carretero}, {Castander}, {Cawthon}, {Chang}, {Chen}, {Conselice}, {Costa}, {Crocce}, {Cunha}, {D'Andrea}, {da Costa}, {Das}, {Daues}, {Davis}, {Davis}, {De Vicente}, {DePoy}, {DeRose}, {Desai}, {Diehl}, {Dietrich}, {Dodelson}, {Doel}, {Drlica-Wagner}, {Eifler}, {Elliott}, {Evrard}, {Farahi}, {Fausti Neto}, {Fernandez}, {Finley}, {Flaugher}, {Foley}, {Fosalba}, {Friedel}, {Frieman}, {Garc{\'\i}a-Bellido}, {Gaztanaga}, {Gerdes}, {Giannantonio}, {Gill}, {Glazebrook}, {Goldstein}, {Gower}, {Gruen}, {Gruendl}, {Gschwend}, {Gupta}, {Gutierrez}, {Hamilton}, {Hartley}, {Hinton}, {Hislop}, {Hollowood}, {Honscheid}, {Hoyle}, {Huterer}, {Jain}, {James}, {Jeltema},
  {Johnson}, {Johnson}, {Kacprzak}, {Kent}, {Khullar}, {Klein}, {Kovacs}, {Koziol}, {Krause}, {Kremin}, {Kron}, {Kuehn}, {Kuhlmann}, {Kuropatkin}, {Lahav}, {Lasker}, {Li}, {Li}, {Liddle}, {Lima}, {Lin}, {L{\'o}pez-Reyes}, {MacCrann}, {Maia}, {Maloney}, {Manera}, {March}, {Marriner}, {Marshall}, {Martini}, {McClintock}, {McKay}, {McMahon}, {Melchior}, {Menanteau}, {Miller}, {Miquel}, {Mohr}, {Morganson}, {Mould}, {Neilsen}, {Nichol}, {Nogueira}, {Nord}, {Nugent}, {Nunes}, {Ogando}, {Old}, {Pace}, {Palmese}, {Paz-Chinch{\'o}n}, {Peiris}, {Percival}, {Petravick}, {Plazas}, {Poh}, {Pond}, {Porredon}, {Pujol}, {Refregier}, {Reil}, {Ricker}, {Rollins}, {Romer}, {Roodman}, {Rooney}, {Ross}, {Rykoff}, {Sako}, {Sanchez}, {Sanchez}, {Santiago}, {Saro}, {Scarpine}, {Scolnic}, {Serrano}, {Sevilla-Noarbe}, {Sheldon}, {Shipp}, {Silveira}, {Smith}, {Smith}, {Smith}, {Soares-Santos}, {Sobreira}, {Song}, {Stebbins}, {Suchyta}, {Sullivan}, {Swanson}, {Tarle}, {Thaler}, {Thomas}, {Thomas}, {Troxel}, {Tucker}, {Vikram}, {Vivas},
  {Walker}, {Wechsler}, {Weller}, {Wester}, {Wolf}, {Wu}, {Yanny}, {Zenteno}, {Zhang}, {Zuntz}, {DES Collaboration}, {Juneau}, {Fitzpatrick}, \& {Nikutta}}]{2018ApJS..239...18A}
{Abbott}, T.~M.~C., {Abdalla}, F.~B., {Allam}, S., {et~al.} 2018, \apjs, 239, 18

\bibitem[{{Aihara} {et~al.}(2018{\natexlab{a}}){Aihara}, {Arimoto}, {Armstrong}, {Arnouts}, {Bahcall}, {Bickerton}, {Bosch}, {Bundy}, {Capak}, {Chan}, {Chiba}, {Coupon}, {Egami}, {Enoki}, {Finet}, {Fujimori}, {Fujimoto}, {Furusawa}, {Furusawa}, {Goto}, {Goulding}, {Greco}, {Greene}, {Gunn}, {Hamana}, {Harikane}, {Hashimoto}, {Hattori}, {Hayashi}, {Hayashi}, {He{\l}miniak}, {Higuchi}, {Hikage}, {Ho}, {Hsieh}, {Huang}, {Huang}, {Ikeda}, {Imanishi}, {Inoue}, {Iwasawa}, {Iwata}, {Jaelani}, {Jian}, {Kamata}, {Karoji}, {Kashikawa}, {Katayama}, {Kawanomoto}, {Kayo}, {Koda}, {Koike}, {Kojima}, {Komiyama}, {Konno}, {Koshida}, {Koyama}, {Kusakabe}, {Leauthaud}, {Lee}, {Lin}, {Lin}, {Lupton}, {Mandelbaum}, {Matsuoka}, {Medezinski}, {Mineo}, {Miyama}, {Miyatake}, {Miyazaki}, {Momose}, {More}, {More}, {Moritani}, {Moriya}, {Morokuma}, {Mukae}, {Murata}, {Murayama}, {Nagao}, {Nakata}, {Niida}, {Niikura}, {Nishizawa}, {Obuchi}, {Oguri}, {Oishi}, {Okabe}, {Okamoto}, {Okura}, {Ono}, {Onodera}, {Onoue}, {Osato}, {Ouchi},
  {Price}, {Pyo}, {Sako}, {Sawicki}, {Shibuya}, {Shimasaku}, {Shimono}, {Shirasaki}, {Silverman}, {Simet}, {Speagle}, {Spergel}, {Strauss}, {Sugahara}, {Sugiyama}, {Suto}, {Suyu}, {Suzuki}, {Tait}, {Takada}, {Takata}, {Tamura}, {Tanaka}, {Tanaka}, {Tanaka}, {Tanaka}, {Terai}, {Terashima}, {Toba}, {Tominaga}, {Toshikawa}, {Turner}, {Uchida}, {Uchiyama}, {Umetsu}, {Uraguchi}, {Urata}, {Usuda}, {Utsumi}, {Wang}, {Wang}, {Wong}, {Yabe}, {Yamada}, {Yamanoi}, {Yasuda}, {Yeh}, {Yonehara}, \& {Yuma}}]{2018PASJ...70S...4A}
{Aihara}, H., {Arimoto}, N., {Armstrong}, R., {et~al.} 2018{\natexlab{a}}, \pasj, 70, S4

\bibitem[{{Aihara} {et~al.}(2018{\natexlab{b}}){Aihara}, {Armstrong}, {Bickerton}, {Bosch}, {Coupon}, {Furusawa}, {Hayashi}, {Ikeda}, {Kamata}, {Karoji}, {Kawanomoto}, {Koike}, {Komiyama}, {Lang}, {Lupton}, {Mineo}, {Miyatake}, {Miyazaki}, {Morokuma}, {Obuchi}, {Oishi}, {Okura}, {Price}, {Takata}, {Tanaka}, {Tanaka}, {Tanaka}, {Uchida}, {Uraguchi}, {Utsumi}, {Wang}, {Yamada}, {Yamanoi}, {Yasuda}, {Arimoto}, {Chiba}, {Finet}, {Fujimori}, {Fujimoto}, {Furusawa}, {Goto}, {Goulding}, {Gunn}, {Harikane}, {Hattori}, {Hayashi}, {He{\l}miniak}, {Higuchi}, {Hikage}, {Ho}, {Hsieh}, {Huang}, {Huang}, {Imanishi}, {Iwata}, {Jaelani}, {Jian}, {Kashikawa}, {Katayama}, {Kojima}, {Konno}, {Koshida}, {Kusakabe}, {Leauthaud}, {Lee}, {Lin}, {Lin}, {Mandelbaum}, {Matsuoka}, {Medezinski}, {Miyama}, {Momose}, {More}, {More}, {Mukae}, {Murata}, {Murayama}, {Nagao}, {Nakata}, {Niida}, {Niikura}, {Nishizawa}, {Oguri}, {Okabe}, {Ono}, {Onodera}, {Onoue}, {Ouchi}, {Pyo}, {Shibuya}, {Shimasaku}, {Simet}, {Speagle}, {Spergel}, {Strauss},
  {Sugahara}, {Sugiyama}, {Suto}, {Suzuki}, {Tait}, {Takada}, {Terai}, {Toba}, {Turner}, {Uchiyama}, {Umetsu}, {Urata}, {Usuda}, {Yeh}, \& {Yuma}}]{2018PASJ...70S...8A}
{Aihara}, H., {Armstrong}, R., {Bickerton}, S., {et~al.} 2018{\natexlab{b}}, \pasj, 70, S8

\bibitem[{{Ay{\c{c}}oberry} {et~al.}(2023){Ay{\c{c}}oberry}, {Ajani}, {Guinot}, {Kilbinger}, {Pettorino}, {Farrens}, {Starck}, {Gavazzi}, \& {Hudson}}]{2023A&A...671A..17A}
{Ay{\c{c}}oberry}, E., {Ajani}, V., {Guinot}, A., {et~al.} 2023, \aap, 671, A17

\bibitem[{{Babusiaux} {et~al.}(2023){Babusiaux}, {Fabricius}, {Khanna}, {Muraveva}, {Reyl{\'e}}, {Spoto}, {Vallenari}, {Luri}, {Arenou}, {{\'A}lvarez}, {Anders}, {Antoja}, {Balbinot}, {Barache}, {Bauchet}, {Bossini}, {Busonero}, {Cantat-Gaudin}, {Carrasco}, {Dafonte}, {Diakit{\'e}}, {Figueras}, {Garcia-Gutierrez}, {Garofalo}, {Helmi}, {Jim{\'e}nez-Arranz}, {Jordi}, {Kervella}, {Kostrzewa-Rutkowska}, {Leclerc}, {Licata}, {Manteiga}, {Masip}, {Mongui{\'o}}, {Ramos}, {Robichon}, {Robin}, {Romero-G{\'o}mez}, {S{\'a}ez}, {Santove{\~n}a}, {Spina}, {Torralba Elipe}, \& {Weiler}}]{2023A&A...674A..32B}
{Babusiaux}, C., {Fabricius}, C., {Khanna}, S., {et~al.} 2023, \aap, 674, A32

\bibitem[{Barbary(2016)}]{Barbary2016}
Barbary, K. 2016, Journal of Open Source Software, 1, 58

\bibitem[{Bennet {et~al.}(2020)Bennet, Sand, Crnojevi{\'c}, Spekkens, Karunakaran, Zaritsky, \& Mutlu-Pakdil}]{bennet2020satellite}
Bennet, P., Sand, D., Crnojevi{\'c}, D., {et~al.} 2020, The Astrophysical Journal Letters, 893, L9

\bibitem[{{Bennet} {et~al.}(2017){Bennet}, {Sand}, {Crnojevi{\'c}}, {Spekkens}, {Zaritsky}, \& {Karunakaran}}]{2017ApJ...850..109B}
{Bennet}, P., {Sand}, D.~J., {Crnojevi{\'c}}, D., {et~al.} 2017, \apj, 850, 109

\bibitem[{Bertin(2010)}]{bertin2010swarp}
Bertin, E. 2010, Astrophysics Source Code Library, ascl

\bibitem[{{Bertin} \& {Arnouts}(1996)}]{1996A&AS..117..393B}
{Bertin}, E. \& {Arnouts}, S. 1996, \aaps, 117, 393

\bibitem[{Bickley {et~al.}(2021)Bickley, Bottrell, Hani, Ellison, Teimoorinia, Yi, Wilkinson, Gwyn, \& Hudson}]{bickley2021convolutional}
Bickley, R.~W., Bottrell, C., Hani, M.~H., {et~al.} 2021, Monthly Notices of the Royal Astronomical Society, 504, 372

\bibitem[{Bickley {et~al.}(2022)Bickley, Ellison, Patton, Bottrell, Gwyn, \& Hudson}]{bickley2022star}
Bickley, R.~W., Ellison, S.~L., Patton, D.~R., {et~al.} 2022, Monthly Notices of the Royal Astronomical Society, 514, 3294

\bibitem[{Bickley {et~al.}(2023)Bickley, Ellison, Patton, \& Wilkinson}]{bickley2023agns}
Bickley, R.~W., Ellison, S.~L., Patton, D.~R., \& Wilkinson, S. 2023, Monthly Notices of the Royal Astronomical Society, 519, 6149

\bibitem[{B{\'\i}lek {et~al.}(2020)B{\'\i}lek, Duc, Cuillandre, Gwyn, Cappellari, Bekaert, Bonfini, Bitsakis, Paudel, Krajnovi{\'c}, {et~al.}}]{bilek2020census}
B{\'\i}lek, M., Duc, P.-A., Cuillandre, J.-C., {et~al.} 2020, Monthly Notices of the Royal Astronomical Society, 498, 2138

\bibitem[{{Binggeli} {et~al.}(1985){Binggeli}, {Sandage}, \& {Tammann}}]{1985AJ.....90.1681B}
{Binggeli}, B., {Sandage}, A., \& {Tammann}, G.~A. 1985, \aj, 90, 1681

\bibitem[{Binggeli {et~al.}(1990)Binggeli, Tarenghi, \& Sandage}]{binggeli1990abundance}
Binggeli, B., Tarenghi, M., \& Sandage, A. 1990, Astronomy and Astrophysics (ISSN 0004-6361), vol. 228, no. 1, Feb. 1990, p. 42-60. Research supported by SNSF., 228, 42

\bibitem[{Bom {et~al.}(2019)Bom, Poh, Nord, Blanco-Valentin, \& Dias}]{bom2019deep}
Bom, C., Poh, J., Nord, B., Blanco-Valentin, M., \& Dias, L. 2019, arXiv preprint arXiv:1911.06341

\bibitem[{{Bosch} {et~al.}(2019){Bosch}, {AlSayyad}, {Armstrong}, {Bellm}, {Chiang}, {Eggl}, {Findeisen}, {Fisher-Levine}, {Guy}, {Guyonnet}, {Ivezi{\'c}}, {Jenness}, {Kov{\'a}cs}, {Krughoff}, {Lupton}, {Lust}, {MacArthur}, {Meyers}, {Moolekamp}, {Morrison}, {Morton}, {O'Mullane}, {Parejko}, {Plazas}, {Price}, {Rawls}, {Reed}, {Schellart}, {Slater}, {Sullivan}, {Swinbank}, {Taranu}, {Waters}, \& {Wood-Vasey}}]{2019ASPC..523..521B}
{Bosch}, J., {AlSayyad}, Y., {Armstrong}, R., {et~al.} 2019, in Astronomical Society of the Pacific Conference Series, Vol. 523, Astronomical Data Analysis Software and Systems XXVII, ed. P.~J. {Teuben}, M.~W. {Pound}, B.~A. {Thomas}, \& E.~M. {Warner}, 521

\bibitem[{{Bosch} {et~al.}(2018){Bosch}, {Armstrong}, {Bickerton}, {Furusawa}, {Ikeda}, {Koike}, {Lupton}, {Mineo}, {Price}, {Takata}, {Tanaka}, {Yasuda}, {AlSayyad}, {Becker}, {Coulton}, {Coupon}, {Garmilla}, {Huang}, {Krughoff}, {Lang}, {Leauthaud}, {Lim}, {Lust}, {MacArthur}, {Mandelbaum}, {Miyatake}, {Miyazaki}, {Murata}, {More}, {Okura}, {Owen}, {Swinbank}, {Strauss}, {Yamada}, \& {Yamanoi}}]{2018PASJ...70S...5B}
{Bosch}, J., {Armstrong}, R., {Bickerton}, S., {et~al.} 2018, \pasj, 70, S5

\bibitem[{Bradley {et~al.}(2024)Bradley, Sip{\H o}cz, Robitaille, Tollerud, Vin{\'{\i}}cius, Deil, Barbary, Wilson, Busko, Donath, G{\"u}nther, Cara, Lim, Me{\ss}linger, Burnett, Conseil, Droettboom, Bostroem, Bray, Bratholm, Jamieson, Ginsburg, Barentsen, Craig, Pascual, Rathi, Perrin, Morris, \& Perren}]{larry_bradley_2024_12585239}
Bradley, L., Sip{\H o}cz, B., Robitaille, T., {et~al.} 2024, astropy/photutils: 1.13.0

\bibitem[{Brier(1950)}]{brier1950verification}
Brier, G.~W. 1950, Monthly weather review, 78, 1

\bibitem[{Bullock \& Boylan-Kolchin(2017)}]{bullock2017small}
Bullock, J.~S. \& Boylan-Kolchin, M. 2017, Annual Review of Astronomy and Astrophysics, 55, 343

\bibitem[{Caldeira {et~al.}(2019)Caldeira, Wu, Nord, Avestruz, Trivedi, \& Story}]{caldeira2019deepcmb}
Caldeira, J., Wu, W.~K., Nord, B., {et~al.} 2019, Astronomy and Computing, 28, 100307

\bibitem[{Carlsten {et~al.}(2019)Carlsten, Beaton, Greco, \& Greene}]{carlsten2019using}
Carlsten, S.~G., Beaton, R.~L., Greco, J.~P., \& Greene, J.~E. 2019, The Astrophysical Journal Letters, 878, L16

\bibitem[{{Carlsten} {et~al.}(2022){Carlsten}, {Greene}, {Beaton}, {Danieli}, \& {Greco}}]{2022ApJ...933...47C}
{Carlsten}, S.~G., {Greene}, J.~E., {Beaton}, R.~L., {Danieli}, S., \& {Greco}, J.~P. 2022, \apj, 933, 47

\bibitem[{Chambers {et~al.}(2016)Chambers, Magnier, Metcalfe, Flewelling, Huber, Waters, Denneau, Draper, Farrow, Finkbeiner, {et~al.}}]{chambers2016pan}
Chambers, K.~C., Magnier, E., Metcalfe, N., {et~al.} 2016, arXiv preprint arXiv:1612.05560

\bibitem[{Chan {et~al.}(2022)Chan, Lemon, Courbin, Gavazzi, Cl{\'e}ment, Millon, Paic, Rojas, Savary, Vernardos, {et~al.}}]{chan2022discovery}
Chan, J., Lemon, C., Courbin, F., {et~al.} 2022, Astronomy \& Astrophysics, 659, A140

\bibitem[{Cheng {et~al.}(2020)Cheng, Conselice, Arag{\'o}n-Salamanca, Li, Bluck, Hartley, Annis, Brooks, Doel, Garc{\'\i}a-Bellido, {et~al.}}]{cheng2020optimizing}
Cheng, T.-Y., Conselice, C.~J., Arag{\'o}n-Salamanca, A., {et~al.} 2020, Monthly Notices of the Royal Astronomical Society, 493, 4209

\bibitem[{Chiboucas {et~al.}(2013)Chiboucas, Jacobs, Tully, \& Karachentsev}]{chiboucas2013confirmation}
Chiboucas, K., Jacobs, B.~A., Tully, R.~B., \& Karachentsev, I.~D. 2013, The Astronomical Journal, 146, 126

\bibitem[{{Chu} {et~al.}(2023){Chu}, {Durret}, {Ellien}, {Sarron}, {Adami}, {M{\'a}rquez}, {Martinet}, {de Boer}, {Chambers}, {Cuillandre}, {Gwyn}, {Magnier}, \& {McConnachie}}]{2023A&A...673A.100C}
{Chu}, A., {Durret}, F., {Ellien}, A., {et~al.} 2023, \aap, 673, A100

\bibitem[{{\'C}iprijanovi{\'c} {et~al.}(2020){\'C}iprijanovi{\'c}, Snyder, Nord, \& Peek}]{ciprijanovic2020deepmerge}
{\'C}iprijanovi{\'c}, A., Snyder, G.~F., Nord, B., \& Peek, J.~E. 2020, Astronomy and Computing, 32, 100390

\bibitem[{Collaboration {et~al.}(2018)Collaboration, Mignard, Klioner, Lindegren, Hern{\'a}ndez, Bastian, Bombrun, Hobbs, Lammers, Michalik, {et~al.}}]{collaboration2018gaia}
Collaboration, G., Mignard, F., Klioner, S., {et~al.} 2018, A\&A, 616, A14

\bibitem[{{Collins} {et~al.}(2020){Collins}, {Tollerud}, {Rich}, {Ibata}, {Martin}, {Chapman}, {Gilbert}, \& {Preston}}]{2020MNRAS.491.3496C}
{Collins}, M. L.~M., {Tollerud}, E.~J., {Rich}, R.~M., {et~al.} 2020, \mnras, 491, 3496

\bibitem[{Cortes(1995)}]{cortes1995support}
Cortes, C. 1995, Machine Learning

\bibitem[{Crnojevi{\'c} {et~al.}(2019)Crnojevi{\'c}, Sand, Bennet, Pasetto, Spekkens, Caldwell, Guhathakurta, McLeod, Seth, Simon, {et~al.}}]{crnojevic2019faint}
Crnojevi{\'c}, D., Sand, D., Bennet, P., {et~al.} 2019, The Astrophysical Journal, 872, 80

\bibitem[{{Crosby} {et~al.}(2023){Crosby}, {Jerjen}, {M{\"u}ller}, {Pawlowski}, {Mateo}, \& {Dirnberger}}]{2023MNRAS.521.4009C}
{Crosby}, E., {Jerjen}, H., {M{\"u}ller}, O., {et~al.} 2023, \mnras, 521, 4009

\bibitem[{{Crosby} {et~al.}(2024){Crosby}, {Jerjen}, {M{\"u}ller}, {Pawlowski}, {Mateo}, \& {Lelli}}]{2024MNRAS.527.9118C}
{Crosby}, E., {Jerjen}, H., {M{\"u}ller}, O., {et~al.} 2024, \mnras, 527, 9118

\bibitem[{{Danieli} \& {van Dokkum}(2019)}]{2019ApJ...875..155D}
{Danieli}, S. \& {van Dokkum}, P. 2019, \apj, 875, 155

\bibitem[{Danieli {et~al.}(2017)Danieli, van Dokkum, Merritt, Abraham, Zhang, Karachentsev, \& Makarova}]{danieli2017dragonfly}
Danieli, S., van Dokkum, P., Merritt, A., {et~al.} 2017, The Astrophysical Journal, 837, 136

\bibitem[{Davies {et~al.}(2019)Davies, Serjeant, \& Bromley}]{davies2019using}
Davies, A., Serjeant, S., \& Bromley, J.~M. 2019, Monthly Notices of the Royal Astronomical Society, 487, 5263

\bibitem[{De~Angeli {et~al.}(2023)De~Angeli, Weiler, Montegriffo, Evans, Riello, Andrae, Carrasco, Busso, Burgess, Cacciari, {et~al.}}]{de2023gaia}
De~Angeli, F., Weiler, M., Montegriffo, P., {et~al.} 2023, Astronomy \& Astrophysics, 674, A2

\bibitem[{{Dey} {et~al.}(2019){Dey}, {Schlegel}, {Lang}, {Blum}, {Burleigh}, {Fan}, {Findlay}, {Finkbeiner}, {Herrera}, {Juneau}, {Landriau}, {Levi}, {McGreer}, {Meisner}, {Myers}, {Moustakas}, {Nugent}, {Patej}, {Schlafly}, {Walker}, {Valdes}, {Weaver}, {Y{\`e}che}, {Zou}, {Zhou}, {Abareshi}, {Abbott}, {Abolfathi}, {Aguilera}, {Alam}, {Allen}, {Alvarez}, {Annis}, {Ansarinejad}, {Aubert}, {Beechert}, {Bell}, {BenZvi}, {Beutler}, {Bielby}, {Bolton}, {Brice{\~n}o}, {Buckley-Geer}, {Butler}, {Calamida}, {Carlberg}, {Carter}, {Casas}, {Castander}, {Choi}, {Comparat}, {Cukanovaite}, {Delubac}, {DeVries}, {Dey}, {Dhungana}, {Dickinson}, {Ding}, {Donaldson}, {Duan}, {Duckworth}, {Eftekharzadeh}, {Eisenstein}, {Etourneau}, {Fagrelius}, {Farihi}, {Fitzpatrick}, {Font-Ribera}, {Fulmer}, {G{\"a}nsicke}, {Gaztanaga}, {George}, {Gerdes}, {Gontcho}, {Gorgoni}, {Green}, {Guy}, {Harmer}, {Hernandez}, {Honscheid}, {Huang}, {James}, {Jannuzi}, {Jiang}, {Joyce}, {Karcher}, {Karkar}, {Kehoe}, {Kneib}, {Kueter-Young}, {Lan},
  {Lauer}, {Le Guillou}, {Le Van Suu}, {Lee}, {Lesser}, {Perreault Levasseur}, {Li}, {Mann}, {Marshall}, {Mart{\'\i}nez-V{\'a}zquez}, {Martini}, {du Mas des Bourboux}, {McManus}, {Meier}, {M{\'e}nard}, {Metcalfe}, {Mu{\~n}oz-Guti{\'e}rrez}, {Najita}, {Napier}, {Narayan}, {Newman}, {Nie}, {Nord}, {Norman}, {Olsen}, {Paat}, {Palanque-Delabrouille}, {Peng}, {Poppett}, {Poremba}, {Prakash}, {Rabinowitz}, {Raichoor}, {Rezaie}, {Robertson}, {Roe}, {Ross}, {Ross}, {Rudnick}, {Safonova}, {Saha}, {S{\'a}nchez}, {Savary}, {Schweiker}, {Scott}, {Seo}, {Shan}, {Silva}, {Slepian}, {Soto}, {Sprayberry}, {Staten}, {Stillman}, {Stupak}, {Summers}, {Sien Tie}, {Tirado}, {Vargas-Maga{\~n}a}, {Vivas}, {Wechsler}, {Williams}, {Yang}, {Yang}, {Yapici}, {Zaritsky}, {Zenteno}, {Zhang}, {Zhang}, {Zhou}, \& {Zhou}}]{2019AJ....157..168D}
{Dey}, A., {Schlegel}, D.~J., {Lang}, D., {et~al.} 2019, \aj, 157, 168

\bibitem[{Dieleman {et~al.}(2015)Dieleman, Willett, \& Dambre}]{dieleman2015rotation}
Dieleman, S., Willett, K.~W., \& Dambre, J. 2015, Monthly notices of the royal astronomical society, 450, 1441

\bibitem[{{Doliva-Dolinsky} {et~al.}(2025){Doliva-Dolinsky}, {Collins}, \& {Martin}}]{2025arXiv250206948D}
{Doliva-Dolinsky}, A., {Collins}, M. L.~M., \& {Martin}, N.~F. 2025, arXiv e-prints, arXiv:2502.06948

\bibitem[{{Domber} {et~al.}(2022){Domber}, {Gygax}, {Aumiller}, {Whipple}, {Walker}, \& {Delker}}]{2022SPIE12180E..1OD}
{Domber}, J.~L., {Gygax}, J.~D., {Aumiller}, P., {et~al.} 2022, in Society of Photo-Optical Instrumentation Engineers (SPIE) Conference Series, Vol. 12180, Space Telescopes and Instrumentation 2022: Optical, Infrared, and Millimeter Wave, ed. L.~E. {Coyle}, S.~{Matsuura}, \& M.~D. {Perrin}, 121801O

\bibitem[{Dom{\'\i}nguez~S{\'a}nchez {et~al.}(2018)Dom{\'\i}nguez~S{\'a}nchez, Huertas-Company, Bernardi, Tuccillo, \& Fischer}]{dominguez2018improving}
Dom{\'\i}nguez~S{\'a}nchez, H., Huertas-Company, M., Bernardi, M., Tuccillo, D., \& Fischer, J. 2018, Monthly Notices of the Royal Astronomical Society, 476, 3661

\bibitem[{{Drlica-Wagner} {et~al.}(2021){Drlica-Wagner}, {Carlin}, {Nidever}, {Ferguson}, {Kuropatkin}, {Adam{\'o}w}, {Cerny}, {Choi}, {Esteves}, {Mart{\'\i}nez-V{\'a}zquez}, {Mau}, {Miller}, {Mutlu-Pakdil}, {Neilsen}, {Olsen}, {Pace}, {Riley}, {Sakowska}, {Sand}, {Santana-Silva}, {Tollerud}, {Tucker}, {Vivas}, {Zaborowski}, {Zenteno}, {Abbott}, {Allam}, {Bechtol}, {Bell}, {Bell}, {Bilaji}, {Bom}, {Carballo-Bello}, {Crnojevi{\'c}}, {Cioni}, {Diaz-Ocampo}, {de Boer}, {Erkal}, {Gruendl}, {Hernandez-Lang}, {Hughes}, {James}, {Johnson}, {Li}, {Mao}, {Mart{\'\i}nez-Delgado}, {Massana}, {McNanna}, {Morgan}, {Nadler}, {No{\"e}l}, {Palmese}, {Peter}, {Rykoff}, {S{\'a}nchez}, {Shipp}, {Simon}, {Smercina}, {Soares-Santos}, {Stringfellow}, {Tavangar}, {van der Marel}, {Walker}, {Wechsler}, {Wu}, {Yanny}, {Fitzpatrick}, {Huang}, {Jacques}, {Nikutta}, {Scott}, \& {Astro Data Lab}}]{2021ApJS..256....2D}
{Drlica-Wagner}, A., {Carlin}, J.~L., {Nidever}, D.~L., {et~al.} 2021, \apjs, 256, 2

\bibitem[{{Drlica-Wagner} {et~al.}(2022){Drlica-Wagner}, {Ferguson}, {Adam{\'o}w}, {Aguena}, {Allam}, {Andrade-Oliveira}, {Bacon}, {Bechtol}, {Bell}, {Bertin}, {Bilaji}, {Bocquet}, {Bom}, {Brooks}, {Burke}, {Carballo-Bello}, {Carlin}, {Carnero Rosell}, {Carrasco Kind}, {Carretero}, {Castander}, {Cerny}, {Chang}, {Choi}, {Conselice}, {Costanzi}, {Crnojevi{\'c}}, {da Costa}, {de Vicente}, {Desai}, {Esteves}, {Everett}, {Ferrero}, {Fitzpatrick}, {Flaugher}, {Friedel}, {Frieman}, {Garc{\'\i}a-Bellido}, {Gatti}, {Gaztanaga}, {Gerdes}, {Gruen}, {Gruendl}, {Gschwend}, {Hartley}, {Hernandez-Lang}, {Hinton}, {Hollowood}, {Honscheid}, {Hughes}, {Jacques}, {James}, {Johnson}, {Kuehn}, {Kuropatkin}, {Lahav}, {Li}, {Lidman}, {Lin}, {March}, {Marshall}, {Mart{\'\i}nez-Delgado}, {Mart{\'\i}nez-V{\'a}zquez}, {Massana}, {Mau}, {McNanna}, {Melchior}, {Menanteau}, {Miller}, {Miquel}, {Mohr}, {Morgan}, {Mutlu-Pakdil}, {Mu{\~n}oz}, {Neilsen}, {Nidever}, {Nikutta}, {Nilo Castellon}, {No{\"e}l}, {Ogando}, {Olsen}, {Pace},
  {Palmese}, {Paz-Chinch{\'o}n}, {Pereira}, {Pieres}, {Plazas Malag{\'o}n}, {Prat}, {Riley}, {Rodriguez-Monroy}, {Romer}, {Roodman}, {Sako}, {Sakowska}, {Sanchez}, {S{\'a}nchez}, {Sand}, {Santana-Silva}, {Santiago}, {Schubnell}, {Serrano}, {Sevilla-Noarbe}, {Simon}, {Smith}, {Soares-Santos}, {Stringfellow}, {Suchyta}, {Suson}, {Tan}, {Tarle}, {Tavangar}, {Thomas}, {To}, {Tollerud}, {Troxel}, {Tucker}, {Varga}, {Vivas}, {Walker}, {Weller}, {Wilkinson}, {Wu}, {Yanny}, {Zaborowski}, {Zenteno}, {Delve Collaboration}, {Des Collaboration}, \& {Astro Data Lab}}]{2022ApJS..261...38D}
{Drlica-Wagner}, A., {Ferguson}, P.~S., {Adam{\'o}w}, M., {et~al.} 2022, \apjs, 261, 38

\bibitem[{Duc {et~al.}(2015)Duc, Cuillandre, Karabal, Cappellari, Alatalo, Blitz, Bournaud, Bureau, Crocker, Davies, {et~al.}}]{duc2015atlas3d}
Duc, P.-A., Cuillandre, J.-C., Karabal, E., {et~al.} 2015, Monthly Notices of the Royal Astronomical Society, 446, 120

\bibitem[{Ellison {et~al.}(2019)Ellison, Viswanathan, Patton, Bottrell, McConnachie, Gwyn, \& Cuillandre}]{ellison2019definitive}
Ellison, S.~L., Viswanathan, A., Patton, D.~R., {et~al.} 2019, Monthly Notices of the Royal Astronomical Society, 487, 2491

\bibitem[{Ellison {et~al.}(2022)Ellison, Wilkinson, Woo, Leung, Wild, Bickley, Patton, Quai, \& Gwyn}]{ellison2022galaxy}
Ellison, S.~L., Wilkinson, S., Woo, J., {et~al.} 2022, Monthly Notices of the Royal Astronomical Society: Letters, 517, L92

\bibitem[{{Euclid Collaboration} {et~al.}(2025){Euclid Collaboration}, {Mellier}, {Abdurro'uf}, {Acevedo Barroso}, {Ach{\'u}carro}, {Adamek}, {Adam}, {Addison}, {Aghanim}, {Aguena}, {Ajani}, {Akrami}, {Al-Bahlawan}, {Alavi}, {Albuquerque}, {Alestas}, {Alguero}, {Allaoui}, {Allen}, {Allevato}, {Alonso-Tetilla}, {Altieri}, {Alvarez-Candal}, {Alvi}, {Amara}, {Amendola}, {Amiaux}, {Andika}, {Andreon}, {Andrews}, {Angora}, {Angulo}, {Annibali}, {Anselmi}, {Anselmi}, {Arcari}, {Archidiacono}, {Aric{\`o}}, {Arnaud}, {Arnouts}, {Asgari}, {Asorey}, {Atayde}, {Atek}, {Atrio-Barandela}, {Aubert}, {Aubourg}, {Auphan}, {Auricchio}, {Aussel}, {Aussel}, {Avelino}, {Avgoustidis}, {Avila}, {Awan}, {Azzollini}, {Baccigalupi}, {Bachelet}, {Bacon}, {Baes}, {Bagley}, {Bahr-Kalus}, {Balaguera-Antolinez}, {Balbinot}, {Balcells}, {Baldi}, {Baldry}, {Balestra}, {Ballardini}, {Ballester}, {Balogh}, {Ba{\~n}ados}, {Barbier}, {Bardelli}, {Baron}, {Barreiro}, {Barrena}, {Barriere}, {Barros}, {Barthelemy}, {Bartolo}, {Basset},
  {Battaglia}, {Battisti}, {Baugh}, {Baumont}, {Bazzanini}, {Beaulieu}, {Beckmann}, {Belikov}, {Bel}, {Bellagamba}, {Bella}, {Bellini}, {Benabed}, {Bender}, {Benevento}, {Bennett}, {Benson}, {Bergamini}, {Bermejo-Climent}, {Bernardeau}, {Bertacca}, {Berthe}, {Berthier}, {Bethermin}, {Beutler}, {Bevillon}, {Bhargava}, {Bhatawdekar}, {Bianchi}, {Bisigello}, {Biviano}, {Blake}, {Blanchard}, {Blazek}, {Blot}, {Bosco}, {Bodendorf}, {Boenke}, {B{\"o}hringer}, {Boldrini}, {Bolzonella}, {Bonchi}, {Bonici}, {Bonino}, {Bonino}, {Bonvin}, {Bon}, {Booth}, {Borgani}, {Borlaff}, {Borsato}, {Bose}, {Botticella}, {Boucaud}, {Bouche}, {Boucher}, {Boutigny}, {Bouvard}, {Bouwens}, {Bouy}, {Bowler}, {Bozza}, {Bozzo}, {Branchini}, {Brando}, {Brau-Nogue}, {Brekke}, {Bremer}, {Brescia}, {Breton}, {Brinchmann}, {Brinckmann}, {Brockley-Blatt}, {Brodwin}, {Brouard}, {Brown}, {Bruton}, {Bucko}, {Buddelmeijer}, {Buenadicha}, {Buitrago}, {Burger}, {Burigana}, {Busillo}, {Busonero}, {Cabanac}, {Cabayol-Garcia}, {Cagliari}, {Caillat},
  {Caillat}, {Calabrese}, {Calabro}, {Calderone}, {Calura}, {Camacho Quevedo}, {Camera}, {Campos}, {Ca{\~n}as-Herrera}, {Candini}, {Cantiello}, {Capobianco}, {Cappellaro}, {Cappelluti}, {Cappi}, {Caputi}, {Cara}, {Carbone}, {Cardone}, {Carella}, {Carlberg}, {Carle}, {Carminati}, {Caro}, {Carrasco}, {Carretero}, {Carrilho}, {Carron Duque}, \& {Carry}}]{2025A&A...697A...1E}
{Euclid Collaboration}, {Mellier}, Y., {Abdurro'uf}, {et~al.} 2025, \aap, 697, A1

\bibitem[{Fantin {et~al.}(2021)Fantin, C{\^o}t{\'e}, McConnachie, Bergeron, Cuillandre, Dufour, Gwyn, Ibata, \& Thomas}]{fantin2021mass}
Fantin, N.~J., C{\^o}t{\'e}, P., McConnachie, A.~W., {et~al.} 2021, The Astrophysical Journal, 913, 30

\bibitem[{Fantin {et~al.}(2019)Fantin, C{\^o}t{\'e}, McConnachie, Bergeron, Cuillandre, Gwyn, Ibata, Thomas, Carlberg, Fabbro, {et~al.}}]{fantin2019canada}
Fantin, N.~J., C{\^o}t{\'e}, P., McConnachie, A.~W., {et~al.} 2019, The Astrophysical Journal, 887, 148

\bibitem[{Ferguson {et~al.}(2002)Ferguson, Irwin, Ibata, Lewis, \& Tanvir}]{ferguson2002evidence}
Ferguson, A.~M., Irwin, M.~J., Ibata, R.~A., Lewis, G.~F., \& Tanvir, N.~R. 2002, The Astronomical Journal, 124, 1452

\bibitem[{{Ferguson}(1989)}]{1989AJ.....98..367F}
{Ferguson}, H.~C. 1989, \aj, 98, 367

\bibitem[{Ferguson \& Binggeli(1994)}]{ferguson1994dwarf}
Ferguson, H.~C. \& Binggeli, B. 1994, The Astronomy and Astrophysics Review, 6, 67

\bibitem[{{Ferguson} \& {Binggeli}(1994)}]{1994A&ARv...6...67F}
{Ferguson}, H.~C. \& {Binggeli}, B. 1994, \aapr, 6, 67

\bibitem[{{Ferguson} \& {Sandage}(1988)}]{1988AJ.....96.1520F}
{Ferguson}, H.~C. \& {Sandage}, A. 1988, \aj, 96, 1520

\bibitem[{{Ferguson} \& {Sandage}(1990)}]{1990AJ....100....1F}
{Ferguson}, H.~C. \& {Sandage}, A. 1990, \aj, 100, 1

\bibitem[{Ferrarese {et~al.}(2012)Ferrarese, Cote, Cuillandre, Gwyn, Peng, MacArthur, Duc, Boselli, Mei, Erben, {et~al.}}]{ferrarese2012next}
Ferrarese, L., Cote, P., Cuillandre, J.-C., {et~al.} 2012, The Astrophysical Journal Supplement Series, 200, 4

\bibitem[{{Ferreira} {et~al.}(2024){Ferreira}, {Bickley}, {Ellison}, {Patton}, {Byrne-Mamahit}, {Wilkinson}, {Bottrell}, {Fabbro}, {Gwyn}, \& {McConnachie}}]{2024MNRAS.533.2547F}
{Ferreira}, L., {Bickley}, R.~W., {Ellison}, S.~L., {et~al.} 2024, \mnras, 533, 2547

\bibitem[{Frenk \& White(2012)}]{frenk2012dark}
Frenk, C.~S. \& White, S.~D. 2012, Annalen der Physik, 524, 507

\bibitem[{{Gaia Collaboration} {et~al.}(2016){Gaia Collaboration}, {Prusti}, {de Bruijne}, {Brown}, {Vallenari}, {Babusiaux}, {Bailer-Jones}, {Bastian}, {Biermann}, {Evans}, {Eyer}, {Jansen}, {Jordi}, {Klioner}, {Lammers}, {Lindegren}, {Luri}, {Mignard}, {Milligan}, {Panem}, {Poinsignon}, {Pourbaix}, {Randich}, {Sarri}, {Sartoretti}, {Siddiqui}, {Soubiran}, {Valette}, {van Leeuwen}, {Walton}, {Aerts}, {Arenou}, {Cropper}, {Drimmel}, {H{\o}g}, {Katz}, {Lattanzi}, {O'Mullane}, {Grebel}, {Holland}, {Huc}, {Passot}, {Bramante}, {Cacciari}, {Casta{\~n}eda}, {Chaoul}, {Cheek}, {De Angeli}, {Fabricius}, {Guerra}, {Hern{\'a}ndez}, {Jean-Antoine-Piccolo}, {Masana}, {Messineo}, {Mowlavi}, {Nienartowicz}, {Ord{\'o}{\~n}ez-Blanco}, {Panuzzo}, {Portell}, {Richards}, {Riello}, {Seabroke}, {Tanga}, {Th{\'e}venin}, {Torra}, {Els}, {Gracia-Abril}, {Comoretto}, {Garcia-Reinaldos}, {Lock}, {Mercier}, {Altmann}, {Andrae}, {Astraatmadja}, {Bellas-Velidis}, {Benson}, {Berthier}, {Blomme}, {Busso}, {Carry}, {Cellino}, {Clementini},
  {Cowell}, {Creevey}, {Cuypers}, {Davidson}, {De Ridder}, {de Torres}, {Delchambre}, {Dell'Oro}, {Ducourant}, {Fr{\'e}mat}, {Garc{\'\i}a-Torres}, {Gosset}, {Halbwachs}, {Hambly}, {Harrison}, {Hauser}, {Hestroffer}, {Hodgkin}, {Huckle}, {Hutton}, {Jasniewicz}, {Jordan}, {Kontizas}, {Korn}, {Lanzafame}, {Manteiga}, {Moitinho}, {Muinonen}, {Osinde}, {Pancino}, {Pauwels}, {Petit}, {Recio-Blanco}, {Robin}, {Sarro}, {Siopis}, {Smith}, {Smith}, {Sozzetti}, {Thuillot}, {van Reeven}, {Viala}, {Abbas}, {Abreu Aramburu}, {Accart}, {Aguado}, {Allan}, {Allasia}, {Altavilla}, {{\'A}lvarez}, {Alves}, {Anderson}, {Andrei}, {Anglada Varela}, {Antiche}, {Antoja}, {Ant{\'o}n}, {Arcay}, {Atzei}, {Ayache}, {Bach}, {Baker}, {Balaguer-N{\'u}{\~n}ez}, {Barache}, {Barata}, {Barbier}, {Barblan}, {Baroni}, {Barrado y Navascu{\'e}s}, {Barros}, {Barstow}, {Becciani}, {Bellazzini}, {Bellei}, {Bello Garc{\'\i}a}, {Belokurov}, {Bendjoya}, {Berihuete}, {Bianchi}, {Bienaym{\'e}}, {Billebaud}, {Blagorodnova}, {Blanco-Cuaresma}, {Boch},
  {Bombrun}, {Borrachero}, {Bouquillon}, {Bourda}, {Bouy}, {Bragaglia}, {Breddels}, {Brouillet}, {Br{\"u}semeister}, {Bucciarelli}, {Budnik}, {Burgess}, {Burgon}, {Burlacu}, {Busonero}, {Buzzi}, {Caffau}, {Cambras}, {Campbell}, {Cancelliere}, {Cantat-Gaudin}, {Carlucci}, {Carrasco}, {Castellani}, {Charlot}, {Charnas}, {Charvet}, {Chassat}, {Chiavassa}, {Clotet}, {Cocozza}, {Collins}, {Collins}, {Costigan}, {Crifo}, {Cross}, {Crosta}, {Crowley}, {Dafonte}, {Damerdji}, {Dapergolas}, {David}, {David}, {De Cat}, {de Felice}, {de Laverny}, {De Luise}, {De March}, {de Martino}, {de Souza}, {Debosscher}, {del Pozo}, {Delbo}, {Delgado}, {Delgado}, {di Marco}, {Di Matteo}, {Diakite}, {Distefano}, {Dolding}, {Dos Anjos}, {Drazinos}, {Dur{\'a}n}, {Dzigan}, {Ecale}, {Edvardsson}, {Enke}, {Erdmann}, {Escolar}, {Espina}, {Evans}, {Eynard Bontemps}, {Fabre}, {Fabrizio}, {Faigler}, {Falc{\~a}o}, {Farr{\`a}s Casas}, {Faye}, {Federici}, {Fedorets}, {Fern{\'a}ndez-Hern{\'a}ndez}, {Fernique}, {Fienga}, {Figueras}, {Filippi},
  {Findeisen}, {Fonti}, {Fouesneau}, {Fraile}, {Fraser}, {Fuchs}, {Furnell}, {Gai}, {Galleti}, {Galluccio}, {Garabato}, {Garc{\'\i}a-Sedano}, {Gar{\'e}}, {Garofalo}, {Garralda}, {Gavras}, {Gerssen}, {Geyer}, {Gilmore}, {Girona}, {Giuffrida}, {Gomes}, {Gonz{\'a}lez-Marcos}, {Gonz{\'a}lez-N{\'u}{\~n}ez}, {Gonz{\'a}lez-Vidal}, {Granvik}, {Guerrier}, {Guillout}, {Guiraud}, {G{\'u}rpide}, {Guti{\'e}rrez-S{\'a}nchez}, {Guy}, {Haigron}, {Hatzidimitriou}, {Haywood}, {Heiter}, {Helmi}, {Hobbs}, {Hofmann}, {Holl}, {Holland}, {Hunt}, {Hypki}, {Icardi}, {Irwin}, {Jevardat de Fombelle}, {Jofr{\'e}}, {Jonker}, {Jorissen}, {Julbe}, {Karampelas}, {Kochoska}, {Kohley}, {Kolenberg}, {Kontizas}, {Koposov}, {Kordopatis}, {Koubsky}, {Kowalczyk}, {Krone-Martins}, {Kudryashova}, {Kull}, {Bachchan}, {Lacoste-Seris}, {Lanza}, {Lavigne}, {Le Poncin-Lafitte}, {Lebreton}, {Lebzelter}, {Leccia}, {Leclerc}, {Lecoeur-Taibi}, {Lemaitre}, {Lenhardt}, {Leroux}, {Liao}, {Licata}, {Lindstr{\o}m}, {Lister}, {Livanou}, {Lobel}, {L{\"o}ffler},
  {L{\'o}pez}, {Lopez-Lozano}, {Lorenz}, {Loureiro}, {MacDonald}, {Magalh{\~a}es Fernandes}, {Managau}, {Mann}, {Mantelet}, {Marchal}, {Marchant}, {Marconi}, {Marie}, {Marinoni}, {Marrese}, {Marschalk{\'o}}, {Marshall}, {Mart{\'\i}n-Fleitas}, {Martino}, {Mary}, {Matijevi{\v{c}}}, {Mazeh}, {McMillan}, {Messina}, {Mestre}, {Michalik}, {Millar}, {Miranda}, {Molina}, {Molinaro}, {Molinaro}, {Moln{\'a}r}, {Moniez}, {Montegriffo}, {Monteiro}, {Mor}, {Mora}, {Morbidelli}, {Morel}, {Morgenthaler}, {Morley}, {Morris}, {Mulone}, {Muraveva}, {Musella}, {Narbonne}, {Nelemans}, {Nicastro}, {Noval}, {Ord{\'e}novic}, {Ordieres-Mer{\'e}}, {Osborne}, {Pagani}, {Pagano}, {Pailler}, {Palacin}, {Palaversa}, {Parsons}, {Paulsen}, {Pecoraro}, {Pedrosa}, {Pentik{\"a}inen}, {Pereira}, {Pichon}, {Piersimoni}, {Pineau}, {Plachy}, {Plum}, {Poujoulet}, {Pr{\v{s}}a}, {Pulone}, {Ragaini}, {Rago}, {Rambaux}, {Ramos-Lerate}, {Ranalli}, {Rauw}, {Read}, {Regibo}, {Renk}, {Reyl{\'e}}, {Ribeiro}, {Rimoldini}, {Ripepi}, {Riva}, {Rixon},
  {Roelens}, {Romero-G{\'o}mez}, {Rowell}, {Royer}, {Rudolph}, {Ruiz-Dern}, {Sadowski}, {Sagrist{\`a} Sell{\'e}s}, {Sahlmann}, {Salgado}, {Salguero}, {Sarasso}, {Savietto}, {Schnorhk}, {Schultheis}, {Sciacca}, {Segol}, {Segovia}, {Segransan}, {Serpell}, {Shih}, {Smareglia}, {Smart}, {Smith}, {Solano}, {Solitro}, {Sordo}, {Soria Nieto}, {Souchay}, {Spagna}, {Spoto}, {Stampa}, {Steele}, {Steidelm{\"u}ller}, {Stephenson}, {Stoev}, {Suess}, {S{\"u}veges}, {Surdej}, {Szabados}, {Szegedi-Elek}, {Tapiador}, {Taris}, {Tauran}, {Taylor}, {Teixeira}, {Terrett}, {Tingley}, {Trager}, {Turon}, {Ulla}, {Utrilla}, {Valentini}, {van Elteren}, {Van Hemelryck}, {van Leeuwen}, {Varadi}, {Vecchiato}, {Veljanoski}, {Via}, {Vicente}, {Vogt}, {Voss}, {Votruba}, {Voutsinas}, {Walmsley}, {Weiler}, {Weingrill}, {Werner}, {Wevers}, {Whitehead}, {Wyrzykowski}, {Yoldas}, {{\v{Z}}erjal}, {Zucker}, {Zurbach}, {Zwitter}, {Alecu}, {Allen}, {Allende Prieto}, {Amorim}, {Anglada-Escud{\'e}}, {Arsenijevic}, {Azaz}, {Balm}, {Beck}, {Bernstein},
  {Bigot}, {Bijaoui}, {Blasco}, {Bonfigli}, {Bono}, {Boudreault}, {Bressan}, {Brown}, {Brunet}, {Bunclark}, {Buonanno}, {Butkevich}, {Carret}, {Carrion}, {Chemin}, {Ch{\'e}reau}, {Corcione}, {Darmigny}, {de Boer}, {de Teodoro}, {de Zeeuw}, {Delle Luche}, {Domingues}, {Dubath}, {Fodor}, {Fr{\'e}zouls}, {Fries}, {Fustes}, {Fyfe}, {Gallardo}, {Gallegos}, {Gardiol}, {Gebran}, {Gomboc}, {G{\'o}mez}, {Grux}, {Gueguen}, {Heyrovsky}, {Hoar}, {Iannicola}, {Isasi Parache}, {Janotto}, {Joliet}, {Jonckheere}, {Keil}, {Kim}, {Klagyivik}, {Klar}, {Knude}, {Kochukhov}, {Kolka}, {Kos}, {Kutka}, {Lainey}, {LeBouquin}, {Liu}, {Loreggia}, {Makarov}, {Marseille}, {Martayan}, {Martinez-Rubi}, {Massart}, {Meynadier}, {Mignot}, {Munari}, {Nguyen}, {Nordlander}, {Ocvirk}, {O'Flaherty}, {Olias Sanz}, {Ortiz}, {Osorio}, {Oszkiewicz}, {Ouzounis}, {Palmer}, {Park}, {Pasquato}, {Peltzer}, {Peralta}, {P{\'e}turaud}, {Pieniluoma}, {Pigozzi}, {Poels}, {Prat}, {Prod'homme}, {Raison}, {Rebordao}, {Risquez}, {Rocca-Volmerange}, {Rosen},
  {Ruiz-Fuertes}, {Russo}, {Sembay}, {Serraller Vizcaino}, {Short}, {Siebert}, {Silva}, {Sinachopoulos}, {Slezak}, {Soffel}, {Sosnowska}, {Strai{\v{z}}ys}, {ter Linden}, {Terrell}, {Theil}, {Tiede}, {Troisi}, {Tsalmantza}, {Tur}, {Vaccari}, {Vachier}, {Valles}, {Van Hamme}, {Veltz}, {Virtanen}, {Wallut}, {Wichmann}, {Wilkinson}, {Ziaeepour}, \& {Zschocke}}]{2016A&A...595A...1G}
{Gaia Collaboration}, {Prusti}, T., {de Bruijne}, J.~H.~J., {et~al.} 2016, \aap, 595, A1

\bibitem[{{Gaia Collaboration} {et~al.}(2023){Gaia Collaboration}, {Vallenari}, {Brown}, {Prusti}, {de Bruijne}, {Arenou}, {Babusiaux}, {Biermann}, {Creevey}, {Ducourant}, {Evans}, {Eyer}, {Guerra}, {Hutton}, {Jordi}, {Klioner}, {Lammers}, {Lindegren}, {Luri}, {Mignard}, {Panem}, {Pourbaix}, {Randich}, {Sartoretti}, {Soubiran}, {Tanga}, {Walton}, {Bailer-Jones}, {Bastian}, {Drimmel}, {Jansen}, {Katz}, {Lattanzi}, {van Leeuwen}, {Bakker}, {Cacciari}, {Casta{\~n}eda}, {De Angeli}, {Fabricius}, {Fouesneau}, {Fr{\'e}mat}, {Galluccio}, {Guerrier}, {Heiter}, {Masana}, {Messineo}, {Mowlavi}, {Nicolas}, {Nienartowicz}, {Pailler}, {Panuzzo}, {Riclet}, {Roux}, {Seabroke}, {Sordo}, {Th{\'e}venin}, {Gracia-Abril}, {Portell}, {Teyssier}, {Altmann}, {Andrae}, {Audard}, {Bellas-Velidis}, {Benson}, {Berthier}, {Blomme}, {Burgess}, {Busonero}, {Busso}, {C{\'a}novas}, {Carry}, {Cellino}, {Cheek}, {Clementini}, {Damerdji}, {Davidson}, {de Teodoro}, {Nu{\~n}ez Campos}, {Delchambre}, {Dell'Oro}, {Esquej},
  {Fern{\'a}ndez-Hern{\'a}ndez}, {Fraile}, {Garabato}, {Garc{\'\i}a-Lario}, {Gosset}, {Haigron}, {Halbwachs}, {Hambly}, {Harrison}, {Hern{\'a}ndez}, {Hestroffer}, {Hodgkin}, {Holl}, {Jan{\ss}en}, {Jevardat de Fombelle}, {Jordan}, {Krone-Martins}, {Lanzafame}, {L{\"o}ffler}, {Marchal}, {Marrese}, {Moitinho}, {Muinonen}, {Osborne}, {Pancino}, {Pauwels}, {Recio-Blanco}, {Reyl{\'e}}, {Riello}, {Rimoldini}, {Roegiers}, {Rybizki}, {Sarro}, {Siopis}, {Smith}, {Sozzetti}, {Utrilla}, {van Leeuwen}, {Abbas}, {{\'A}brah{\'a}m}, {Abreu Aramburu}, {Aerts}, {Aguado}, {Ajaj}, {Aldea-Montero}, {Altavilla}, {{\'A}lvarez}, {Alves}, {Anders}, {Anderson}, {Anglada Varela}, {Antoja}, {Baines}, {Baker}, {Balaguer-N{\'u}{\~n}ez}, {Balbinot}, {Balog}, {Barache}, {Barbato}, {Barros}, {Barstow}, {Bartolom{\'e}}, {Bassilana}, {Bauchet}, {Becciani}, {Bellazzini}, {Berihuete}, {Bernet}, {Bertone}, {Bianchi}, {Binnenfeld}, {Blanco-Cuaresma}, {Blazere}, {Boch}, {Bombrun}, {Bossini}, {Bouquillon}, {Bragaglia}, {Bramante}, {Breedt},
  {Bressan}, {Brouillet}, {Brugaletta}, {Bucciarelli}, {Burlacu}, {Butkevich}, {Buzzi}, {Caffau}, {Cancelliere}, {Cantat-Gaudin}, {Carballo}, {Carlucci}, {Carnerero}, {Carrasco}, {Casamiquela}, {Castellani}, {Castro-Ginard}, {Chaoul}, {Charlot}, {Chemin}, {Chiaramida}, {Chiavassa}, {Chornay}, {Comoretto}, {Contursi}, {Cooper}, {Cornez}, {Cowell}, {Crifo}, {Cropper}, {Crosta}, {Crowley}, {Dafonte}, {Dapergolas}, {David}, {David}, {de Laverny}, {De Luise}, {De March}, {De Ridder}, {de Souza}, {de Torres}, {del Peloso}, {del Pozo}, {Delbo}, {Delgado}, {Delisle}, {Demouchy}, {Dharmawardena}, {Di Matteo}, {Diakite}, {Diener}, {Distefano}, {Dolding}, {Edvardsson}, {Enke}, {Fabre}, {Fabrizio}, {Faigler}, {Fedorets}, {Fernique}, {Fienga}, {Figueras}, {Fournier}, {Fouron}, {Fragkoudi}, {Gai}, {Garcia-Gutierrez}, {Garcia-Reinaldos}, {Garc{\'\i}a-Torres}, {Garofalo}, {Gavel}, {Gavras}, {Gerlach}, {Geyer}, {Giacobbe}, {Gilmore}, {Girona}, {Giuffrida}, {Gomel}, {Gomez}, {Gonz{\'a}lez-N{\'u}{\~n}ez},
  {Gonz{\'a}lez-Santamar{\'\i}a}, {Gonz{\'a}lez-Vidal}, {Granvik}, {Guillout}, {Guiraud}, {Guti{\'e}rrez-S{\'a}nchez}, {Guy}, {Hatzidimitriou}, {Hauser}, {Haywood}, {Helmer}, {Helmi}, {Sarmiento}, {Hidalgo}, {Hilger}, {H{\l}adczuk}, {Hobbs}, {Holland}, {Huckle}, {Jardine}, {Jasniewicz}, {Jean-Antoine Piccolo}, {Jim{\'e}nez-Arranz}, {Jorissen}, {Juaristi Campillo}, {Julbe}, {Karbevska}, {Kervella}, {Khanna}, {Kontizas}, {Kordopatis}, {Korn}, {K{\'o}sp{\'a}l}, {Kostrzewa-Rutkowska}, {Kruszy{\'n}ska}, {Kun}, {Laizeau}, {Lambert}, {Lanza}, {Lasne}, {Le Campion}, {Lebreton}, {Lebzelter}, {Leccia}, {Leclerc}, {Lecoeur-Taibi}, {Liao}, {Licata}, {Lindstr{\o}m}, {Lister}, {Livanou}, {Lobel}, {Lorca}, {Loup}, {Madrero Pardo}, {Magdaleno Romeo}, {Managau}, {Mann}, {Manteiga}, {Marchant}, {Marconi}, {Marcos}, {Marcos Santos}, {Mar{\'\i}n Pina}, {Marinoni}, {Marocco}, {Marshall}, {Martin Polo}, {Mart{\'\i}n-Fleitas}, {Marton}, {Mary}, {Masip}, {Massari}, {Mastrobuono-Battisti}, {Mazeh}, {McMillan}, {Messina}, {Michalik},
  {Millar}, {Mints}, {Molina}, {Molinaro}, {Moln{\'a}r}, {Monari}, {Mongui{\'o}}, {Montegriffo}, {Montero}, {Mor}, {Mora}, {Morbidelli}, {Morel}, {Morris}, {Muraveva}, {Murphy}, {Musella}, {Nagy}, {Noval}, {Oca{\~n}a}, {Ogden}, {Ordenovic}, {Osinde}, {Pagani}, {Pagano}, {Palaversa}, {Palicio}, {Pallas-Quintela}, {Panahi}, {Payne-Wardenaar}, {Pe{\~n}alosa Esteller}, {Penttil{\"a}}, {Pichon}, {Piersimoni}, {Pineau}, {Plachy}, {Plum}, {Poggio}, {Pr{\v{s}}a}, {Pulone}, {Racero}, {Ragaini}, {Rainer}, {Raiteri}, {Rambaux}, {Ramos}, {Ramos-Lerate}, {Re Fiorentin}, {Regibo}, {Richards}, {Rios Diaz}, {Ripepi}, {Riva}, {Rix}, {Rixon}, {Robichon}, {Robin}, {Robin}, {Roelens}, {Rogues}, {Rohrbasser}, {Romero-G{\'o}mez}, {Rowell}, {Royer}, {Ruz Mieres}, {Rybicki}, {Sadowski}, {S{\'a}ez N{\'u}{\~n}ez}, {Sagrist{\`a} Sell{\'e}s}, {Sahlmann}, {Salguero}, {Samaras}, {Sanchez Gimenez}, {Sanna}, {Santove{\~n}a}, {Sarasso}, {Schultheis}, {Sciacca}, {Segol}, {Segovia}, {S{\'e}gransan}, {Semeux}, {Shahaf}, {Siddiqui}, {Siebert},
  {Siltala}, {Silvelo}, {Slezak}, {Slezak}, {Smart}, {Snaith}, {Solano}, {Solitro}, {Souami}, {Souchay}, {Spagna}, {Spina}, {Spoto}, {Steele}, {Steidelm{\"u}ller}, {Stephenson}, {S{\"u}veges}, {Surdej}, {Szabados}, {Szegedi-Elek}, {Taris}, {Taylor}, {Teixeira}, {Tolomei}, {Tonello}, {Torra}, {Torra}, {Torralba Elipe}, {Trabucchi}, {Tsounis}, {Turon}, {Ulla}, {Unger}, {Vaillant}, {van Dillen}, {van Reeven}, {Vanel}, {Vecchiato}, {Viala}, {Vicente}, {Voutsinas}, {Weiler}, {Wevers}, {Wyrzykowski}, {Yoldas}, {Yvard}, {Zhao}, {Zorec}, {Zucker}, \& {Zwitter}}]{2023A&A...674A...1G}
{Gaia Collaboration}, {Vallenari}, A., {Brown}, A.~G.~A., {et~al.} 2023, \aap, 674, A1

\bibitem[{{Geha} {et~al.}(2017){Geha}, {Wechsler}, {Mao}, {Tollerud}, {Weiner}, {Bernstein}, {Hoyle}, {Marchi}, {Marshall}, {Mu{\~n}oz}, \& {Lu}}]{2017ApJ...847....4G}
{Geha}, M., {Wechsler}, R.~H., {Mao}, Y.-Y., {et~al.} 2017, \apj, 847, 4

\bibitem[{{Goto} {et~al.}(2023){Goto}, {Zaritsky}, {Karunakaran}, {Donnerstein}, \& {Sand}}]{2023AJ....166..185G}
{Goto}, H., {Zaritsky}, D., {Karunakaran}, A., {Donnerstein}, R., \& {Sand}, D.~J. 2023, \aj, 166, 185

\bibitem[{{Greco} {et~al.}(2018){Greco}, {Greene}, {Strauss}, {Macarthur}, {Flowers}, {Goulding}, {Huang}, {Kim}, {Komiyama}, {Leauthaud}, {Leisman}, {Lupton}, {Sif{\'o}n}, \& {Wang}}]{2018ApJ...857..104G}
{Greco}, J.~P., {Greene}, J.~E., {Strauss}, M.~A., {et~al.} 2018, \apj, 857, 104

\bibitem[{{Guerrini} {et~al.}(2024){Guerrini}, {Kilbinger}, {Leterme}, {Guinot}, {Wang}, {Hervas Peters}, {Hildebrandt}, {Hudson}, \& {McConnachie}}]{2024arXiv241214666G}
{Guerrini}, S., {Kilbinger}, M., {Leterme}, H., {et~al.} 2024, arXiv e-prints, arXiv:2412.14666

\bibitem[{{Guinot} {et~al.}(2022){Guinot}, {Kilbinger}, {Farrens}, {Peel}, {Pujol}, {Schmitz}, {Starck}, {Erben}, {Gavazzi}, {Gwyn}, {Hudson}, {Hildebrandt}, {Tobias}, {Miller}, {Spitzer}, {Van Waerbeke}, {Cuillandre}, {Fabbro}, {McConnachie}, \& {Mellier}}]{2022A&A...666A.162G}
{Guinot}, A., {Kilbinger}, M., {Farrens}, S., {et~al.} 2022, \aap, 666, A162

\bibitem[{Gwyn(2019)}]{gwyn2019megapipe}
Gwyn, S. 2019, Astronomical Data Analysis Software and Systems XXVII, 523, 649

\bibitem[{{Gwyn} {et~al.}(2025){Gwyn}, {McConnachie}, {Cuillandre}, {Chambers}, {Magnier}, {Hudson}, {Oguri}, {Furusawa}, {Hildebrandt}, {Carlberg}, {Ellison}, {Furusawa}, {Gavazzi}, {Ibata}, {Mellier}, {Osato}, {Aussel}, {Baumont}, {Bayer}, {Boulade}, {C{\^o}t{\'e}}, {Chemaly}, {Daley}, {Duc}, {Ellien}, {Fabbro}, {Ferreira}, {Fitriana}, {Le Floc'h}, {Hammer}, {Francois}, {Fudamoto}, {Gao}, {Goh}, {Goto}, {Guerrini}, {Guinot}, {H{\'e}nault-Brunet}, {Harikane}, {Hayashi}, {Heesters}, {Ichikawa}, {Kilbinger}, {Kuzma}, {Li}, {Liaudat}, {Lin}, {M{\"u}ller}, {Martin}, {Matsuoka}, {Medina}, {Miyatake}, {Miyazaki}, {Mpetha}, {Nagao}, {Navarro}, {Niwano}, {Ogami}, {Okabe}, {Onoue}, {Paek}, {Parker}, {Patton}, {Hervas Peters}, {Prunet}, {S{\'a}nchez-Janssen}, {Schultheis}, {Sestito}, {Smith}, {Starck}, {Starkenburg}, {Stone}, {Storfer}, {Suzuki}, {Erben}, {T.}, {Taibi}, {Thomas}, {TianFang}, {Toba}, {Uchiyama}, {Valls-Gabaud}, {Venn}, {Van Waerbeke}, {Wainscoat}, {Wilkinson}, {Wittje}, {Yoshida}, \&
  {Zhong}}]{2025arXiv250313783G}
{Gwyn}, S., {McConnachie}, A.~W., {Cuillandre}, J.-C., {et~al.} 2025, arXiv e-prints, arXiv:2503.13783

\bibitem[{Gwyn(2008)}]{gwyn2008megapipe}
Gwyn, S.~D. 2008, Publications of the Astronomical Society of the Pacific, 120, 212

\bibitem[{{Habas} {et~al.}(2020){Habas}, {Marleau}, {Duc}, {Durrell}, {Paudel}, {Poulain}, {S{\'a}nchez-Janssen}, {Sreejith}, {Ramasawmy}, {Stemock}, {Leach}, {Cuillandre}, {Gwyn}, {Agnello}, {B{\'\i}lek}, {Fensch}, {M{\"u}ller}, {Peng}, \& {van der Burg}}]{2020MNRAS.491.1901H}
{Habas}, R., {Marleau}, F.~R., {Duc}, P.-A., {et~al.} 2020, \mnras, 491, 1901

\bibitem[{{Ibata} {et~al.}(2007){Ibata}, {Martin}, {Irwin}, {Chapman}, {Ferguson}, {Lewis}, \& {McConnachie}}]{2007ApJ...671.1591I}
{Ibata}, R., {Martin}, N.~F., {Irwin}, M., {et~al.} 2007, \apj, 671, 1591

\bibitem[{Ibata {et~al.}(1994)Ibata, Gilmore, \& Irwin}]{ibata1994dwarf}
Ibata, R.~A., Gilmore, G., \& Irwin, M. 1994, Nature, 370, 194

\bibitem[{Ibata {et~al.}(2017{\natexlab{a}})Ibata, McConnachie, Cuillandre, Fantin, Haywood, Martin, Bergeron, Beckmann, Bernard, Bonifacio, {et~al.}}]{ibata2017canada}
Ibata, R.~A., McConnachie, A., Cuillandre, J.-C., {et~al.} 2017{\natexlab{a}}, The Astrophysical Journal, 848, 128

\bibitem[{Ibata {et~al.}(2017{\natexlab{b}})Ibata, McConnachie, Cuillandre, Fantin, Haywood, Martin, Bergeron, Beckmann, Bernard, Bonifacio, {et~al.}}]{ibata2017chemical}
Ibata, R.~A., McConnachie, A., Cuillandre, J.-C., {et~al.} 2017{\natexlab{b}}, The Astrophysical Journal, 848, 129

\bibitem[{Irwin {et~al.}(1990)Irwin, Bunclark, Bridgeland, \& McMahon}]{irwin1990new}
Irwin, M., Bunclark, P., Bridgeland, M., \& McMahon, R. 1990, Monthly Notices of the Royal Astronomical Society (ISSN 0035-8711), vol. 244, May 15, 1990, p. 16P-19P., 244, 16P

\bibitem[{Irwin {et~al.}(2008)Irwin, Ferguson, Huxor, Tanvir, Ibata, \& Lewis}]{irwin2008andromeda}
Irwin, M., Ferguson, A., Huxor, A., {et~al.} 2008, The Astrophysical Journal, 676, L17

\bibitem[{{Ivezi{\'c}} {et~al.}(2019){Ivezi{\'c}}, {Kahn}, {Tyson}, {Abel}, {Acosta}, {Allsman}, {Alonso}, {AlSayyad}, {Anderson}, {Andrew}, {Angel}, {Angeli}, {Ansari}, {Antilogus}, {Araujo}, {Armstrong}, {Arndt}, {Astier}, {Aubourg}, {Auza}, {Axelrod}, {Bard}, {Barr}, {Barrau}, {Bartlett}, {Bauer}, {Bauman}, {Baumont}, {Bechtol}, {Bechtol}, {Becker}, {Becla}, {Beldica}, {Bellavia}, {Bianco}, {Biswas}, {Blanc}, {Blazek}, {Blandford}, {Bloom}, {Bogart}, {Bond}, {Booth}, {Borgland}, {Borne}, {Bosch}, {Boutigny}, {Brackett}, {Bradshaw}, {Brandt}, {Brown}, {Bullock}, {Burchat}, {Burke}, {Cagnoli}, {Calabrese}, {Callahan}, {Callen}, {Carlin}, {Carlson}, {Chandrasekharan}, {Charles-Emerson}, {Chesley}, {Cheu}, {Chiang}, {Chiang}, {Chirino}, {Chow}, {Ciardi}, {Claver}, {Cohen-Tanugi}, {Cockrum}, {Coles}, {Connolly}, {Cook}, {Cooray}, {Covey}, {Cribbs}, {Cui}, {Cutri}, {Daly}, {Daniel}, {Daruich}, {Daubard}, {Daues}, {Dawson}, {Delgado}, {Dellapenna}, {de Peyster}, {de Val-Borro}, {Digel}, {Doherty}, {Dubois},
  {Dubois-Felsmann}, {Durech}, {Economou}, {Eifler}, {Eracleous}, {Emmons}, {Fausti Neto}, {Ferguson}, {Figueroa}, {Fisher-Levine}, {Focke}, {Foss}, {Frank}, {Freemon}, {Gangler}, {Gawiser}, {Geary}, {Gee}, {Geha}, {Gessner}, {Gibson}, {Gilmore}, {Glanzman}, {Glick}, {Goldina}, {Goldstein}, {Goodenow}, {Graham}, {Gressler}, {Gris}, {Guy}, {Guyonnet}, {Haller}, {Harris}, {Hascall}, {Haupt}, {Hernandez}, {Herrmann}, {Hileman}, {Hoblitt}, {Hodgson}, {Hogan}, {Howard}, {Huang}, {Huffer}, {Ingraham}, {Innes}, {Jacoby}, {Jain}, {Jammes}, {Jee}, {Jenness}, {Jernigan}, {Jevremovi{\'c}}, {Johns}, {Johnson}, {Johnson}, {Jones}, {Juramy-Gilles}, {Juri{\'c}}, {Kalirai}, {Kallivayalil}, {Kalmbach}, {Kantor}, {Karst}, {Kasliwal}, {Kelly}, {Kessler}, {Kinnison}, {Kirkby}, {Knox}, {Kotov}, {Krabbendam}, {Krughoff}, {Kub{\'a}nek}, {Kuczewski}, {Kulkarni}, {Ku}, {Kurita}, {Lage}, {Lambert}, {Lange}, {Langton}, {Le Guillou}, {Levine}, {Liang}, {Lim}, {Lintott}, {Long}, {Lopez}, {Lotz}, {Lupton}, {Lust}, {MacArthur}, {Mahabal},
  {Mandelbaum}, {Markiewicz}, {Marsh}, {Marshall}, {Marshall}, {May}, {McKercher}, {McQueen}, {Meyers}, {Migliore}, {Miller}, \& {Mills}}]{2019ApJ...873..111I}
{Ivezi{\'c}}, {\v{Z}}., {Kahn}, S.~M., {Tyson}, J.~A., {et~al.} 2019, \apj, 873, 111

\bibitem[{Jacobs {et~al.}(2019)Jacobs, Collett, Glazebrook, McCarthy, Qin, Abbott, Abdalla, Annis, Avila, Bechtol, {et~al.}}]{jacobs2019finding}
Jacobs, C., Collett, T., Glazebrook, K., {et~al.} 2019, Monthly Notices of the Royal Astronomical Society, 484, 5330

\bibitem[{{Javanmardi} {et~al.}(2016){Javanmardi}, {Martinez-Delgado}, {Kroupa}, {Henkel}, {Crawford}, {Teuwen}, {Gabany}, {Hanson}, {Chonis}, \& {Neyer}}]{2016A&A...588A..89J}
{Javanmardi}, B., {Martinez-Delgado}, D., {Kroupa}, P., {et~al.} 2016, \aap, 588, A89

\bibitem[{{Jerjen} \& {Dressler}(1997)}]{1997A&AS..124....1J}
{Jerjen}, H. \& {Dressler}, A. 1997, \aaps, 124, 1

\bibitem[{Kaiser {et~al.}(2002)Kaiser, Aussel, Burke, Boesgaard, Chambers, Chun, Heasley, Hodapp, Hunt, Jedicke, {et~al.}}]{kaiser2002pan}
Kaiser, N., Aussel, H., Burke, B.~E., {et~al.} 2002, in Survey and Other Telescope Technologies and Discoveries, Vol. 4836, SPIE, 154--164

\bibitem[{Kim \& Brunner(2016)}]{kim2016star}
Kim, E.~J. \& Brunner, R.~J. 2016, Monthly Notices of the Royal Astronomical Society, stw2672

\bibitem[{Kingma \& Ba(2014)}]{Kingma2014AdamAM}
Kingma, D.~P. \& Ba, J. 2014, CoRR, abs/1412.6980

\bibitem[{Klypin {et~al.}(1999)Klypin, Kravtsov, Valenzuela, \& Prada}]{klypin1999missing}
Klypin, A., Kravtsov, A.~V., Valenzuela, O., \& Prada, F. 1999, The Astrophysical Journal, 522, 82

\bibitem[{{Kovlakas} {et~al.}(2021){Kovlakas}, {Zezas}, {Andrews}, {Basu-Zych}, {Fragos}, {Hornschemeier}, {Kouroumpatzakis}, {Lehmer}, \& {Ptak}}]{2021MNRAS.506.1896K}
{Kovlakas}, K., {Zezas}, A., {Andrews}, J.~J., {et~al.} 2021, \mnras, 506, 1896

\bibitem[{Lanusse {et~al.}(2018)Lanusse, Ma, Li, Collett, Li, Ravanbakhsh, Mandelbaum, \& P{\'o}czos}]{lanusse2018cmu}
Lanusse, F., Ma, Q., Li, N., {et~al.} 2018, Monthly Notices of the Royal Astronomical Society, 473, 3895

\bibitem[{LeCun {et~al.}(1998)LeCun, Bottou, Bengio, \& Haffner}]{lecun1998gradient}
LeCun, Y., Bottou, L., Bengio, Y., \& Haffner, P. 1998, Proceedings of the IEEE, 86, 2278

\bibitem[{{Li} {et~al.}(2024){Li}, {Kilbinger}, {Luo}, {Wang}, {Wang}, {Wittje}, {Hildebrandt}, {Van Waerbeke}, {Hudson}, {Farrens}, {Liaudat}, {Liu}, {Zhang}, {Wang}, {Russier}, {Guinot}, {Baumont}, {Hervas Peters}, {de Boer}, {Wang}, {McConnachie}, {Cuillandre}, \& {Fabbro}}]{2024ApJ...969L..25L}
{Li}, Q., {Kilbinger}, M., {Luo}, W., {et~al.} 2024, \apjl, 969, L25

\bibitem[{{Lim} {et~al.}(2023){Lim}, {Hill}, {Scott}, {van Waerbeke}, {Cuillandre}, {Carlberg}, {Chisari}, {Dvornik}, {Erben}, {Gwyn}, {McConnachie}, {Miville-Desch{\^e}nes}, {Wright}, \& {Duc}}]{2023MNRAS.525.1443L}
{Lim}, S., {Hill}, R., {Scott}, D., {et~al.} 2023, \mnras, 525, 1443

\bibitem[{Lintott {et~al.}(2011)Lintott, Schawinski, Bamford, Slosar, Land, Thomas, Edmondson, Masters, Nichol, Raddick, {et~al.}}]{lintott2011galaxy}
Lintott, C., Schawinski, K., Bamford, S., {et~al.} 2011, Monthly Notices of the Royal Astronomical Society, 410, 166

\bibitem[{Lintott {et~al.}(2008)Lintott, Schawinski, Slosar, Land, Bamford, Thomas, Raddick, Nichol, Szalay, Andreescu, {et~al.}}]{lintott2008galaxy}
Lintott, C.~J., Schawinski, K., Slosar, A., {et~al.} 2008, Monthly Notices of the Royal Astronomical Society, 389, 1179

\bibitem[{Liu {et~al.}(2022)Liu, Mao, Wu, Feichtenhofer, Darrell, \& Xie}]{liu2022convnet2020s}
Liu, Z., Mao, H., Wu, C.-Y., {et~al.} 2022, A ConvNet for the 2020s

\bibitem[{Loshchilov \& Hutter(2019)}]{loshchilov2018decoupled}
Loshchilov, I. \& Hutter, F. 2019, in International Conference on Learning Representations

\bibitem[{{Lupton} {et~al.}(2004){Lupton}, {Blanton}, {Fekete}, {Hogg}, {O'Mullane}, {Szalay}, \& {Wherry}}]{2004PASP..116..133L}
{Lupton}, R., {Blanton}, M.~R., {Fekete}, G., {et~al.} 2004, \pasp, 116, 133

\bibitem[{MacGillivray {et~al.}(1987)MacGillivray, Bhatia, Beard, \& Dodd}]{macgillivray1987automated}
MacGillivray, H., Bhatia, R., Beard, S., \& Dodd, R. 1987, in ESO Conference Workshop Proceedings, No. 27, p. 477-479, Vol.~27, 477--479

\bibitem[{{Magnier} {et~al.}(2020{\natexlab{a}}){Magnier}, {Chambers}, {Flewelling}, {Hoblitt}, {Huber}, {Price}, {Sweeney}, {Waters}, {Denneau}, {Draper}, {Hodapp}, {Jedicke}, {Kaiser}, {Kudritzki}, {Metcalfe}, {Stubbs}, \& {Wainscoat}}]{2020ApJS..251....3M}
{Magnier}, E.~A., {Chambers}, K.~C., {Flewelling}, H.~A., {et~al.} 2020{\natexlab{a}}, \apjs, 251, 3

\bibitem[{{Magnier} {et~al.}(2020{\natexlab{b}}){Magnier}, {Schlafly}, {Finkbeiner}, {Tonry}, {Goldman}, {R{\"o}ser}, {Schilbach}, {Casertano}, {Chambers}, {Flewelling}, {Huber}, {Price}, {Sweeney}, {Waters}, {Denneau}, {Draper}, {Hodapp}, {Jedicke}, {Kaiser}, {Kudritzki}, {Metcalfe}, {Stubbs}, \& {Wainscoat}}]{2020ApJS..251....6M}
{Magnier}, E.~A., {Schlafly}, E.~F., {Finkbeiner}, D.~P., {et~al.} 2020{\natexlab{b}}, \apjs, 251, 6

\bibitem[{{Mao} {et~al.}(2024){Mao}, {Geha}, {Wechsler}, {Asali}, {Wang}, {Kado-Fong}, {Kallivayalil}, {Nadler}, {Tollerud}, {Weiner}, {de los Reyes}, \& {Wu}}]{2024ApJ...976..117M}
{Mao}, Y.-Y., {Geha}, M., {Wechsler}, R.~H., {et~al.} 2024, \apj, 976, 117

\bibitem[{{Mao} {et~al.}(2021){Mao}, {Geha}, {Wechsler}, {Weiner}, {Tollerud}, {Nadler}, \& {Kallivayalil}}]{2021ApJ...907...85M}
{Mao}, Y.-Y., {Geha}, M., {Wechsler}, R.~H., {et~al.} 2021, \apj, 907, 85

\bibitem[{Martin {et~al.}(2006)Martin, Ibata, Irwin, Chapman, Lewis, Ferguson, Tanvir, \& McConnachie}]{martin2006discovery}
Martin, N., Ibata, R., Irwin, M., {et~al.} 2006, Monthly Notices of the Royal Astronomical Society, 371, 1983

\bibitem[{{Martin} {et~al.}(2013){Martin}, {Ibata}, {McConnachie}, {Mackey}, {Ferguson}, {Irwin}, {Lewis}, \& {Fardal}}]{2013ApJ...776...80M}
{Martin}, N.~F., {Ibata}, R.~A., {McConnachie}, A.~W., {et~al.} 2013, \apj, 776, 80

\bibitem[{{Martin} {et~al.}(2009){Martin}, {McConnachie}, {Irwin}, {Widrow}, {Ferguson}, {Ibata}, {Dubinski}, {Babul}, {Chapman}, {Fardal}, {Lewis}, {Navarro}, \& {Rich}}]{2009ApJ...705..758M}
{Martin}, N.~F., {McConnachie}, A.~W., {Irwin}, M., {et~al.} 2009, \apj, 705, 758

\bibitem[{{Martinez-Delgado} {et~al.}(2024){Martinez-Delgado}, {Stein}, {Pawlowski}, {Makarov}, {Makarova}, {Donatiello}, \& {Lang}}]{2024arXiv240503769M}
{Martinez-Delgado}, D., {Stein}, M., {Pawlowski}, M.~S., {et~al.} 2024, arXiv e-prints, arXiv:2405.03769

\bibitem[{Mateo {et~al.}(1991)Mateo, Olszewski, Welch, Fischer, \& Kunkel}]{mateo1991kinematic}
Mateo, M., Olszewski, E., Welch, D.~L., Fischer, P., \& Kunkel, W. 1991, Astronomical Journal (ISSN 0004-6256), vol. 102, Sept. 1991, p. 914-926., 102, 914

\bibitem[{McConnachie {et~al.}(2008)McConnachie, Huxor, Martin, Irwin, Chapman, Fahlman, Ferguson, Ibata, Lewis, Richer, {et~al.}}]{mcconnachie2008trio}
McConnachie, A.~W., Huxor, A., Martin, N.~F., {et~al.} 2008, The Astrophysical Journal, 688, 1009

\bibitem[{{McConnachie} {et~al.}(2009){McConnachie}, {Irwin}, {Ibata}, {Dubinski}, {Widrow}, {Martin}, {C{\^o}t{\'e}}, {Dotter}, {Navarro}, {Ferguson}, {Puzia}, {Lewis}, {Babul}, {Barmby}, {Bienaym{\'e}}, {Chapman}, {Cockcroft}, {Collins}, {Fardal}, {Harris}, {Huxor}, {Mackey}, {Pe{\~n}arrubia}, {Rich}, {Richer}, {Siebert}, {Tanvir}, {Valls-Gabaud}, \& {Venn}}]{2009Natur.461...66M}
{McConnachie}, A.~W., {Irwin}, M.~J., {Ibata}, R.~A., {et~al.} 2009, \nat, 461, 66

\bibitem[{{Merritt} {et~al.}(2014){Merritt}, {van Dokkum}, \& {Abraham}}]{2014ApJ...787L..37M}
{Merritt}, A., {van Dokkum}, P., \& {Abraham}, R. 2014, \apjl, 787, L37

\bibitem[{{Merritt} {et~al.}(2016){Merritt}, {van Dokkum}, {Danieli}, {Abraham}, {Zhang}, {Karachentsev}, \& {Makarova}}]{2016ApJ...833..168M}
{Merritt}, A., {van Dokkum}, P., {Danieli}, S., {et~al.} 2016, \apj, 833, 168

\bibitem[{Moore {et~al.}(1999)Moore, Ghigna, Governato, Lake, Quinn, Stadel, \& Tozzi}]{moore1999dark}
Moore, B., Ghigna, S., Governato, F., {et~al.} 1999, The Astrophysical Journal, 524, L19

\bibitem[{{Mosby} {et~al.}(2020){Mosby}, {Rauscher}, {Bennett}, {Cheng}, {Cheung}, {Cillis}, {Content}, {Cottingham}, {Foltz}, {Gygax}, {Hill}, {Kruk}, {Mah}, {Meier}, {Merchant}, {Miko}, {Piquette}, {Waczynski}, \& {Wen}}]{2020JATIS...6d6001M}
{Mosby}, G., {Rauscher}, B.~J., {Bennett}, C., {et~al.} 2020, Journal of Astronomical Telescopes, Instruments, and Systems, 6, 046001

\bibitem[{{Mpetha} {et~al.}(2024){Mpetha}, {Taylor}, {Amoura}, \& {Haggar}}]{2024MNRAS.532.2521M}
{Mpetha}, C.~T., {Taylor}, J.~E., {Amoura}, Y., \& {Haggar}, R. 2024, \mnras, 532, 2521

\bibitem[{{Mpetha} {et~al.}(2025){Mpetha}, {Taylor}, {Amoura}, {Haggar}, {de Boer}, {Guerrini}, {Guinot}, {Hervas Peters}, {Hildebrandt}, {Hudson}, {Kilbinger}, {Liaudat}, {McConnachie}, {Van Waerbeke}, \& {Wittje}}]{2025arXiv250109147M}
{Mpetha}, C.~T., {Taylor}, J.~E., {Amoura}, Y., {et~al.} 2025, arXiv e-prints, arXiv:2501.09147

\bibitem[{{M{\"u}ller} \& {Jerjen}(2020)}]{2020A&A...644A..91M}
{M{\"u}ller}, O. \& {Jerjen}, H. 2020, \aap, 644, A91

\bibitem[{{M{\"u}ller} {et~al.}(2017{\natexlab{a}}){M{\"u}ller}, {Jerjen}, \& {Binggeli}}]{2017A&A...597A...7M}
{M{\"u}ller}, O., {Jerjen}, H., \& {Binggeli}, B. 2017{\natexlab{a}}, \aap, 597, A7

\bibitem[{{M{\"u}ller} {et~al.}(2017{\natexlab{b}}){M{\"u}ller}, {Scalera}, {Binggeli}, \& {Jerjen}}]{2017A&A...602A.119M}
{M{\"u}ller}, O., {Scalera}, R., {Binggeli}, B., \& {Jerjen}, H. 2017{\natexlab{b}}, \aap, 602, A119

\bibitem[{{M{\"u}ller} \& {Schnider}(2021)}]{2021OJAp....4E...3M}
{M{\"u}ller}, O. \& {Schnider}, E. 2021, The Open Journal of Astrophysics, 4, 3

\bibitem[{{Pace}(2024)}]{Pace2024arXiv241107424P}
{Pace}, A.~B. 2024, arXiv e-prints, arXiv:2411.07424

\bibitem[{Paranjpye {et~al.}(2020)Paranjpye, Mahabal, Ramaprakash, Panopoulou, Cleary, Readhead, Blinov, \& Tassis}]{paranjpye2020eliminating}
Paranjpye, D., Mahabal, A., Ramaprakash, A., {et~al.} 2020, Monthly Notices of the Royal Astronomical Society, 491, 5151

\bibitem[{{Paudel} {et~al.}(2023){Paudel}, {Yoon}, {Yoo}, {Smith}, {Chhatkuli}, {Kumar Bachchan}, {Aryal}, {Adhikari}, {Adhikari}, {Sedain}, {Sheikh}, {Dhital}, {Giri}, \& {Baral}}]{2023ApJS..265...57P}
{Paudel}, S., {Yoon}, S.-J., {Yoo}, J., {et~al.} 2023, \apjs, 265, 57

\bibitem[{Payerne {et~al.}(2024)Payerne, Doumerg, Y{\`e}che, Ruhlmann-Kleider, Raichoor, Lang, Aguilar, Ahlen, Bianchi, Brooks, {et~al.}}]{payerne2024high}
Payerne, C., Doumerg, W., Y{\`e}che, C., {et~al.} 2024, arXiv preprint arXiv:2410.08062

\bibitem[{Peng {et~al.}(2010)Peng, Ho, Impey, \& Rix}]{peng2010detailed}
Peng, C.~Y., Ho, L.~C., Impey, C.~D., \& Rix, H.-W. 2010, The Astronomical Journal, 139, 2097

\bibitem[{{Poulain} {et~al.}(2021){Poulain}, {Marleau}, {Habas}, {Duc}, {S{\'a}nchez-Janssen}, {Durrell}, {Paudel}, {Ahad}, {Chougule}, {M{\"u}ller}, {Lim}, {B{\'\i}lek}, \& {Fensch}}]{2021MNRAS.506.5494P}
{Poulain}, M., {Marleau}, F.~R., {Habas}, R., {et~al.} 2021, \mnras, 506, 5494

\bibitem[{Revaz \& Jablonka(2018)}]{revaz2018pushing}
Revaz, Y. \& Jablonka, P. 2018, Astronomy \& Astrophysics, 616, A96

\bibitem[{Ribli {et~al.}(2019)Ribli, Pataki, Zorrilla~Matilla, Hsu, Haiman, \& Csabai}]{ribli2019weak}
Ribli, D., Pataki, B.~{\'A}., Zorrilla~Matilla, J.~M., {et~al.} 2019, Monthly Notices of the Royal Astronomical Society, 490, 1843

\bibitem[{Richardson {et~al.}(2011)Richardson, Irwin, McConnachie, Martin, Dotter, Ferguson, Ibata, Chapman, Lewis, Tanvir, {et~al.}}]{richardson2011pandas}
Richardson, J.~C., Irwin, M.~J., McConnachie, A.~W., {et~al.} 2011, The Astrophysical Journal, 732, 76

\bibitem[{Roberts {et~al.}(2022)Roberts, Parker, Gwyn, Hudson, Carlberg, McConnachie, Cuillandre, Chambers, Duc, Furusawa, {et~al.}}]{roberts2022ram}
Roberts, I.~D., Parker, L.~C., Gwyn, S., {et~al.} 2022, Monthly Notices of the Royal Astronomical Society, 509, 1342

\bibitem[{{Robison} {et~al.}(2023){Robison}, {Hudson}, {Cuillandre}, {Erben}, {Fabbro}, {Gavazzi}, {Guinot}, {Gwyn}, {Hildebrandt}, {Kilbinger}, {McConnachie}, {Miller}, {Spitzer}, \& {van Waerbeke}}]{2023MNRAS.523.1614R}
{Robison}, B., {Hudson}, M.~J., {Cuillandre}, J.-C., {et~al.} 2023, \mnras, 523, 1614

\bibitem[{{Sandage} \& {Binggeli}(1984)}]{1984AJ.....89..919S}
{Sandage}, A. \& {Binggeli}, B. 1984, \aj, 89, 919

\bibitem[{Savary {et~al.}(2022)Savary, Rojas, Maus, Cl{\'e}ment, Courbin, Gavazzi, Chan, Lemon, Vernardos, Ca{\~n}ameras, {et~al.}}]{savary2022strong}
Savary, E., Rojas, K., Maus, M., {et~al.} 2022, Astronomy \& Astrophysics, 666, A1

\bibitem[{Smith {et~al.}(2024)Smith, Cerny, Hayes, Sestito, Jensen, McConnachie, Geha, Navarro, Li, Cuillandre, {et~al.}}]{smith2024discovery}
Smith, S.~E., Cerny, W., Hayes, C.~R., {et~al.} 2024, The Astrophysical Journal, 961, 92

\bibitem[{Smith {et~al.}(2023)Smith, Jensen, Roediger, Sestito, Hayes, McConnachie, Cuillandre, Gwyn, Magnier, Chambers, {et~al.}}]{smith2023discovery}
Smith, S.~E., Jensen, J., Roediger, J., {et~al.} 2023, The Astronomical Journal, 166, 76

\bibitem[{Sola {et~al.}(2022)Sola, Duc, Richards, Paiement, Urbano, Klehammer, B{\'\i}lek, Cuillandre, Gwyn, \& McConnachie}]{sola2022characterization}
Sola, E., Duc, P.-A., Richards, F., {et~al.} 2022, Astronomy \& Astrophysics, 662, A124

\bibitem[{Szegedy {et~al.}(2016)Szegedy, Vanhoucke, Ioffe, Shlens, \& Wojna}]{szegedy2016rethinking}
Szegedy, C., Vanhoucke, V., Ioffe, S., Shlens, J., \& Wojna, Z. 2016, in Proceedings of the IEEE conference on computer vision and pattern recognition, 2818--2826

\bibitem[{{Tanoglidis} {et~al.}(2021{\natexlab{a}}){Tanoglidis}, {{\'C}iprijanovi{\'c}}, \& {Drlica-Wagner}}]{2021bA&C....3500469T}
{Tanoglidis}, D., {{\'C}iprijanovi{\'c}}, A., \& {Drlica-Wagner}, A. 2021{\natexlab{a}}, Astronomy and Computing, 35, 100469

\bibitem[{{Tanoglidis} {et~al.}(2021{\natexlab{b}}){Tanoglidis}, {Drlica-Wagner}, {Wei}, {Li}, {S{\'a}nchez}, {Zhang}, {Peter}, {Feldmeier-Krause}, {Prat}, {Casey}, {Palmese}, {S{\'a}nchez}, {DeRose}, {Conselice}, {Gagnon}, {Abbott}, {Aguena}, {Allam}, {Avila}, {Bechtol}, {Bertin}, {Bhargava}, {Brooks}, {Burke}, {Rosell}, {Kind}, {Carretero}, {Chang}, {Costanzi}, {da Costa}, {De Vicente}, {Desai}, {Diehl}, {Doel}, {Eifler}, {Everett}, {Evrard}, {Flaugher}, {Frieman}, {Garc{\'\i}a-Bellido}, {Gerdes}, {Gruendl}, {Gschwend}, {Gutierrez}, {Hartley}, {Hollowood}, {Huterer}, {James}, {Krause}, {Kuehn}, {Kuropatkin}, {Maia}, {March}, {Marshall}, {Menanteau}, {Miquel}, {Ogando}, {Paz-Chinch{\'o}n}, {Romer}, {Roodman}, {Sanchez}, {Scarpine}, {Serrano}, {Sevilla-Noarbe}, {Smith}, {Suchyta}, {Tarle}, {Thomas}, {Tucker}, {Walker}, \& {DES Collaboration}}]{2021aApJS..252...18T}
{Tanoglidis}, D., {Drlica-Wagner}, A., {Wei}, K., {et~al.} 2021{\natexlab{b}}, \apjs, 252, 18

\bibitem[{Teeninga {et~al.}(2015)Teeninga, Moschini, Trager, \& Wilkinson}]{teeninga2015improved}
Teeninga, P., Moschini, U., Trager, S.~C., \& Wilkinson, M.~H. 2015, in International Symposium on Mathematical Morphology and Its Applications to Signal and Image Processing, Springer, 157--168

\bibitem[{Teeninga {et~al.}(2016)Teeninga, Moschini, Trager, \& Wilkinson}]{teeninga2016statistical}
Teeninga, P., Moschini, U., Trager, S.~C., \& Wilkinson, M.~H. 2016, Mathematical Morphology-Theory and Applications, 1

\bibitem[{Thomas {et~al.}(2019{\natexlab{a}})Thomas, Annau, McConnachie, Fabbro, Teimoorinia, C{\^o}t{\'e}, Cuillandre, Gwyn, Ibata, Starkenburg, {et~al.}}]{thomas2019dwarfs}
Thomas, G.~F., Annau, N., McConnachie, A., {et~al.} 2019{\natexlab{a}}, The Astrophysical Journal, 886, 10

\bibitem[{Thomas {et~al.}(2020)Thomas, Jensen, McConnachie, C{\^o}t{\'e}, Venn, Longeard, Carlberg, Chapman, Cuillandre, Famaey, {et~al.}}]{thomas2020hidden}
Thomas, G.~F., Jensen, J., McConnachie, A., {et~al.} 2020, The Astrophysical Journal, 902, 89

\bibitem[{Thomas {et~al.}(2019{\natexlab{b}})Thomas, Laporte, McConnachie, Famaey, Ibata, Martin, Starkenburg, Carlberg, Malhan, \& Venn}]{thomas2019type}
Thomas, G.~F., Laporte, C.~F., McConnachie, A.~W., {et~al.} 2019{\natexlab{b}}, Monthly Notices of the Royal Astronomical Society, 483, 3119

\bibitem[{{Thomas} {et~al.}(2018){Thomas}, {McConnachie}, {Ibata}, {C{\^o}t{\'e}}, {Martin}, {Starkenburg}, {Carlberg}, {Chapman}, {Fabbro}, {Famaey}, {Fantin}, {Gwyn}, {H{\'e}nault-Brunet}, {Malhan}, {Navarro}, {Robin}, \& {Scott}}]{2018MNRAS.481.5223T}
{Thomas}, G.~F., {McConnachie}, A.~W., {Ibata}, R.~A., {et~al.} 2018, \mnras, 481, 5223

\bibitem[{{van Dokkum} {et~al.}(2015){van Dokkum}, {Abraham}, {Merritt}, {Zhang}, {Geha}, \& {Conroy}}]{2015ApJ...798L..45V}
{van Dokkum}, P.~G., {Abraham}, R., {Merritt}, A., {et~al.} 2015, \apjl, 798, L45

\bibitem[{Walker {et~al.}(2009)Walker, Mateo, Olszewski, Penarrubia, Evans, \& Gilmore}]{walker2009universal}
Walker, M.~G., Mateo, M., Olszewski, E.~W., {et~al.} 2009, The Astrophysical Journal, 704, 1274

\bibitem[{Walmsley {et~al.}(2023)Walmsley, Allen, Aussel, Bowles, Gregorowicz, Slijepcevic, Lintott, Scaife, Jab{\l}o{\'n}ska, Karchev, {et~al.}}]{walmsley2023zoobot}
Walmsley, M., Allen, C., Aussel, B., {et~al.} 2023, Journal of Open Source Software, 8

\bibitem[{Wilkinson {et~al.}(2022)Wilkinson, Ellison, Bottrell, Bickley, Gwyn, Cuillandre, \& Wild}]{wilkinson2022merger}
Wilkinson, S., Ellison, S.~L., Bottrell, C., {et~al.} 2022, Monthly Notices of the Royal Astronomical Society, 516, 4354

\bibitem[{Willett {et~al.}(2013)Willett, Lintott, Bamford, Masters, Simmons, Casteels, Edmondson, Fortson, Kaviraj, Keel, {et~al.}}]{willett2013galaxy}
Willett, K.~W., Lintott, C.~J., Bamford, S.~P., {et~al.} 2013, Monthly Notices of the Royal Astronomical Society, 435, 2835

\bibitem[{Willman(2010)}]{willman2010pursuit}
Willman, B. 2010, Advances in Astronomy, 2010, 285454

\bibitem[{York {et~al.}(2000)York, Adelman, Anderson~Jr, Anderson, Annis, Bahcall, Bakken, Barkhouser, Bastian, Berman, {et~al.}}]{york2000sloan}
York, D.~G., Adelman, J., Anderson~Jr, J.~E., {et~al.} 2000, The Astronomical Journal, 120, 1579

\bibitem[{{Zaritsky} {et~al.}(2019){Zaritsky}, {Donnerstein}, {Dey}, {Kadowaki}, {Zhang}, {Karunakaran}, {Mart{\'\i}nez-Delgado}, {Rahman}, \& {Spekkens}}]{2019ApJS..240....1Z}
{Zaritsky}, D., {Donnerstein}, R., {Dey}, A., {et~al.} 2019, \apjs, 240, 1

\bibitem[{{Zaritsky} {et~al.}(2023){Zaritsky}, {Donnerstein}, {Dey}, {Karunakaran}, {Kadowaki}, {Khim}, {Spekkens}, \& {Zhang}}]{2023ApJS..267...27Z}
{Zaritsky}, D., {Donnerstein}, R., {Dey}, A., {et~al.} 2023, \apjs, 267, 27

\bibitem[{{Zhang} {et~al.}(2024){Zhang}, {Kilbinger}, {Peters}, {Li}, {Luo}, {Baumont}, {Cuillandre}, {Fabbro}, {Gwyn}, {McConnachie}, \& {Wittje}}]{2024A&A...691A..75Z}
{Zhang}, Z., {Kilbinger}, M., {Peters}, F.~H., {et~al.} 2024, \aap, 691, A75

\bibitem[{Zhu {et~al.}(2019)Zhu, Dai, Bian, Chen, Chen, \& Hu}]{zhu2019galaxy}
Zhu, X.-P., Dai, J.-M., Bian, C.-J., {et~al.} 2019, Astrophysics and Space Science, 364, 1

\end{thebibliography}
%

\begin{appendix}
\section{Enhancing UDG Detectability through Binning}
\label{sec:appendix_udgs}

\begin{figure*}[!htb]
\centering
\includegraphics[width=\linewidth]{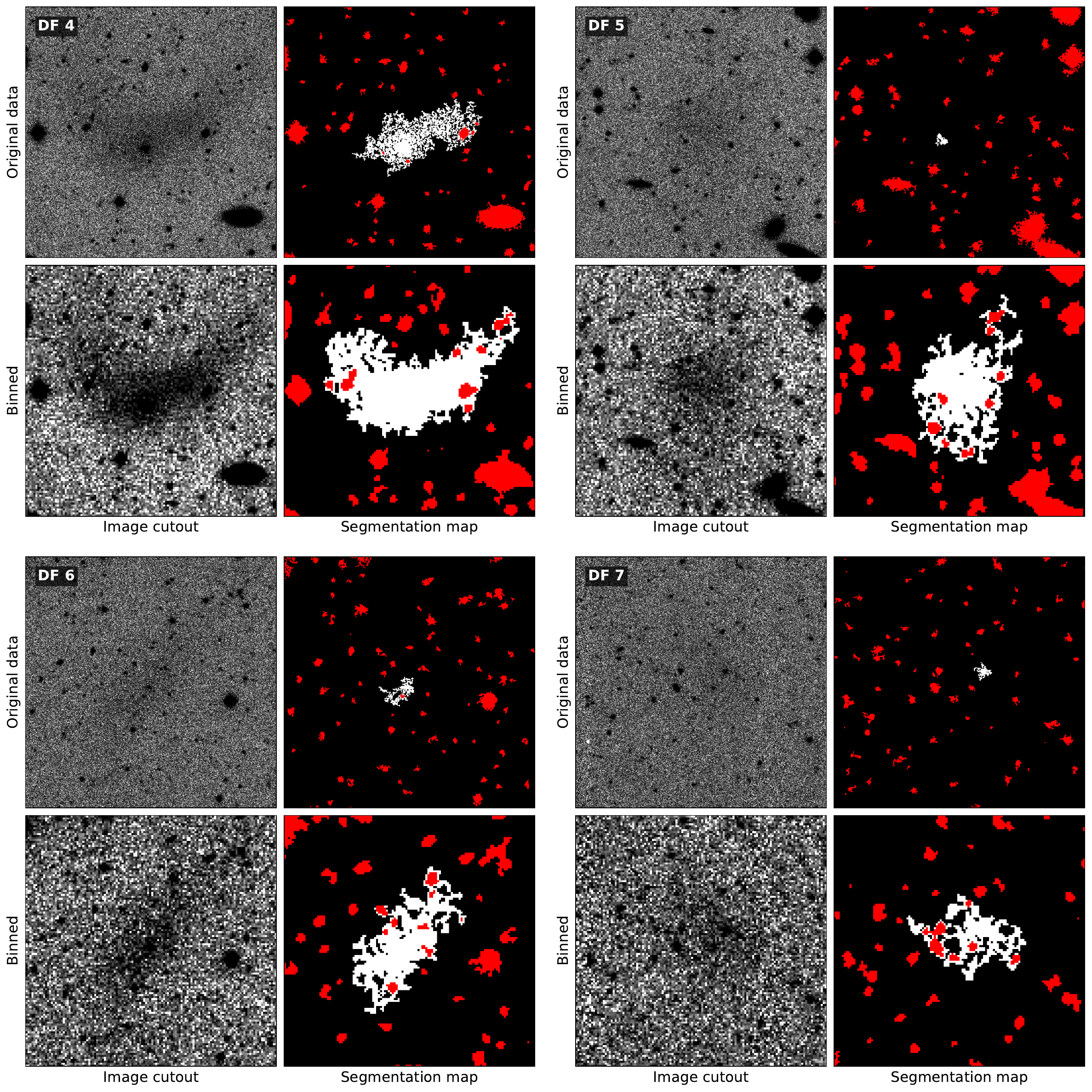}
\caption{Detectability of four UDGs detected around NGC\,5485 in \citet{2014ApJ...787L..37M} before and after 4$\times$4 binning. Each quadrant presents one source with four panels arranged as follows: original data (upper left) with its corresponding segmentation map (upper right), and 4$\times$4 binned data (lower left) with its corresponding segmentation map (lower right). The segmentation map shows the UDG segment in white and all other sources in red.}
\label{fig:appendix_udgs}
\end{figure*}

The 4$\times$4 binning scheme we applied as our first preprocessing step of the original images is motivated by the detectability gains for some of the faintest objects we gathered from the literature. In Figure \ref{fig:appendix_udgs} we illustrate this sensitivity enhancement by showing the full resolution images vs. binned images and the corresponding \textsc{MTO} segmentation maps.

\section{Object Filtering Criteria}
\label{sec:appendix_obj_filters}

\begin{figure*}[!htb]
\centering
\includegraphics[width=0.49\linewidth]{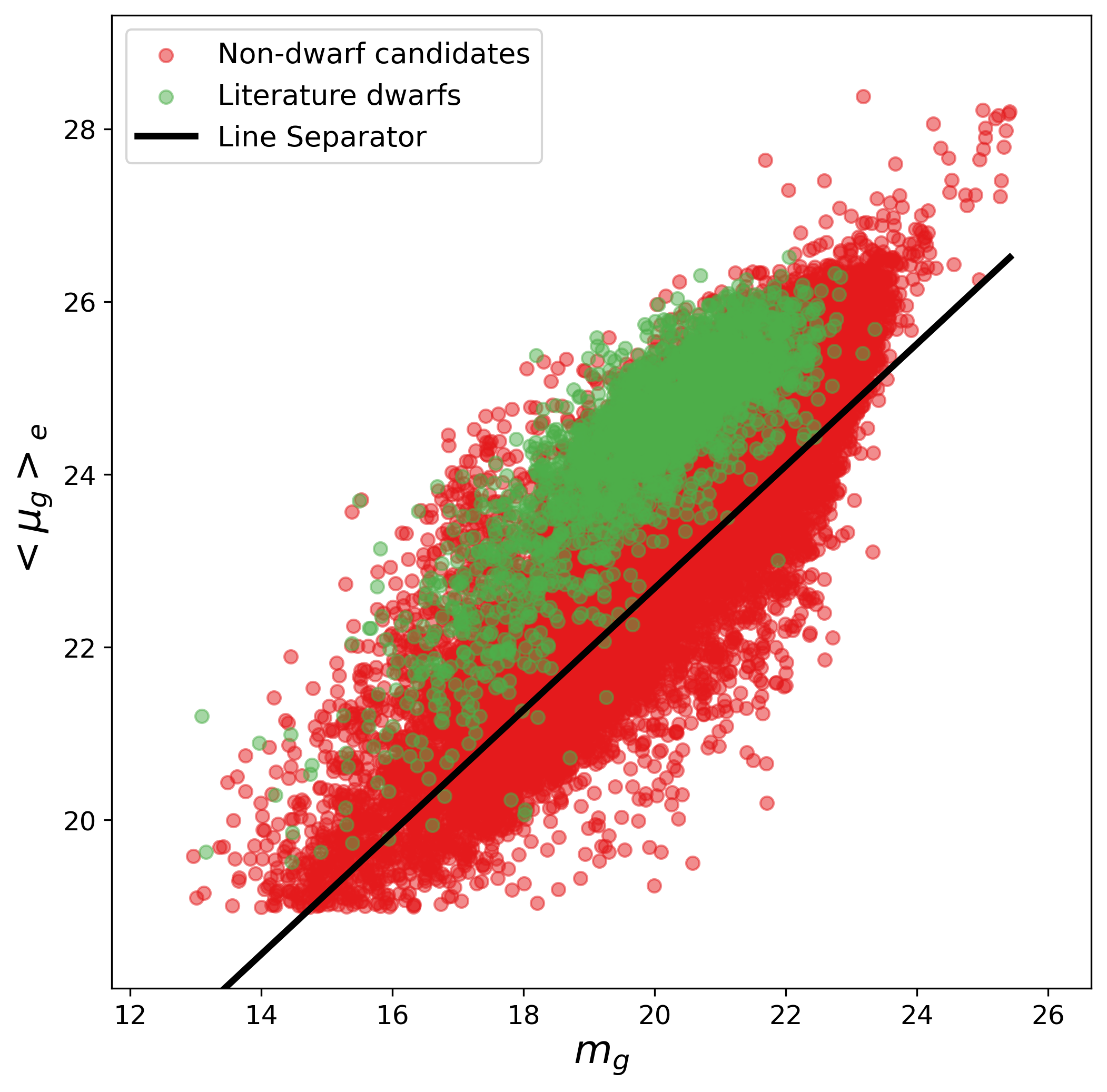}
\includegraphics[width=0.49\linewidth]{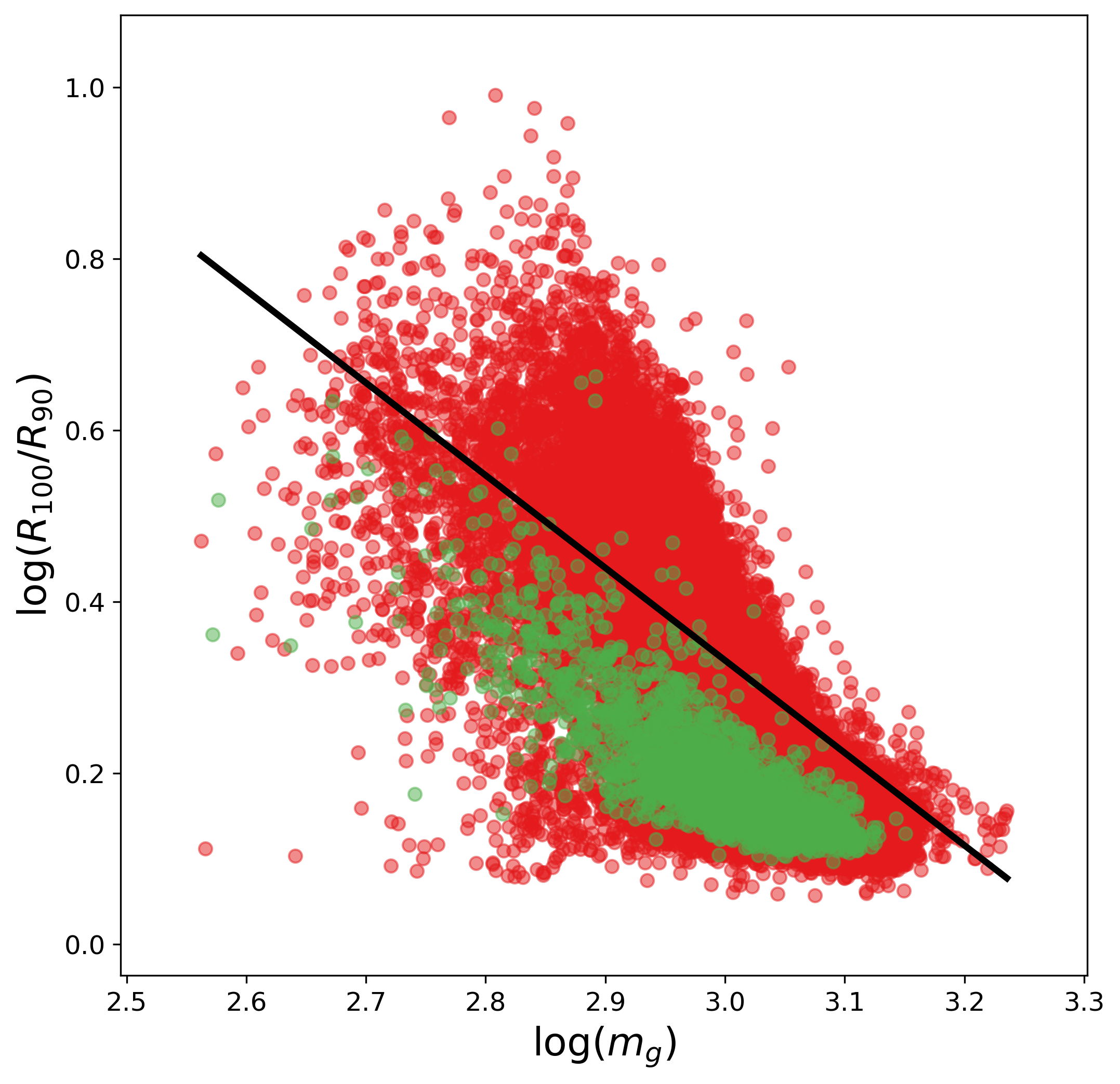}
\includegraphics[width=0.49\linewidth]{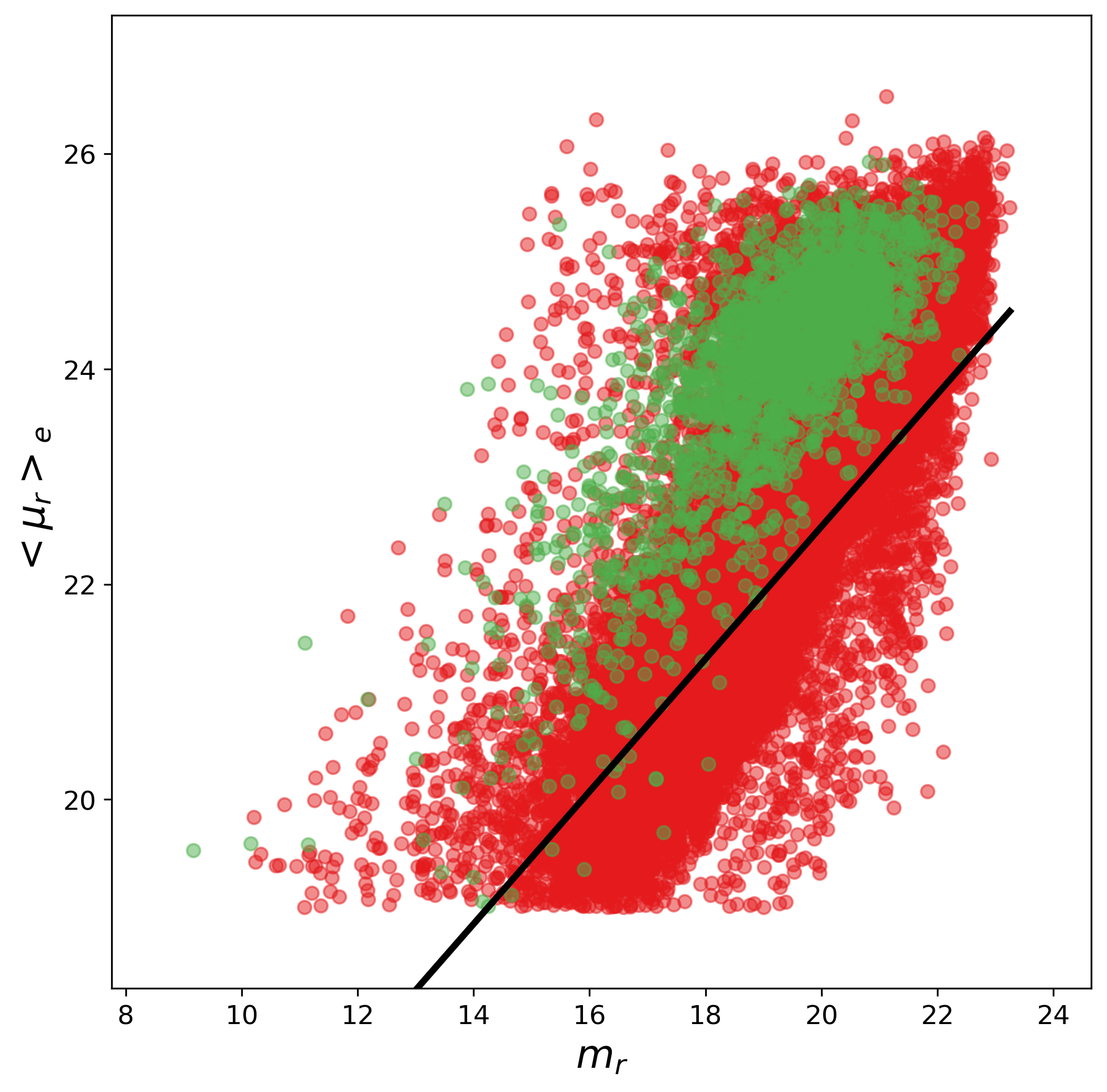}
\includegraphics[width=0.49\linewidth]{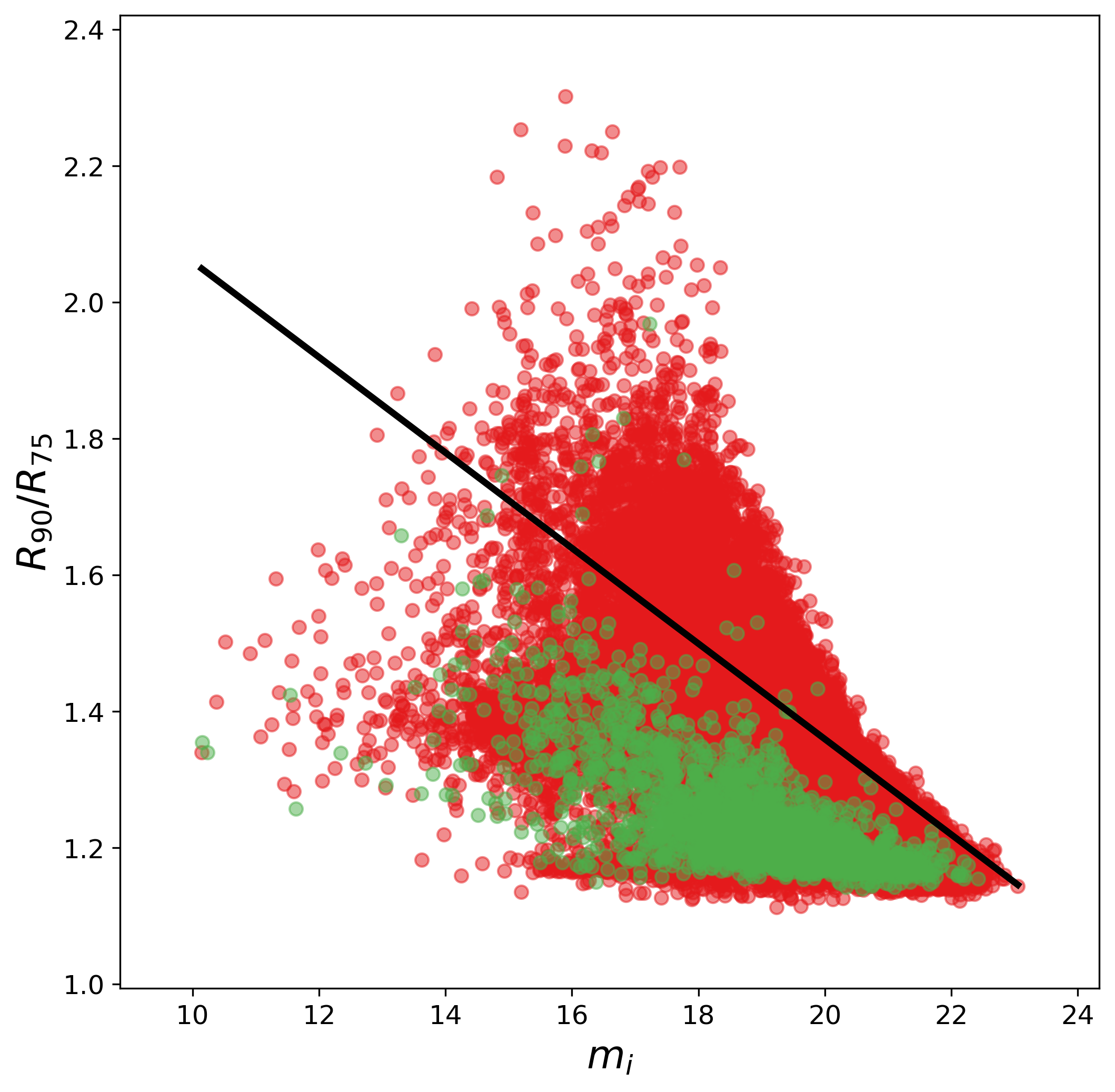}
\caption{Examples of parameter spaces used to filter dwarf galaxy candidates. These four panels show the clearest separations we could empirically determine for our data between known dwarf galaxies from the literature (green points) and their nearest neighbors assumed to be non-dwarfs (red points). Top left: Mean effective surface brightness $<\mu_{g}>_{e}$ vs. apparent magnitude $m_g$ in the \emph{g}-band. Top right: $log(R_{100}/R_{90}$) vs. $log(m_g$) in the \emph{g}-band. Bottom left: mean effective surface brightness $<\mu_{r}>_{e}$ vs. $m_r$ in the \emph{r}-band. Bottom right: $R_{90}/R_{75}$ vs. $m_i$ in the \emph{i}-band. The solid black line in each panel indicates the selection boundary we applied to filter out likely non-dwarf objects while preserving the majority of known dwarfs. The lines follow the relations \ref{eq:mu_mag_g}, \ref{eq:log_r_100_r_90_log_mag_g}, \ref{eq:mu_mag_r} and \ref{eq:r_90_r_75_mag_i}, respectively.}
\label{fig:filter_cireria}
\end{figure*}

To identify potential dwarf galaxy candidates, we implemented a multi-stage filtering process across the three different photometric bands we used in this work. For each band, we applied both basic threshold filtering and complex relational criteria designed to isolate low surface brightness galaxies. Tables \ref{tab:basic_r}, \ref{tab:basic_g}, and \ref{tab:basic_i} present the basic selection thresholds applied to our catalog in the \emph{r}, \emph{g}, and \emph{i} bands, respectively. Beyond basic thresholds, we implemented more complex relational criteria to further refine our selection. These criteria were empirically derived to maximize the inclusion of known dwarf galaxies from the literature while minimizing contamination from other astronomical sources. We report these relational criteria for each band in the following subsections. In addition to the parameters already described in Table \ref{tab:master_cat}, $\mu_{\text{median}}$, $\mu_{\text{mean}}$, $\mu_{\text{max}}$ are the median, mean and maximum pixel values of the objects provided by \textsc{MTO}. Dashes in the tables indicate the absence of a limiting value. In Figure \ref{fig:filter_cireria} we show examples of separation criteria for the different bands that allowed for the clearest separations between dwarfs and likely non-dwarfs we could determine for our dataset.

\subsection{\emph{r}-band Filtering Criteria}
\vspace{-10pt}
\begin{table}[!htb]
\centering
\caption{Basic selection criteria for the \emph{r}-band}
\vspace{-10pt}
\begin{threeparttable}
\label{tab:basic_r}
\begin{tabular}{lcc}
\toprule
Parameter & Min & Max \\
\midrule
$<\mu_{r}>_{e}$ (mag/arcsec$^{2}$) & 19.0 & -- \\
R$_{e}$ (arcsec) & 1.6 & 55.0 \\
B/A & 0.17 & -- \\
R$_{10}$ (arcsec) & 0.353 & 18.2 \\
R$_{25}$ (arcsec) & 0.4 & 32.1 \\
R$_{75}$ (arcsec) & 2.16 & 102.1 \\
R$_{90}$ (arcsec) & 2.5 & 145.1 \\
R$_{100}$ (arcsec) & 2.8 & 254.1 \\
R$_{\text{FWHM}}$ (arcsec) & 0.4 & 13.8 \\
$\mu_{\text{median}}$ (ADU) & 5.44 & 459.2 \\
$\mu_{\text{mean}}$ (ADU) & 6.4 & 1041.6 \\
$\mu_{\text{max}}$ (ADU) & 32.0 & 100080 \\
Flux (ADU) & 880 & -- \\
$m_{r}$ (mag) & 9.16 & 22.69 \\
\bottomrule
\end{tabular}
\end{threeparttable}
\end{table}
\vspace{-20pt}

\begin{equation}
<\mu_{r}>_{e}\, > 0.6155 \times m_{r} + 10.2257
\label{eq:mu_mag_r}
\end{equation}
\vspace{-20pt}
\begin{equation}
<\mu_{r}>_{e}\, > 1.35 \times m_r - 5.3464
\end{equation}
\vspace{-20pt}
\begin{equation}
\frac{<\mu_{r}>_{e}}{R_{75}} < \max(1.3611 \times m_r - 18.2513, \,2.51)
\end{equation}
\vspace{-20pt}
\begin{equation}
\log\left(\frac{R_{90}}{R_{75}}\right) < 0.295 \times \log(R_{100}) - 0.13
\end{equation}
\vspace{-20pt}

\subsection{\emph{g}-band Filtering Criteria}

\begin{table}[!htb]
\centering
\caption{Basic selection criteria for the WHIGS g-band}
\vspace{-10pt}
\begin{threeparttable}
\label{tab:basic_g}
\begin{tabular}{lcc}
\toprule
Parameter & Min & Max \\
\midrule
$<\mu_{g}>_{e}$ (mag/arcsec$^{2}$) & 19.5 & -- \\
R$_e$ (arcsec) & 1.6 & -- \\
B/A & 0.17 & -- \\
R$_{10}$ (arcsec) & 0.4 & 8.1 \\
R$_{25}$ (arcsec) & 0.72 & 15.0 \\
R$_{75}$ (arcsec) & 2.17 & 42.0 \\
R$_{90}$ (arcsec) & 2.47 & 54.0 \\
R$_{100}$ (arcsec) & 2.8 & 79.0 \\
R$_{\text{FWHM}}$ (arcsec) & 0.417 & 13.5 \\
$\mu_{\text{median}}$ (ADU) & 0.1152 & 12.8 \\
$\mu_{\text{mean}}$ (ADU) & 0.272 & 32.16 \\
$\mu_{\text{max}}$ (ADU) & 1.056 & 3616.0 \\
Flux (ADU) & 28.32 & -- \\
$m_{g}$ (mag) & 13.0 & 23.49 \\
\bottomrule
\end{tabular}
\end{threeparttable}
\end{table}

\begin{equation}
<\mu_{g}>_{e}\,\, > 0.7063 \times m_g + 8.5579
\label{eq:mu_mag_g}
\end{equation}
\vspace{-20pt}
\begin{equation}
<\mu_{g}>_{e}\,\, > 4.1 \times m_g - 68.6681
\end{equation}
\vspace{-20pt}
\begin{equation}
\log\left(\frac{R_{100}}{R_{90}}\right) < -1.0795 \times \log(m_g) + 3.57
\label{eq:log_r_100_r_90_log_mag_g}
\end{equation}
\vspace{-20pt}
\begin{equation}
\log\left(\frac{m_g}{R_{75}}\right) < \max(5.6017 \times \log(m_g) - 14.6037,\, 0.57)
\end{equation}
\vspace{-20pt}

\subsection{\emph{i}-band Filtering Criteria}

\begin{table}[!htb]
\centering
\caption{Basic selection criteria for the \emph{i}-band}
\vspace{-10pt}
\begin{threeparttable}
\label{tab:basic_i}
\begin{tabular}{lcc}
\toprule
Parameter & Min & Max \\
\midrule
$<\mu_{i}>_{e}$ (mag/arcsec$^{2}$) & 19.0 & 26.49 \\
R$_e$ (arcsec) & 1.6 & 40.0 \\
B/A & 0.17 & -- \\
R$_{10}$ (arcsec) & 0.4 & 14.95 \\
R$_{25}$ (arcsec) & 0.58 & 25.5 \\
R$_{75}$ (arcsec) & 2.1 & 58.0 \\
R$_{90}$ (arcsec) & 2.5 & 77.0 \\
R$_{100}$ (arcsec) & 2.7 & 117.0 \\
R$_{\text{FWHM}}$ (arcsec) & 0.4 & 20.8 \\
$\mu_{\text{median}}$ (ADU) & 7.12 & 651.2 \\
$\mu_{\text{mean}}$ (ADU) & 9.04 & 1612.8 \\
$\mu_{\text{max}}$ (ADU) & 39.52 & 71440 \\
Flux (ADU) & 1056 & -- \\
$m_{i}$ (mag) & 10.1 & 23.0 \\
\bottomrule
\end{tabular}
\end{threeparttable}
\end{table}

\begin{equation}
\frac{R_{90}}{R_{75}} < 0.01 \times R_{100} + 1.4
\end{equation}
\vspace{-5pt}
\begin{equation}
\frac{R_{100}}{R_{90}} < -0.085 \times m_i + 3.0441
\end{equation}
\vspace{-5pt}
\begin{equation}
\frac{m_i}{R_{75}} < \max(1.2479 \times m_i - 16.9164,\, 2.08)
\end{equation}
\vspace{-5pt}
\begin{equation}
\frac{R_{90}}{R_{75}} < -0.07 \times m_i + 2.7593
\label{eq:r_90_r_75_mag_i}
\end{equation}
\vspace{-5pt}
\begin{equation}
\frac{R_{75}}{R_{25}} < 0.2 \times m_i + 1.6021
\end{equation}
\vspace{-5pt}
\begin{equation}
\frac{R_{75}}{R_{25}} < -1.1 \times m_i + 27.1887
\end{equation}
\vspace{-5pt}
\begin{equation}
\log\left(\frac{R_{100}}{R_{75}}\right) < -2.3353 \times \log(m_i) + 7.5703
\end{equation}
\vspace{-5pt}
\begin{equation}
\log\left(\frac{R_{75}}{R_{25}}\right) < 0.6606 \times \log(m_i) - 0.2820
\end{equation}

\subsection{Examples of band filter plots}

\section{Classification tool}
\label{sec:appendix_class_ui}

\begin{figure*}[!htb]
\centering
\includegraphics[width=1\linewidth]{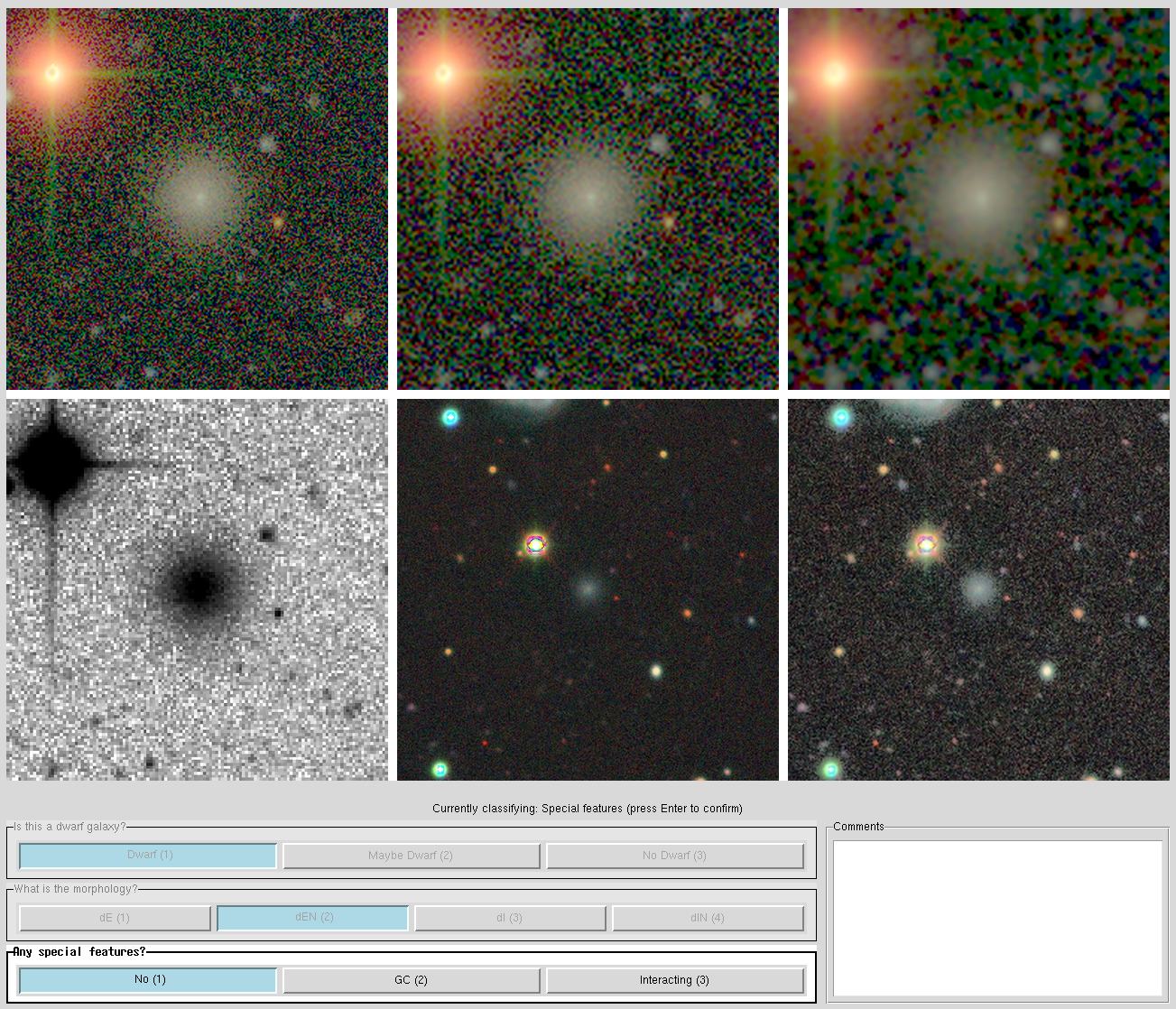}
\caption{Classification tool used to label our training data. The UI shows six different views of the same object. Top left: full resolution RGB image cutout, top center: 2$\times$2 binned RGB cutout, top right: 2$\times$2 binned and smoothed RGB cutout, bottom left: \emph{r}-band image cutout, bottom center: Legacy Survey cutout, bottom right: 2$\times$2 binned and CLAHE enhanced Legacy Survey cutout. The tool asks the questions: 1) Is this a dwarf galaxy? If yes or maybe: 2) What is the morphology? and 3) Any special feature? Finally, a comment box is available to note any additional features.}
\label{fig:appendix_class_ui}
\end{figure*}

In Figure \ref{fig:appendix_class_ui} we show the user interface (UI) for the classification tool we used to label our training data. It supports both single-view (single RGB cutout) and multi-view (six different versions of the same cutout) classification. The first question asks about the nature of the object, with the possible answers `dwarf' (1), `maybe dwarf' (0.5), and `no dwarf' (0). If the first question is answered with `dwarf' or `maybe dwarf', the user is asked about the morphology of the candidate. The possible choices are dwarf elliptical (dE), nucleated dwarf elliptical (dEN), dwarf irregular (dI), and nucleated dwarf irregular (dIN). The third question asks about special features, such as globular clusters or whether the object is disturbed due to gravitational interaction. In this work, we only utilized the answers to the first question about the dwarf nature, but we plan to use the additional classifications in future work.

\section{Visually classified non-dwarfs}
\label{sec:appendix_non_dwarfs}

\begin{figure*}[!htb]
\centering
\includegraphics[width=1\linewidth]{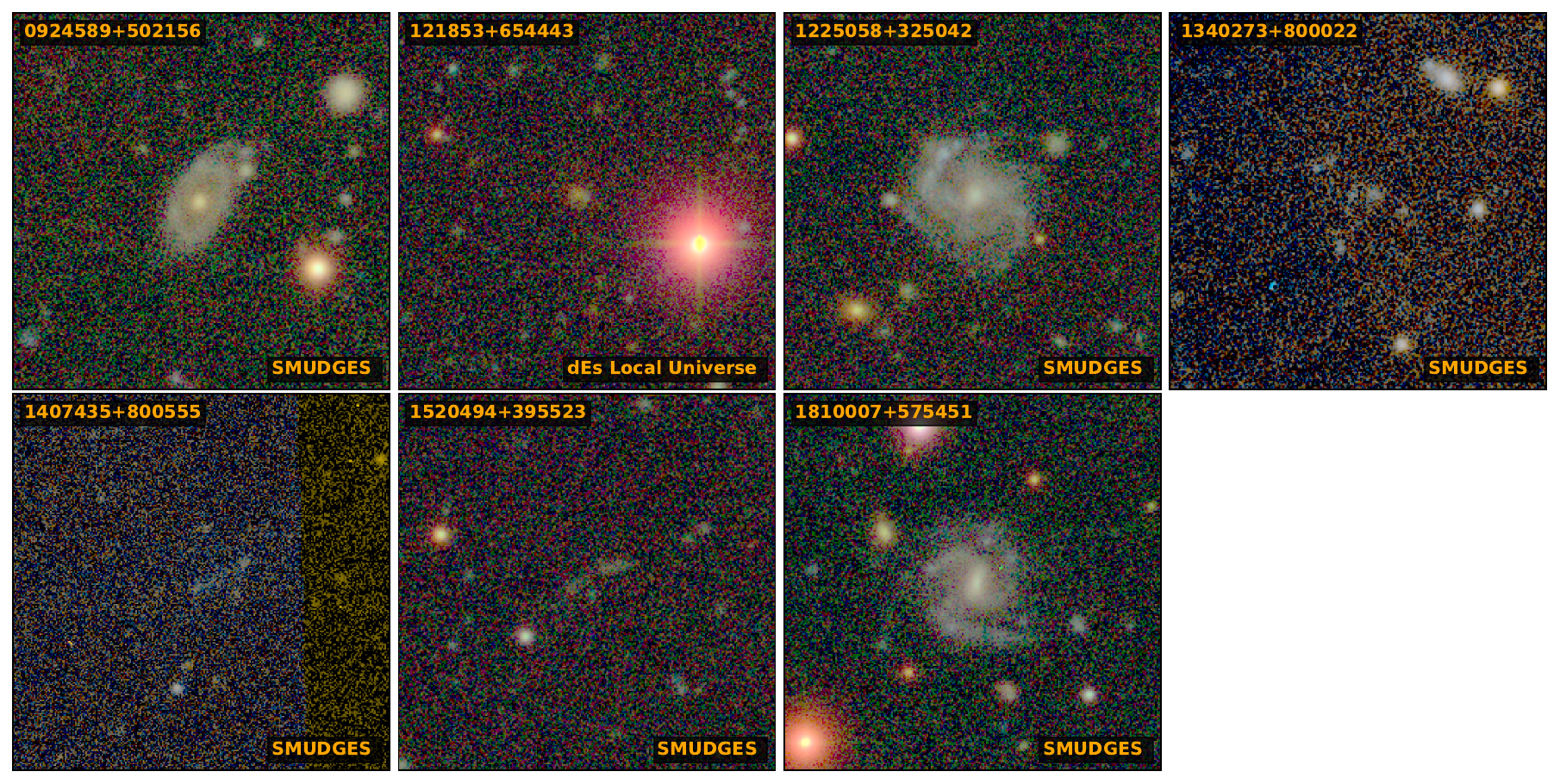}
\caption{RGB image cutouts of objects we unanimously visually classified as non-dwarfs while generating labels to train our dwarf classifier. These objects were reported as dwarf galaxy candidates in the literature. We show their ID in the top left corner and the source catalog in the bottom right corner of every cutout. The references for the source papers are SMUDGES \citep{2023AJ....166..185G,2023ApJS..267...27Z} and dEs in the Local Universe \citep{2023ApJS..265...57P}.}
\label{fig:appendix_non_dwarfs}
\end{figure*}

During the visual classification of dwarf galaxies presented in the literature, we found 7 objects that we unanimously classified as non-dwarfs during all three classification runs. For reference, we show image cutouts of these objects along with their source catalog in Figure \ref{fig:appendix_non_dwarfs}.

\section{Comparison between \textsc{MTO} and literature parameters}
\label{sec:appendix_param_comp}

\begin{figure*}[!htb]
\centering
\includegraphics[width=0.4\linewidth]{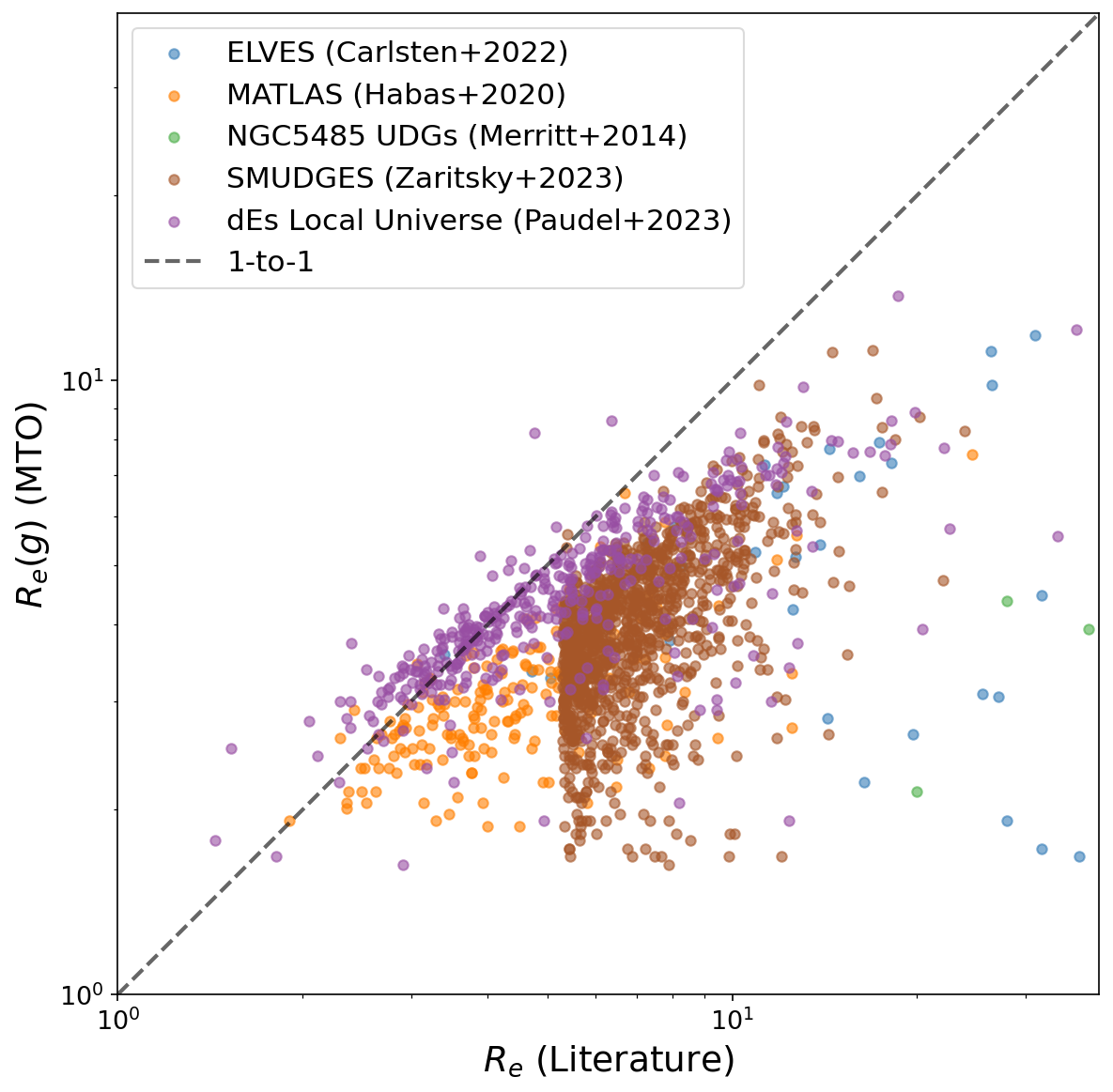} \\
\vspace{3mm}
\includegraphics[width=0.4\linewidth]{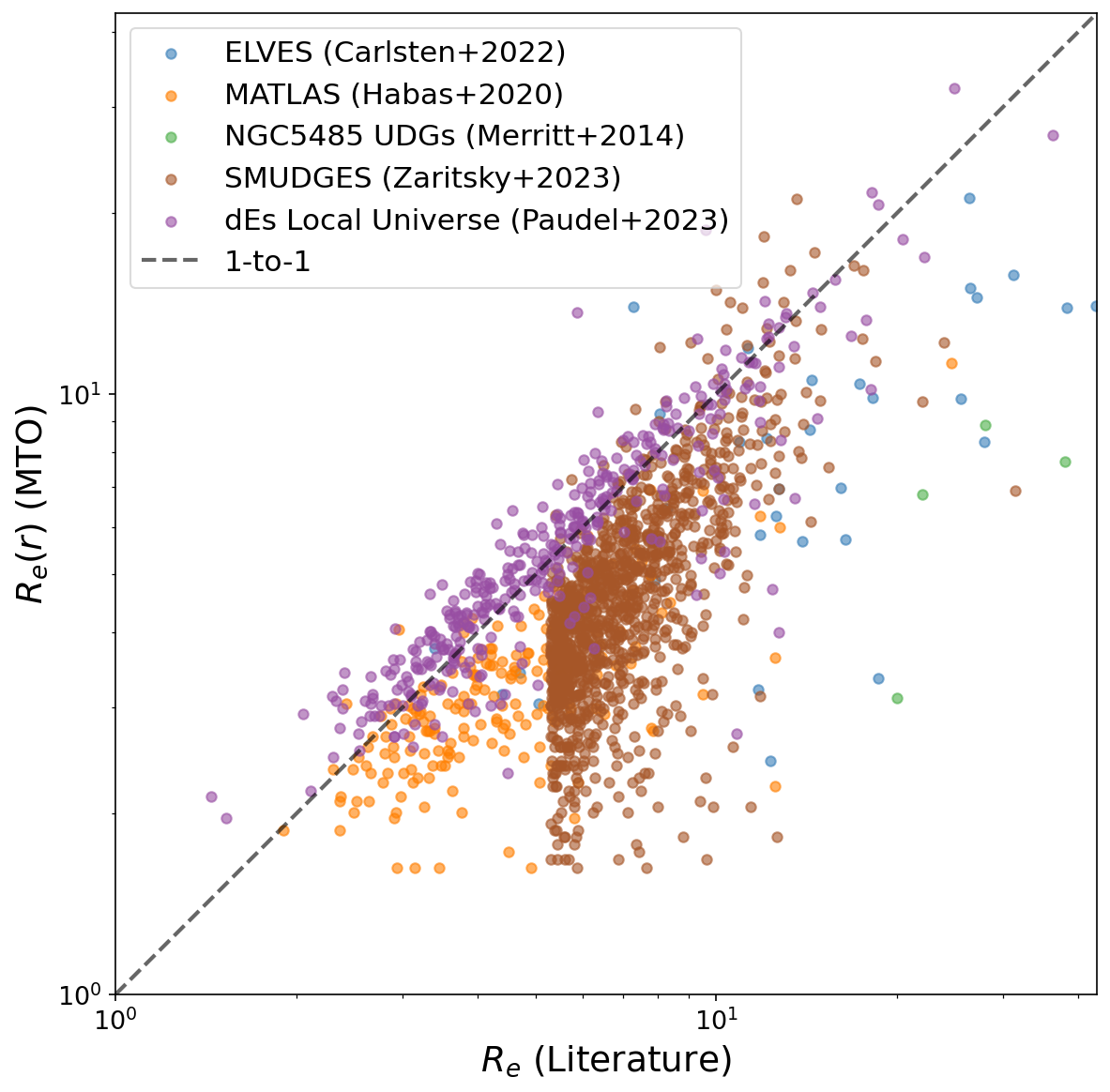} \\
\vspace{3mm}
\includegraphics[width=0.4\linewidth]{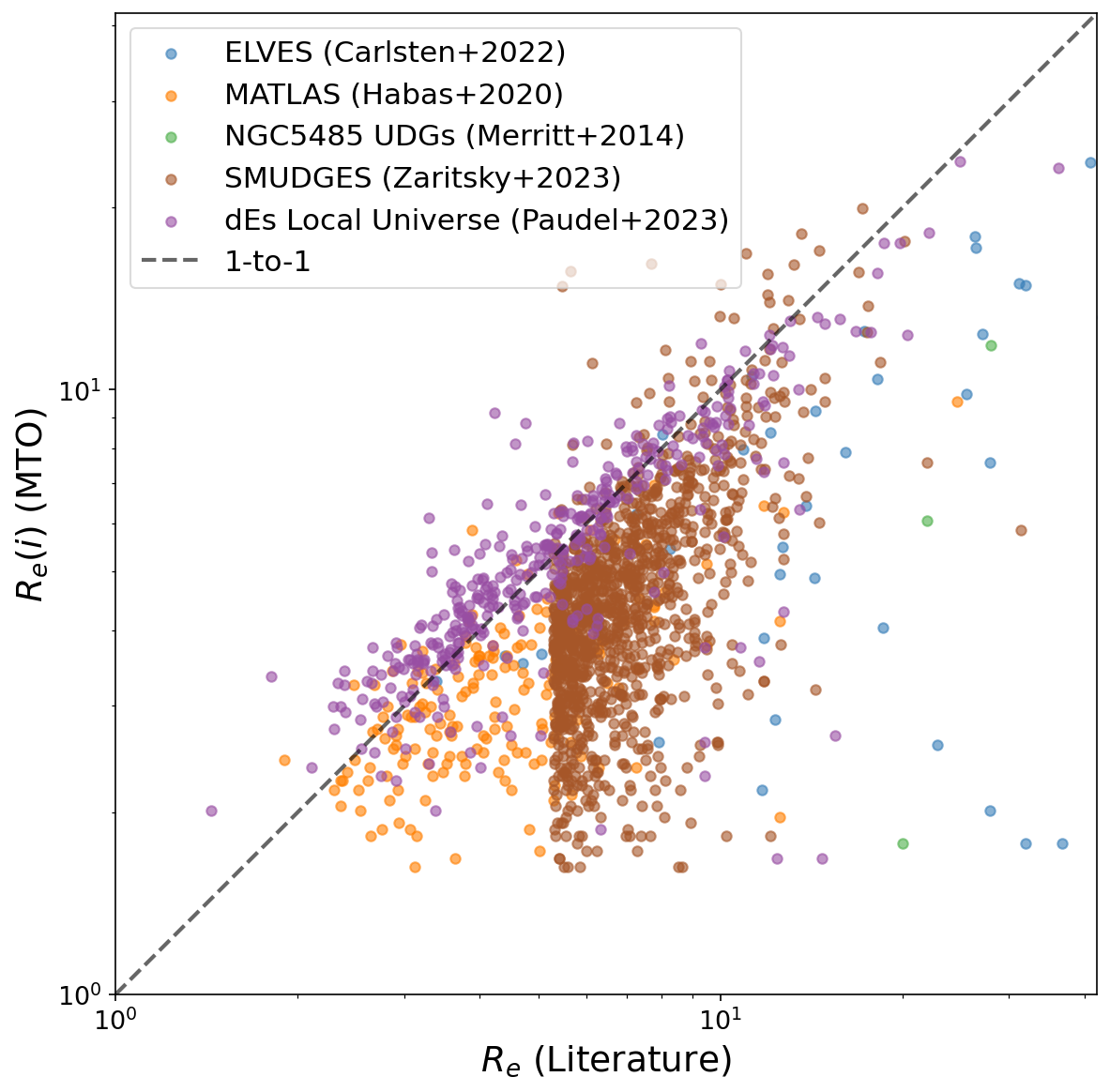} \\
\caption{Comparison between the effective radii $R_{e}$ measured by \textsc{MTO} and the values reported in the literature for the dwarf candidates we used to train our model. We show the literature values on the x-axis and the values measured by \textsc{MTO} on the y-axis. Both axes are shown on logarithmic scales. The data points are color coded by the survey they were identified in as shown in the figure legends. The one-to-one relation is plotted as the gray dashed line. We compare our measurements in the individual bands with the single values reported in the literature. Top: \emph{g}-band, center: \emph{r}-band, and bottom: \emph{i}-band.}
\label{fig:appendix_param_comp_reff}
\end{figure*}

\begin{figure*}[!htb]
\centering
\includegraphics[width=0.49\linewidth]{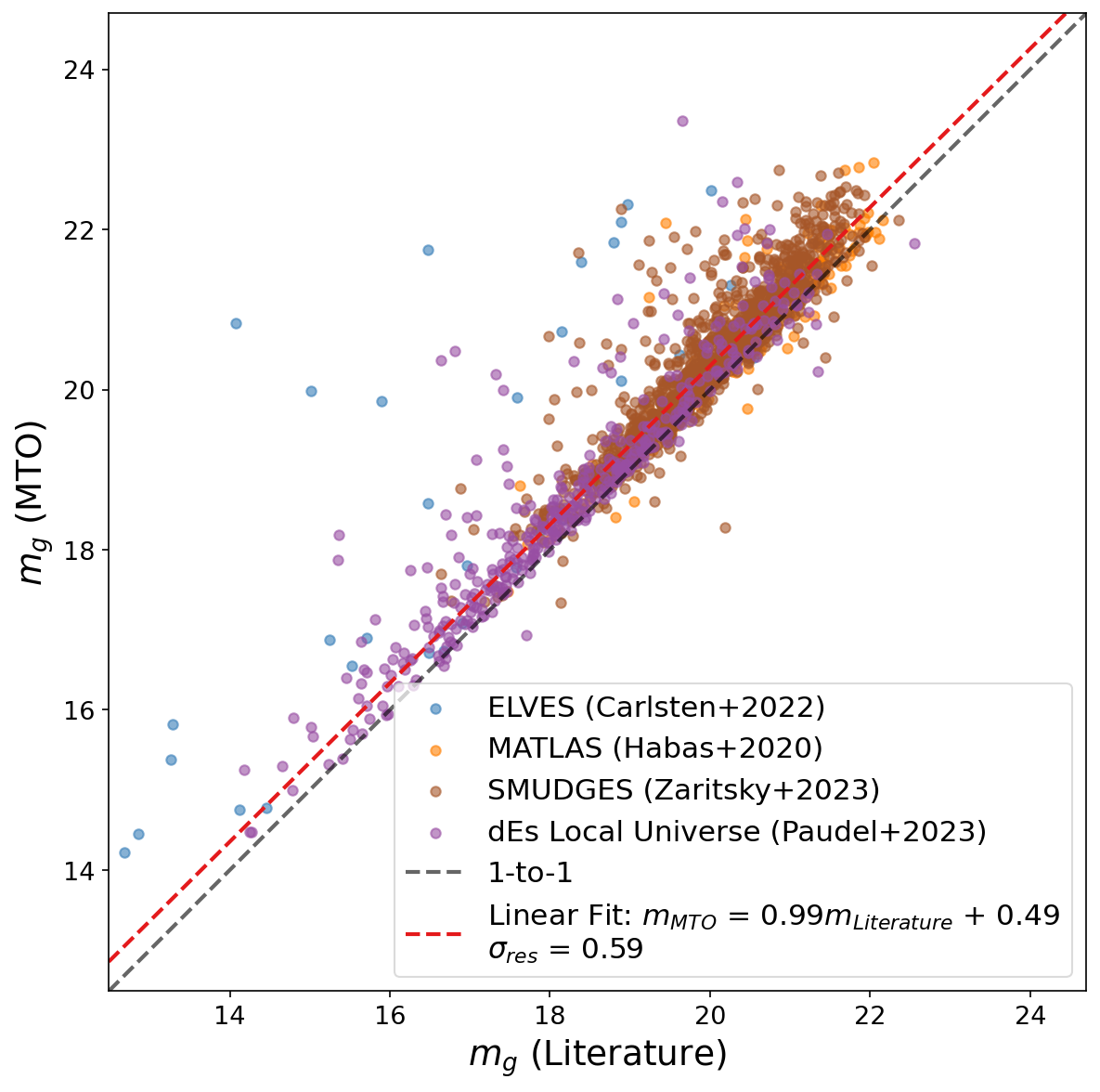}
\includegraphics[width=0.49\linewidth]{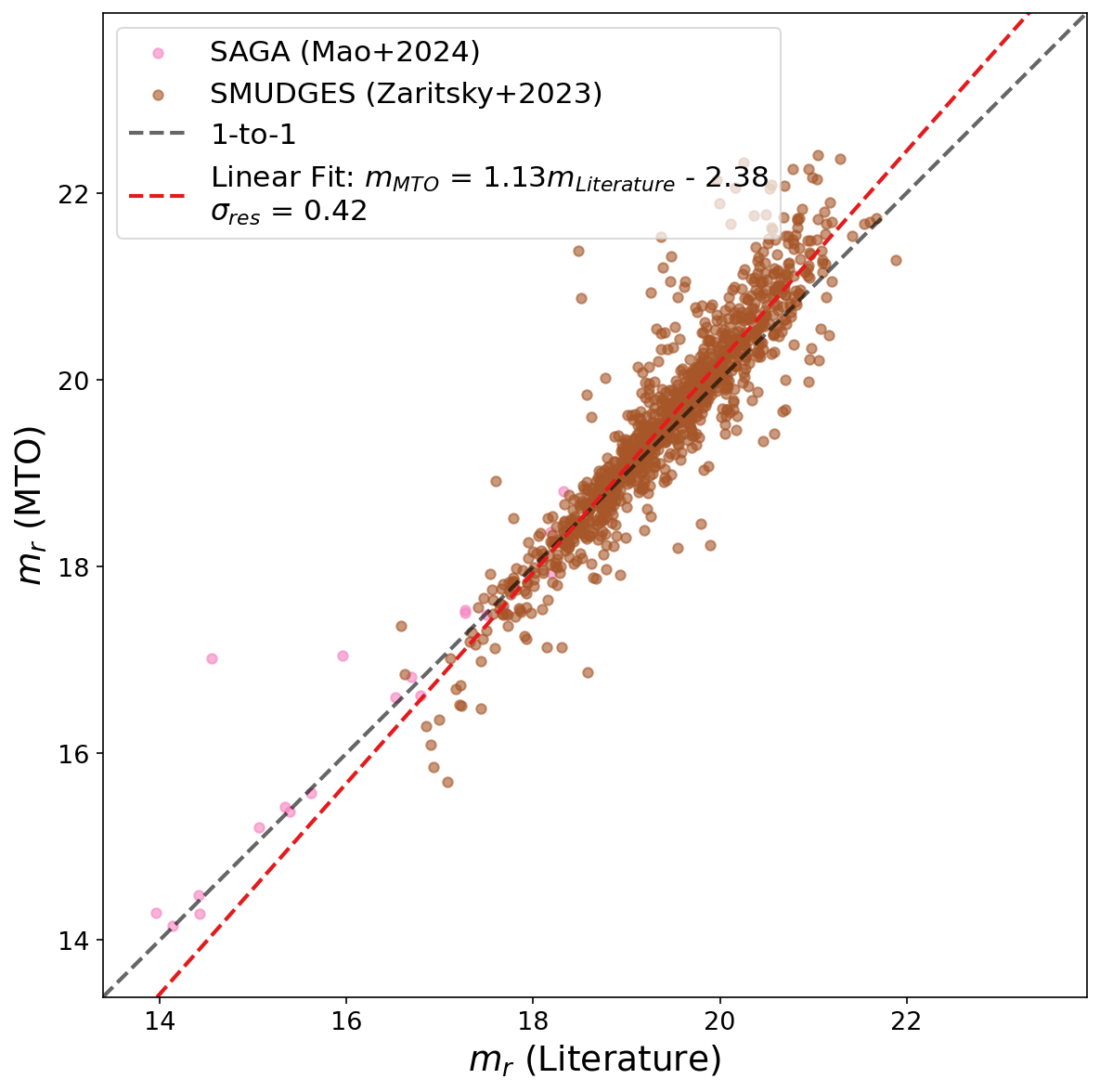}
\includegraphics[width=0.49\linewidth]{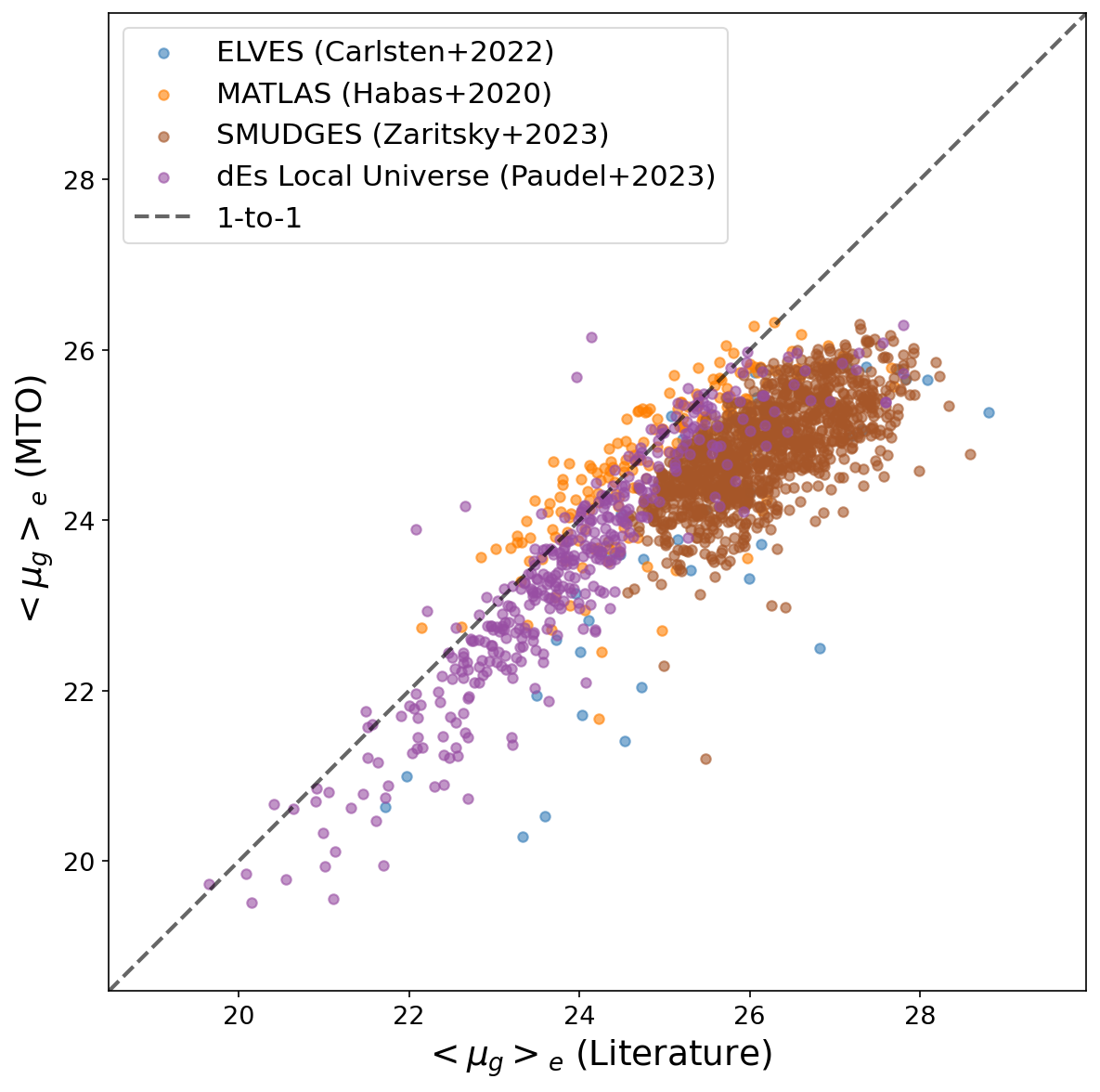}
\includegraphics[width=0.49\linewidth]{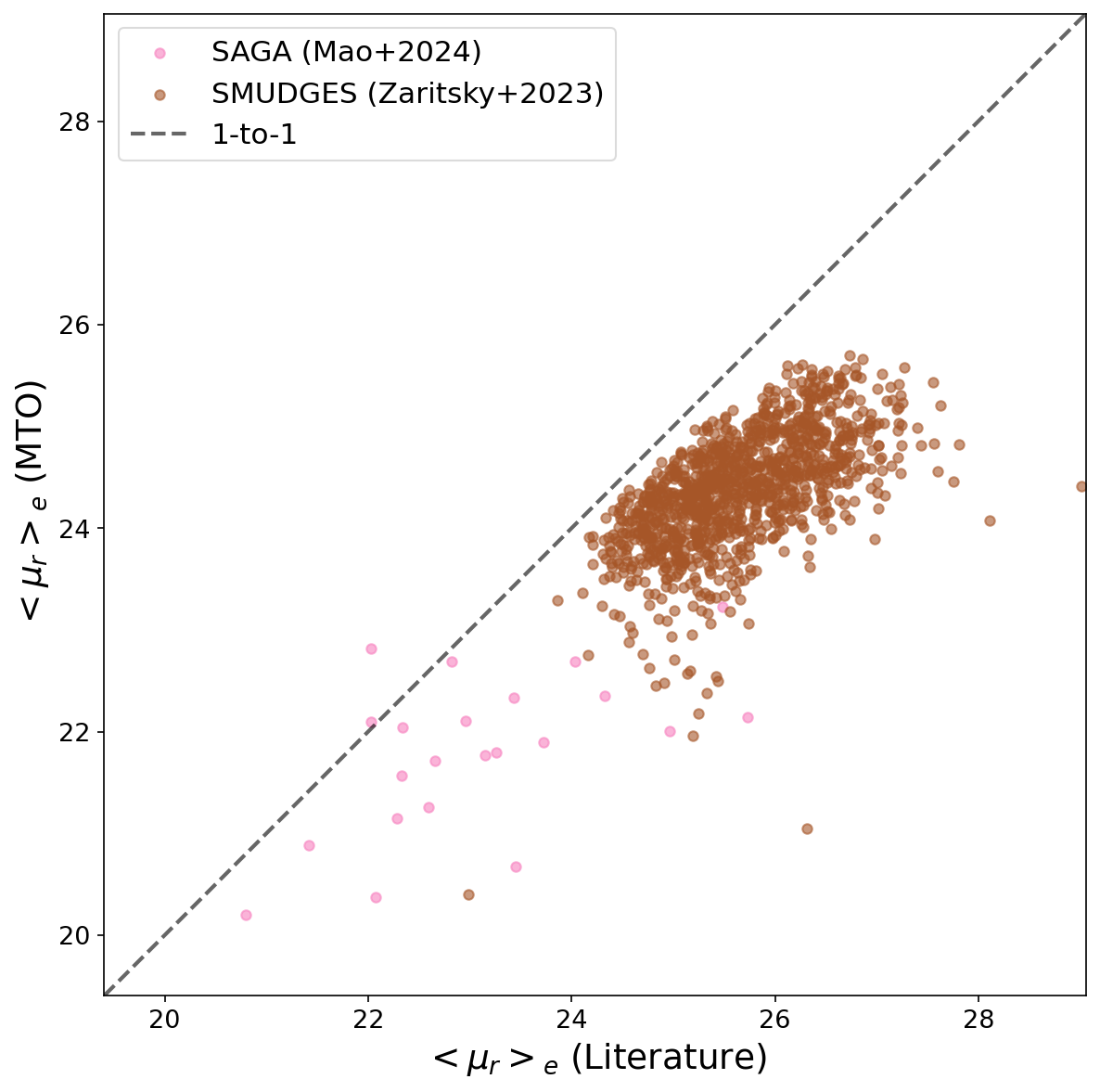}
\caption{Comparison between the apparent magnitudes $m$ (top row) and the mean effective surface brightnesses $<\mu>_{e}$ (bottom row) derived from the \textsc{MTO} measurements and the values reported in the literature for the dwarf candidates we used to train our model. We show the literature values on the x-axis and the values measured by \textsc{MTO} on the y-axis. The data points are color coded by the survey they were identified in as shown in the figure legends. The one-to-one relation is plotted as the gray dashed line. The red dashed line in the top row plots shows a linear fit to the data that can be used to correct the magnitude offset from the one-to-one relation. We report the standard deviation of the residuals with respect to this linear fit $\sigma_{res}$ in the legend. Left column: \emph{g}-band, right column: \emph{r}-band.}
\label{fig:appendix_param_comp_mag_mu}
\end{figure*}

This section compares \textsc{MTO}-derived parameters with published literature values for the dwarf galaxy candidates used in our training sample. In Figure \ref{fig:appendix_param_comp_reff} we show the comparison for the effective radius $R_{e}$. We note that $R_{e}$ is mostly underestimated by \textsc{MTO}, especially towards larger radii. We attribute this discrepancy to the fact that \textsc{MTO} does not calculate $R_{e}$ via a radial profile, but sorts the pixels, regardless of their location from the center of the object to calculate the effective radius. In Figure \ref{fig:appendix_param_comp_mag_mu} we show the comparisons for the absolute magnitude $m$ and the mean effective surface brightness $<\mu>_{e}$ in the \emph{g} and \emph{r}-bands; corresponding \emph{i}-band parameters are not available in the literature sample. For the magnitudes, there is a positive bias for the \emph{g}-band, i.e., \textsc{MTO} systematically overestimates the magnitude. For the \emph{r}-band, there is a scaling bias, i.e., the data points follow a different slope than the one-to-one relation. These two biases can be corrected via simple linear shifts. Therefore, we report the corrected along with the measured values for $m_{g}$ and $m_{r}$ in the final catalog. The standard deviations of the residuals after this correction are $\sigma_{res} (g)$ = 0.59 and $\sigma_{res} (r)$ = 0.42. Due to the discrepancy in the determination of $R_{e}$, $<\mu>_{e}$ is also largely underestimated but cannot be addressed via a linear correction.

\end{appendix}

\end{document}